\documentclass[fleqn,usenatbib]{mnras}

\usepackage{newtxtext,newtxmath}

\usepackage[T1]{fontenc}
\usepackage{ae,aecompl}

\usepackage{graphicx}	
\usepackage{amsmath}	
\usepackage{amssymb}	
\usepackage{comment}
\usepackage{tabularx}

\newcommand{\Msun}{\mbox{$M_{\odot}$}}

\title[Asteroseismology of 36 {\em Kepler} subgiants -- II]{Asteroseismology of 36 {\em Kepler} subgiants -- II. Determining ages from detailed modelling}

\author[Tanda Li et al.]{%
Tanda Li,$^{1,2,3,4}$\thanks{E-mail: tanda.li@sydney.edu.au}
Timothy R. Bedding,$^{1,2}$
J\o{}rgen Christensen-Dalsgaard,$^{2}$
Dennis Stello,$^{5,2}$
\newauthor
Yaguang Li$^{3,1,2}$ and 
Matthew A. Keen$^{1}$
\\
$^{1}$Sydney Institute for Astronomy (SIfA), School of Physics, University of Sydney, NSW 2006, Australia\\
$^{2}$Stellar Astrophysics Centre, Department of Physics and Astronomy, Aarhus University, Ny Munkegade 120, DK-8000 Aarhus C, Denmark\\
$^{3}$Department of Astronomy, Beijing Normal University, Beijing 100875, China\\
$^{4}$School of Physics and Astronomy, University of Birmingham, Birmingham B15 2TT, UK\\
$^{5}$School of Physics, University of New South Wales, NSW 2052, Australia\\
}

\date{Accepted XXX. Received YYY; in original form ZZZ}

\pubyear{2015}

\begin{document}
\label{firstpage}
\pagerange{\pageref{firstpage}--\pageref{lastpage}}
\maketitle

\begin{abstract}
Detailed modelling of stellar oscillations is able to give precise estimates for stellar ages, but the inferred results typically depend on the adopted model parameters used for the age inference. 
High-quality asteroseismic data with precise measurements of mixed modes are available for 36 {\em Kepler} subgiants. To obtain a handle on the robustness of the ages for these stars, 
we first study the dependencies of seismic ages on three model input parameters.
We find that inferred ages do not change systematically with the helium fraction ($Y$) or the mixing-length parameter ($\alpha_{\rm MLT}$) but depend strongly on the metallicity ([M/H]) of the model. 
The results indicate that age estimates of subgiants have less model dependence and hence are more reliable than those of main-sequence stars or red giants.
We then model individual oscillation frequencies of the same 36 {\em Kepler} subgiants, using observed metallicities, and obtain their ages with an average precision of $\sim 15\%$. 
The comparison with previous age estimates with different stellar codes or input physics show good agreement (mostly within 2$\sigma$). We hence suggest that seismology-determined ages of subgiants are not greatly model-dependent.
%
\end{abstract}





\section{Introduction}\label{sec:intro}

Asteroseismology using solar-like oscillations is a powerful method for studying stars.
Detailed modelling of individual oscillation frequencies is able to give precise estimates of stellar parameters, but these are model dependent.    
Obvious offsets have been found between the inferred parameters determined with different stellar codes and different input physics \citep[e.g.][]{2017ApJ...835..173S}. 
For example, the mixing length parameter ($\alpha_{\rm MLT}$) describes the efficiency of the convection and  varies for different types of stars, but has only been well calibrated for the Sun and a small number of stars \citep{2011A&A...535A..91D,Li2018sixredgiants,Ball2018redgiants,Bazot201818sco}.
Although hydrodynamic simulations \citep{2014MNRAS.445.4366T, 2015A&A...573A..89M} could provide $\alpha_{\rm MLT}$ for a given set of surface properties, their predictions for evolved stars have shown significant disagreements with the $\alpha_{\rm{MLT}}$ that is needed for stellar models to fit observations.
The value of the parameter is hence still under debate. 
The helium abundance is another uncertain model input. It is difficult to measure from stellar spectra and so an approximate value given by the Galactic element enrichment law is widely adopted in theoretical models. Recent findings showed that analysing acoustic glitches in seismic frequencies could constrain the surface helium abundances, but the analysis requires very high-quality asteroseismic data \citep{2014ApJ...790..138V,2017ApJ...837...47V}. 
Moreover, systematic offsets can also occur when chemical compositions are different from the typically assumed solar mixture, especially for the $\alpha$-enhanced stars \citep{2015MNRAS.447..680G}. Spectroscopy surveys have show that the [$\alpha$/Fe] of stars in the Galaxy spreads widely for a given [Fe/H]  \citep[e.g.][]{BuderGalahDR2}. Hence, model input metallicities derived with observed [Fe/H] and the solar composition are biased to some extent.

Subgiants are good laboratories to study stellar physics because of the presence of mixed modes.
When a star exhausts its core hydrogen and becomes a subgiant, the pressure and gravity modes (p- and g-modes) couple to form mixed oscillation modes that are shifted from the regular spacing (known as mode bumping).
The bumping of mixed dipole ($\ell =1$) modes in subgiants was firstly observed with ground-based telescopes in $\eta$~Boo \citep{1995AJ....109.1313K,2003AJ....126.1483K} and $\beta$~Hyi \citep{2007ApJ...663.1315B}, and then modelled as mixed modes \citep{1995ApJ...443L..29C,1996ApJ...456..798G,2003A&A...404..341D,2003A&A...399..243F}.
Recent space missions (CoRoT, {\em Kepler}, TESS) and new ground-based telescopes (e.g. SONG) have provided more examples with high-quality asteroseismic data. 
Compared with dwarfs, subgiants present mixed modes that provide additional information about the core. Moreover, their g-mode period spacings are indicators of stellar masses, unlike those of red giants \citep[Fig. 2]{2014A&A...572L...5M}. These advantages make subgiants a very special sample for studying stellar physics. 
As discussed by \citet{2011A&A...535A..91D}, the avoided-crossing features causing the bumping of mixed modes are sensitive to the stellar properties and can narrow down the dimensions of the model space. 
They used mixed modes to constrain the convective parameters of the CoRoT subgiant HD 49385.
Moreover, \citet{2012ApJ...745L..33B} inferred that the coupling strength of the $\ell = 1$ mixed modes is predominantly a function of stellar mass and appears to be independent of the metallicity.
\citet{2014MNRAS.445.2999T} found that the frequency differences between the mixed modes and the nearest p-mode ($d\nu_{m-p}$) are useful for constraining stellar models when they studied two {\em Kepler} subgiants KIC 6442183 ('Dougal') and KIC 11137075 ('Zebedee').
Using information of g-modes, 
\citet{bedding2014asteroseismology} suggested a new asteroseismic p-g diagram, in which the frequencies of the avoided crossings (corresponding to the so-called '$\gamma$ modes', i.e, the pure g modes had there been no mixing with the p modes) are plotted against the large separation of the p modes. The diagram can be used to make a first comparison with theoretical models to determine stellar parameters \citep[e.g.][]{campante++2011-subgiants-kic10273246-kic10920273}.
Asteroseismology has also been used to determine the chemical composition of subgiants. 
The modelling of KIC 7976303 indicates that the $\ell = 1$ mixed modes are able to tell the $\alpha$-elements enhancement of this star \citep{2015MNRAS.447..680G}.

With the high-quality data from {\em Kepler}, 
\cite{2011ApJ...732L...5C} and \citet{2010ApJ...723.1583M} reaffirmed the interesting possibility that detailed modelling of subgiants could provide a very precise determination of their age. 
Moreover, some previous studies also found that the age estimates are less model-dependent for subgiants than for main-sequence stars. For example, \citet{2003A&A...399..243F} and \citet{2010Ap&SS.328...73P} found that model solutions for the age of $\beta$ Hydri and $\mu$ Herculis are insensitive to convection parameters (mixing-length and overshooting parameters). 
\citet{2011A&A...535A..91D} adopted two different solar mixtures when modelling the CoRoT subgiant HD 49385 but obtained very similar inferred ages, which differed by only 0.5$\sigma$ ($\sim$0.05Gyr). Recently, we also found that the age determination of the SONG subgiant $\mu$ Herculis does not greatly depend on the initial chemical composition or the mixing-length parameter of the model \citep{LiTmuher}. Thus, subgiants turn out to be a very special sample, in which we could potentially obtain good ages. In this work, we modelled 36 {\em Kepler} subgiants for two purposes. One was to test the model dependency of the inferred age found in our previous work on $\mu$ Her by putting it into context of {\em Kepler} subgiants that cover from the early- to the late-subgiant phases. The other was to give good age estimates for all 36 stars analysed in this paper.
The rest of the paper is organised as follow. We describe adopted observational inputs in Section \ref{sec:obs}. We then introduce the physics of theoretical models and the details of the grid modelling in Section \ref{sec:model}, and present the method of interpolating oscillation modes in Section \ref{sec:interpolation}. The method and procedure of fitting are mentioned in Section \ref{sec:fitting}. We analyse and discuss the dependencies of age determinations on three model inputs ($Y_{\rm init}$, [Fe/H], and $\alpha_{\rm MLT}$) in Section \ref{sec:dependence}. The ages of 36 subgiants are estimated in Section \ref{sec:results}. Section \ref{sec:p-g-diagram} includes discussions about the 'p-g diagram'. Lastly, we summarise our results and give conclusions in Section \ref{sec:conclusions}.

\section{Observations of 36 subgiants}\label{sec:obs}

\begin{table*}
	\centering
	\caption{Observed stellar parameters of 36 {\em Kepler} subgiants.}
	\label{tab:obs}
	\begin{tabular}{rlrrrc} 
		\hline
		KIC & $T_{\rm{eff}}$ (K) & $\Delta \nu$ ($\mu$Hz)& $\nu_{\rm{max}}$ ($\mu$Hz)& [M/H] (dex) & Ref. for [M/H]$^{a}$ \\
        \hline
   2991448 &  $5623^{+92}_{-67}$ &  61.7 &  1138   & $-0.10\pm0.08$ &  2  \\
   3852594 &  $6296^{+82}_{-75}$ &  52.3 &   873   & $-0.40\pm0.15$ &  1  \\
   4346201 &  $6058^{+90}_{-72}$ &  56.2 &  1020   & $-0.21\pm0.08$ &  2  \\
   5108214 &  $5799^{+78}_{-78}$ &  41.2 &   700   & $ 0.17\pm0.08$ &  2  \\
   5607242 &  $5485^{+81}_{-81}$ &  40.5 &   665   & $-0.06\pm0.08$ &  2  \\
   5689820 &  $5037^{+70}_{-81}$ &  41.1 &   680   & $ 0.22\pm0.08$ &  2  \\
   5955122 &  $5877^{+70}_{-88}$ &  49.4 &   854   & $-0.20\pm0.08$ &  2  \\
   6064910 &  $6376^{+83}_{-76}$ &  44.4 &   826   & $-0.27\pm0.08$ &  2  \\
   6370489 &  $6184^{+80}_{-80}$ &  51.6 &   893   & $-0.40\pm0.08$ &  2  \\
   6688822 &  $5593^{+82}_{-74}$ &  47.8 &   774   & $ 0.34\pm0.08$ &  2  \\
   6442183 &  $5702^{+76}_{-76}$ &  64.6 &  1068   & $-0.20\pm0.08$ &  2  \\
   6693861 &  $5626^{+67}_{-101}$ &  47.1 &  753   & $-0.37\pm0.08$ &  2  \\
   6766513 &  $6227^{+81}_{-75}$ &  51.1 &   872   & $-0.17\pm0.08$ &  2  \\
   7174707 &  $5168^{+72}_{-92}$ &  47.2 &   803   & $ 0.07\pm0.15$ &  1  \\
   7199397 &  $5903^{+79}_{-79}$ &  38.7 &   632   & $-0.13\pm0.08$ &  2  \\
   7448374 &  $5626^{+84}_{-75}$ &  44.3 &   706   & $ 0.16\pm0.08$ &  2  \\
   7747078 &  $5903^{+88}_{-61}$ &  53.6 &   905   & $-0.24\pm0.08$ &  2  \\
   7976303 &  $6079^{+72}_{-90}$ &  51.1 &   832   & $-0.50\pm0.08$ &  2  \\
   8524425 &  $5543^{+116}_{-58}$ &  59.9 &  1040  & $ 0.07\pm0.08$ &  2  \\
   8702606 &  $5529^{+74}_{-90}$ &  39.4 &   631   & $-0.15\pm0.08$ &  2  \\
   8738809 &  $6045^{+72}_{-72}$ &  49.5 &   875   & $ 0.20\pm0.08$ &  2  \\
   9414381 &  $5861^{+96}_{-79}$ &  48.5 &   870   & $ 0.09\pm0.08$ &  2  \\
   9512063 &  $5838^{+78}_{-78}$ &  49.5 &   860   & $-0.20\pm0.08$ &  2  \\
  10018963 &  $6177^{+73}_{-92}$ &  55.2 &   977   & $-0.25\pm0.08$ &  2  \\
  10147635 &  $5941^{+89}_{-71}$ &  37.2 &   587   & $-0.02\pm0.08$ &  2  \\
  10273246 &  $6269^{+109}_{-140}$ &  48.3 & 789   & $ 0.21\pm0.15$ &  1  \\
  10593351 &  $5754^{+86}_{-77}$ &  31.4 &   495   & $ 0.15\pm0.08$ &  2  \\
  10920273 &  $5365^{+85}_{-85}$ &  57.1 &   969   & $-0.16\pm0.15$ &  1  \\
  10972873 &  $5705^{+85}_{-77}$ &  58.1 &   987   & $-0.08\pm0.08$ &  2  \\
  11026764 &  $5636^{+84}_{-75}$ &  50.2 &   850   & $ 0.04\pm0.08$ &  2  \\
  11137075 &  $5510^{+91}_{-58}$ &  65.0 &  1082   & $-0.13\pm0.08$ &  2  \\
  11193681 &  $5575^{+83}_{-75}$ &  42.6 &   702   & $ 0.23\pm0.08$ &  2  \\
  11395018 &  $5753^{+103}_{-126}$ &  47.5 & 805   & $ 0.02\pm0.15$ &  1  \\
  11414712 &  $5622^{+84}_{-75}$ &  43.7 &   697   & $-0.14\pm0.08$ &  2  \\
  11771760 &  $5796^{+78}_{-78}$ &  32.3 &   539   & $-0.05\pm0.08$ &  2  \\
  12508433 &  $5303^{+84}_{-73}$ &  44.8 &   762   & $ 0.23\pm0.08$ &  2  \\
  \hline
     \multicolumn{6}{l}{Observed $T_{\rm{eff}}$, $\Delta \nu$, and $\nu_{\rm{max}}$ are from \citet{2017ApJS..229...30M}.}\\  
      \multicolumn{6}{l}{$^a$ References: 1--\citet{2017ApJS..229...30M}; 2--\citet{2015ApJ...808..187B}.} \\  
	\end{tabular}
\end{table*}
We selected the 36 {\em Kepler} subgiants that have good asteroseismic data, as studied by Li Y. et al. (submitted, Paper~I). They selected all {\em Kepler} targets observed in short cadence mode for at least 110 days. They fitted the modes with Lorenztian profiles in the Bayesian framework with a Markov Chain Monte Carlo (MCMC) algorithm \citep{2010CAMCS...5...65G,2013PASP..125..306F}. The significance of each mode was characterized by a Bayes factor, $\ln K$, the logarithm of the probability of detection over null detection. Modes with $\ln K>1$ were selected for this work, suggesting a positive detection of the signal according to the \cite{kass&raftery} scale.


The observed global stellar parameters of the 36 stars are listed in Table~\ref{tab:obs}. We show the effective temperature ($T_{\rm{eff}}$), the large separation ($\Delta \nu$), and the frequency of the maximum amplitude ($\nu_{\rm{max}}$) given by \citet{2017ApJS..229...30M}, although we note that $\Delta \nu$ and $\nu_{\rm{max}}$ were not used in our model fitting. The metallicities are mostly from \citet{2015ApJ...808..187B}, who analysed medium-resolution spectra and reported the total abundance of heavy elements ([M/H]) of these stars. The advantage of using [M/H] instead of [Fe/H] is a reduction of the systematic effects caused by the differences in chemical mixtures. Detailed discussions about the effect of different composition on the estimates of stellar mass and age have been given by \citet{2015MNRAS.447..680G}.  

\section{Theoretical models}\label{sec:model}

\subsection{Stellar models and input physics}

We used Modules for Experiments in Stellar Astrophysics
(\textsc{MESA}, version 8118) to compute stellar evolutionary tracks and
structural models. \textsc{MESA} is an open-source stellar evolution package
that is undergoing active development. Detailed descriptions of the physics
and the numerical methods can be found in \citet{2011ApJS..192....3P,2013ApJS..208....4P, 2015ApJS..220...15P}.

We adopted the solar chemical mixture [$(Z/X)_{\odot}$ = 0.0229]
provided by \citet{1998SSRv...85..161G}, because the
sound speed profile of the calibrated solar model with this mixture shows better agreement
with that given by helioseismology than those with other mixtures \citep[e.g.][]{2011ApJ...731L..42B}. 
The initial chemical abundances of hydrogen, helium, and heavy elements
($X_{\rm init}$, $Y_{\rm init}$, and $Z_{\rm init}$)
for each star were then calculated by: 
\begin{equation}
\log (Z_{\rm{init}}/X_{\rm{init}}) = \log (Z/X)_{\odot} + \rm{[M/H]}.  \\
\end{equation}
We used the \textsc{MESA} $\rho-T$ tables based on the 2005
update of OPAL EOS tables \citep{2002ApJ...576.1064R} and OPAL opacity for the
solar composition of \citet{1998SSRv...85..161G} supplemented by
low-temperature opacity \citep{2005ApJ...623..585F}. The MESA photosphere tables were used as the set of boundary conditions for modelling the atmosphere.
The mixing-length theory of convection was implemented and the mixing-length parameter is descried as $\alpha_{\rm MLT} = \ell_{\rm MLT}/H_p$, where $\ell_{\rm MLT}$ and $H_p$ refer to the mixing length and the pressure scale height.
The exponential scheme by \citet{2000A&A...360..952H} was adopted for the convective overshooting, where the diffusion coefficient in the overshoot region is given as  
\begin{equation}
{D_{\rm{OV}}} = {D_{\rm{conv,0}}}\exp \left( - \frac{{2(r-r_{0})}}{{(f_0 + f_{\rm{ov}}){H
_{p}}}}\right).
\end{equation}
Here, $D_{\rm{conv,0}}$ is the diffusion coefficient from the mixing-length theory at a user-defined location near the Schwarzschild boundary. The switch from convection to overshooting is set to occur at $r_0$. To consider the step taken inside the convective region, $(f_0 + f_{\rm{ov}}){H_{p}}$ is used. Both $f_0$ and $f_{\rm{ov}}$ are free parameters in \textsc{MESA}  and we adopted a fixed $f_{\rm{ov}}$ at 0.018 and  $f_{\rm 0}$ = 0.5$f_{\rm{ov}}$ for all of the computations in this work. Note that we considered the overshooting at boundaries of convective cores and hydrogen-burning shells but not at the bottom of convective envelope.     
The mixing profiles at the boundary of the convective region given by the exponential overshooting can be found in \citet{2018A&A...614A.128P}. The \textsc{MESA} inlist used for the computation is available on \url{https://github.com/litanda/mesa_inlist}.   

Theoretical stellar oscillations were calculated with the \textsc{GYRE} code (version 4.4), which was developed by \citet{2013MNRAS.435.3406T}. For each structural model generated by \textsc{MESA}, we computed $\ell$ = 0, 1, and 2 modes by solving the adiabatic stellar pulsation equations. 

\subsection{The Surface Correction} 
Due to the improper modelling of the surface layers, theoretical oscillation frequencies have systematic offsets from observations, which is known as the surface effect.
For correcting it, we used the expression described by \citet{2014A&A...568A.123B}. 
The formula is a combination of an inverse 
term and a cubic term:     
\begin{equation}\label{eq:ball_sc}
\delta \nu  = ({a_{-1}}{(\nu /{\nu _{ac}})^{-1}} + {a_3}{(\nu
/{\nu _{ac}})^3})/I,
\end{equation}
where $I$ is the normalised mode inertia, ${a_{-1}}$
and ${a_3}$ are coefficients determined with observed ($\ell$ = 0, 1, and 2) and theoretical modes to obtain the best frequency correction ($\delta \nu$). 
The acoustic cut-off frequency ${\nu _{\rm{ac}}}$ is derived from the scaling relation \citep{1991ApJ...368..599B}:
\begin{equation}
\frac{\nu _{\rm{ac}}}{\nu _{\rm{ac, \odot}}} \approx \frac{g}{{{g_
\odot}}}\left(\frac{{{T_{\rm{eff}}}}}{{{T_{\rm{eff, \odot }}}}}\right)^{-1/2}.
\end{equation}
Here we take $\nu _{\rm{ac, \odot}}$ = 5000 $\mu {\rm Hz}$ following \citet{2014A&A...568A.123B}. Solar values of effective temperature and surface gravity are $T_{\rm{eff},\odot}$ = 5777 K and $\log g$$_{\odot}= 4.44$ \citep{cox2015allen}.


\subsection{Model Grids}\label{subsec:grid1}

To study the dependencies of inferred stellar parameters on the model inputs ($Y_{\rm init}$, $[M/H]$, and $\alpha_{\rm MLT}$),
we calculated three different model grids, which we refer to as `Grid-Y', `Grid-Z', and `Grid-$\alpha$'. 
For each grid, we adjusted one of the three parameters and fixed the other two. The details of the three grids can be seen in Table~\ref{tab:grid1}.

\begin{figure}
	\includegraphics[width=\columnwidth]{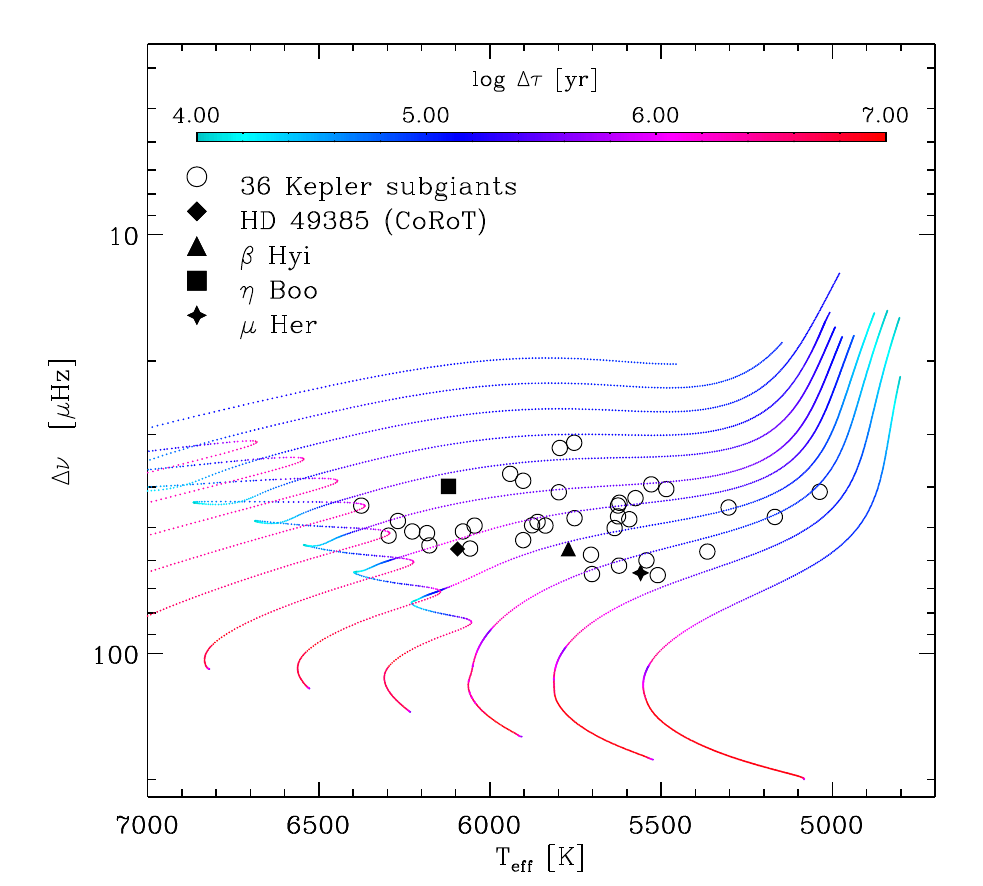}
    \caption{Grid models and locations of 36 {\em Kepler} subgiants on the $T_{\rm{eff}}$-$\Delta \nu$ diagram. Each small dot on the evolutionary track represent one oscillating model that was computed by \textsc{MESA} and \textsc{GYRE}. The colour of the dots indicates the time steps between consecutive models. Open symbols represent 36 {\em Kepler} stars, and filled symbols indicate one CoRoT star and three subgiants observed by ground-based telescopes.}
    \label{fig:hr}
\end{figure}

We calculated stellar models from the Hayashi line to the bottom of the red-giant branch (the computation stopped when $\nu_{\rm{max}}$ went below $\sim$ 200 $\mu$Hz) and saved the structural model at every time step after the zero-age main sequence. For each evolutionary track, $\sim$900 structural models were saved, including $200-300$ at the main-sequence stage and $600-700$ at the subgiant phase. 
The evolution time step after central hydrogen exhaustion was mainly controlled by the set-up tolerances on changes in surface effective temperature and luminosity, which are $\Delta \log T_{\rm eff}$ = 0.001 and $\Delta \log L$ = 0.004. We show our evolutionary tracks in Figure~\ref{fig:hr}, where the time steps are indicated by the colour-code. Typical time steps ranged from $10^6$ to $10^5$ years for the early to the late subgiant phases.  

\begin{table*}
	\centering
	\caption{Three grids calculated in a mass range of 0.80 -- 1.90$M_{\odot}$ with a mass step of 0.02$M_{\odot}$ for studying the model dependencies.}
	\label{tab:grid1}
	\begin{tabular}{llll} 
		\hline
		Grid name & Adjusted input & Grid points & Fixed inputs  \\
		\hline
		Grid-Y&Helium & $Y_{\rm{init}}$ = 0.25, 0.27, 0.29, 0.31, 0.33 & [M/H] = 0.0, $\alpha_{\rm{MLT}}$ = 1.9, GS98, Photosphere Tables\\
		Gird-Z&Heavy elements & [M/H] = -0.5, -0.3, 0.0, 0.3, 0.5 & $Y_{\rm{init}}$ = 0.29, $\alpha_{\rm{MLT}}$ = 1.9, GS98, Photosphere Tables \\
		Grid-$\alpha$&The mixing-length parameter &  $\alpha_{\rm{MLT}}$ = 1.7, 1.9, 2.1 & $Y_{\rm{init}}$ = 0.29, [M/H] = 0.0, GS98, Photosphere Tables \\
		\hline
        \end{tabular}
\end{table*}

\section{Interpolation}\label{sec:interpolation}

Computing a grid of oscillation models is time-consuming for subgiants because their mixed modes vary rapidly with age. Hence, interpolating models as a function of the age will significantly improve the efficiency of detailed modelling. 

We first considered interpolation for five key stellar parameters, namely, the effective temperature, the gravity, the radius, the luminosity, and the helium-core mass. 
During the subgiant phase, a star cools and becomes larger in radius. Meanwhile, its helium core mass keeps increasing because it burns hydrogen in the shell. These five parameters all change monotonically with the age and are hence not difficult to interpolate. 
For the typical time steps as presented in Figure~\ref{fig:hr}, these parameters change approximately linearly in the small time interval between consecutive models and we hence interpolated them with a linear function. 

\begin{figure}
    \includegraphics[width=1.0\columnwidth]{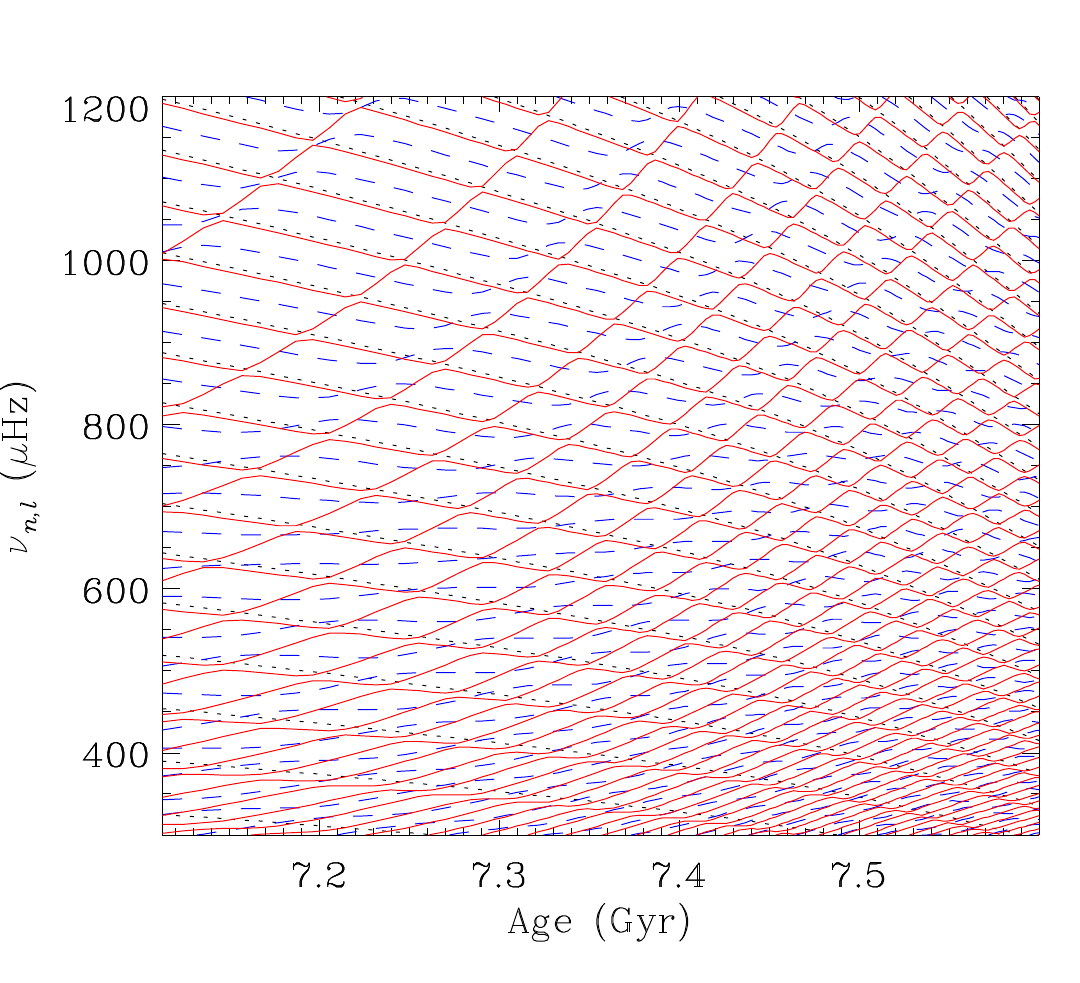}
    \caption{The evolution of theoretical oscillation frequencies during the subgiant phase of a 1.1$M_{\odot}$ model. Black dotted, blue dashed, and red solid lines indicate $\ell$ = 0, 1, and 2 modes.} 
    \label{fig:mode_features}
\end{figure}

We then investigated interpolation of the oscillation modes, which is challenging because of the rapid changes of mixed modes.   
As discussed by \citet[Fig. 13, 15]{2017A&ARv..25....1H}, the frequencies and inertias of $\ell = 0$ modes vary approximately linearly with time over small time intervals and are easy to deal with.
However, the evolution of mixed modes tends to be complicated because they have characteristics of both p- and g-modes. 
To work out a proper approach, we examined modes evolving from the early to the late subgiant phases ($\nu_{\rm max}$ = 1100 to 400 $\mu$Hz), as presented in Figure \ref{fig:mode_features}. 
We connected modes at different time with the same radial order, which is defined as $n_{\rm pg}$ = $n_{\rm p}$ where $n_{\rm p}$ and $n_{\rm g}$ are the acoustic- and gravity-wave winding numbers \citep{1974A&A....36..107S,1975PASJ...27..237O,2005PASJ...57..375T}. The radial order is invariant for a given mode as the star evolves.   
Unlike the $\ell = 0$ mode frequencies, which maintain roughly equal spacing, mixed-mode frequencies with adjacent $n_{\rm pg}$ undergo avoided crossings.
These modes have p-like behaviour in the envelope and the g-like behaviour in the core.
%
The evolution of the mode inertia also depends on the character of a mode. It changes approximately linearly with the age when the mode is predominately acoustic, but increases and then decreases when the mode goes through an avoided crossing.  
These features do not follow any analytical mathematical functions.
However, a frequency or a mode inertia at a given time strongly correlates with those before and after it, because the transition between mode characters is progressive with the age. Thus, we used cubic splines for interpolating frequencies and inertias of mixed modes.     

\subsection{Interpolation for $\ell = 1$ mixed modes}

\begin{figure}
    \includegraphics[width=\columnwidth]{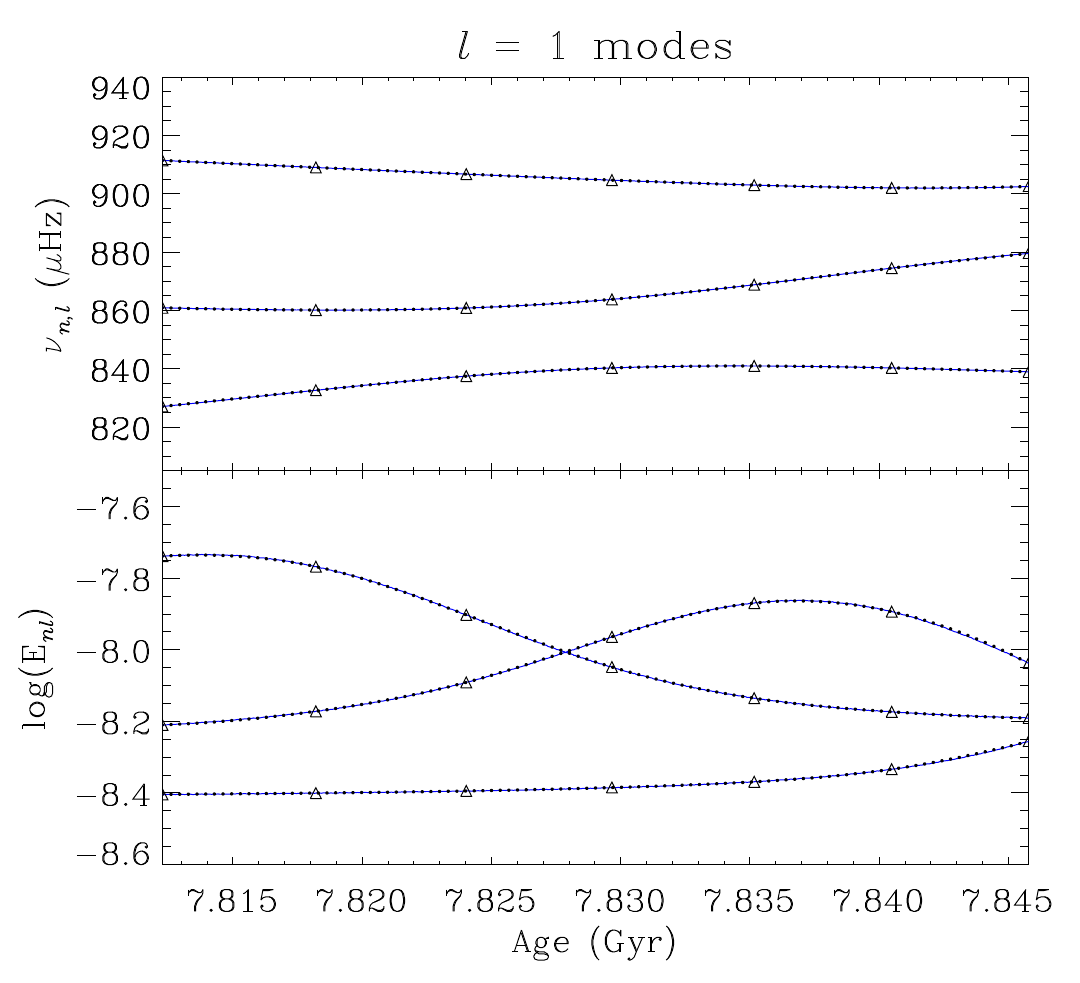}
    \includegraphics[width=\columnwidth]{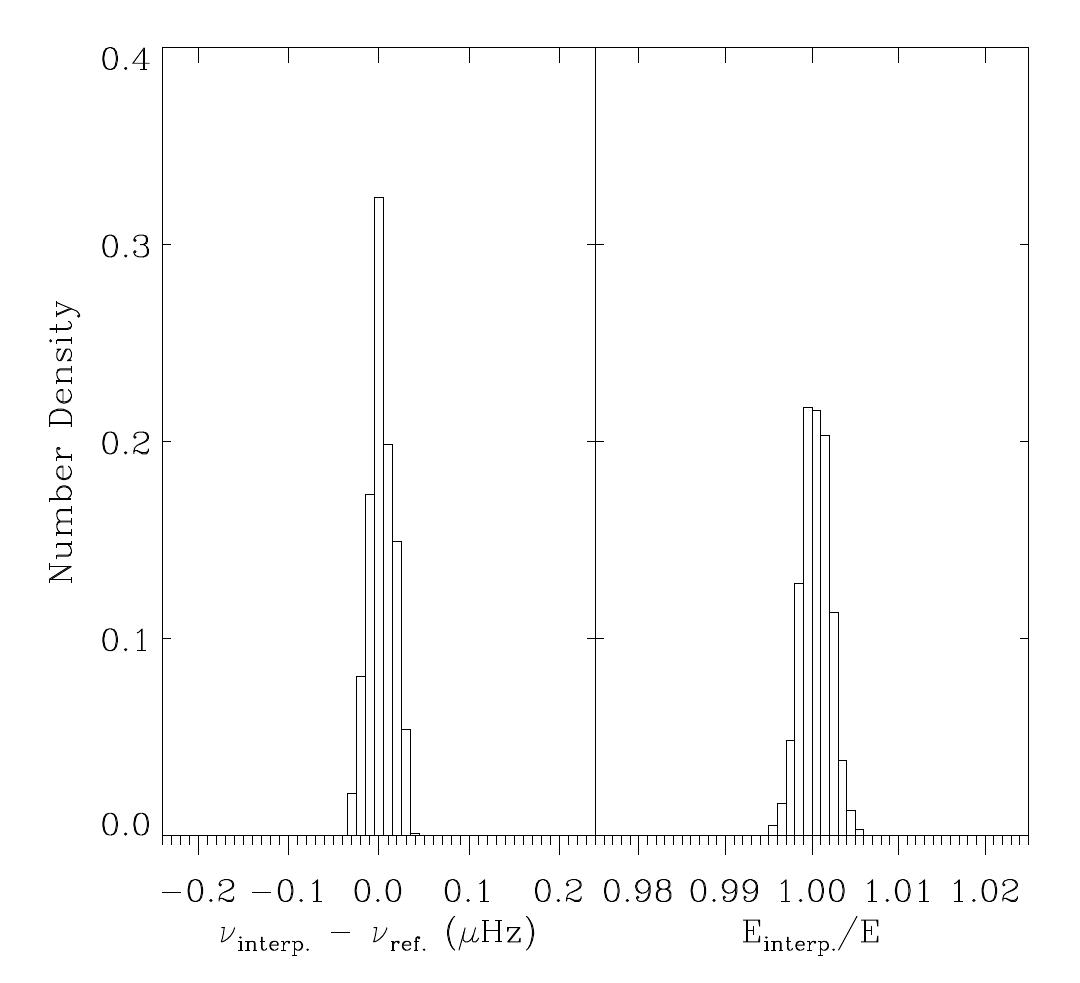}
    \caption{Top: interpolations for frequencies and inertias of three $\ell = 1$ modes of a 1.2$M_{\odot}$ model during the early subgiant phase ($\nu _{\rm max}$ = 1100 $\mu$Hz). Blue solid lines represent modes calculated with very small time steps ($\sim 10^5$ yr) as a check; black triangles indicate calculated modes with typical time steps of the our grid ($\sim 5 \times 10^6$ yr); small black dots are interpolated modes base on the cubic-spline interpolation. Bottom: histograms of the differences between interpolated and reference modes.} 
    \label{fig:inter_l1}
\end{figure}

The accuracy of the interpolation strongly depends on the sampling. We investigated whether the cubic spline gave reasonable interpolations for our grid models whose time steps ranged from $\sim 5 \times  10^6$yr early-subgiant to $\sim 5 \times 10^5$ yr for late-subgiant phases. To do this, we computed three reference models (1.0, 1.2, and 1.5M$_{\odot}$) with very small time steps ($10^5$ yr during the subgiant phase) as a check. We first plotted all reference modes in a small time interval, then picked out modes with the typical time step of the grid for interpolating; and lastly compared interpolated results with references. 
An example is illustrated in Figure ~\ref{fig:inter_l1}, where we show frequencies and inertias of three mixed modes at the early-subgiant phase ($\nu _{\rm max}$ = 1100 $\mu$Hz) of an 1.2$M_{\odot}$ models (solid lines). The typical time step for this mass and age in our grid is about $6 \times 10^6$ yr, and we therefore picked out one model every $6 \times 10^6$ yr (open triangles) and then interpolated with the cubic spline (small dots). Note that the mode inertia was interpolated on a logarithmic scale. To estimate the quality of the interpolation, we show at the bottom of Figure~\ref{fig:inter_l1} histograms of all the differences between interpolated modes and references spanning the range plotted in the top panel. 
The histograms are Gaussian-like with small widths and centred at 0.0 and 1.0, indicating good agreement.    
With this method, we examined dipole modes in the frequency range of 0.5 -- 1.5 $\nu_{\rm max}$ for different evolutionary stages ($\sim5 \times 10^6$ yr, $\sim1 \times 10^6$ yr, and $\sim5 \times 10^5$ yr for the early, the middle, and the late-subgiants). 
The interpolated $\ell = 1$ modes were found to agree well with the references. The differences between the references and interpolations were mostly below $\sim$0.1$\mu$Hz for the frequencies and $\sim$0.5\% for the mode inertias. The histograms generally had quite small widths ($\leqslant$0.05$\mu$Hz for the frequencies and $\leqslant$0.2\% for the inertias). 

\subsection{Interpolation for $\ell = 2$ mixed modes}

\begin{figure}
    \includegraphics[width=\columnwidth]{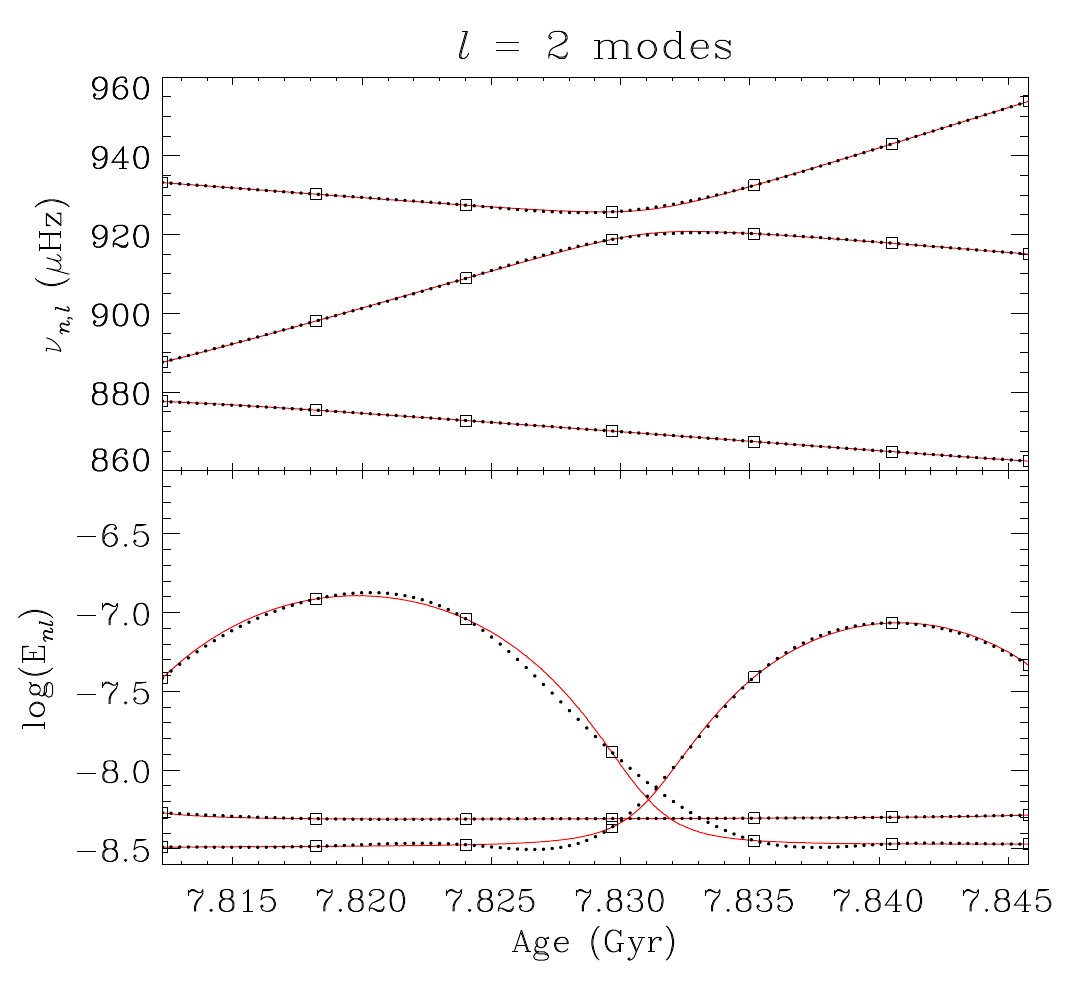}
    \includegraphics[width=\columnwidth]{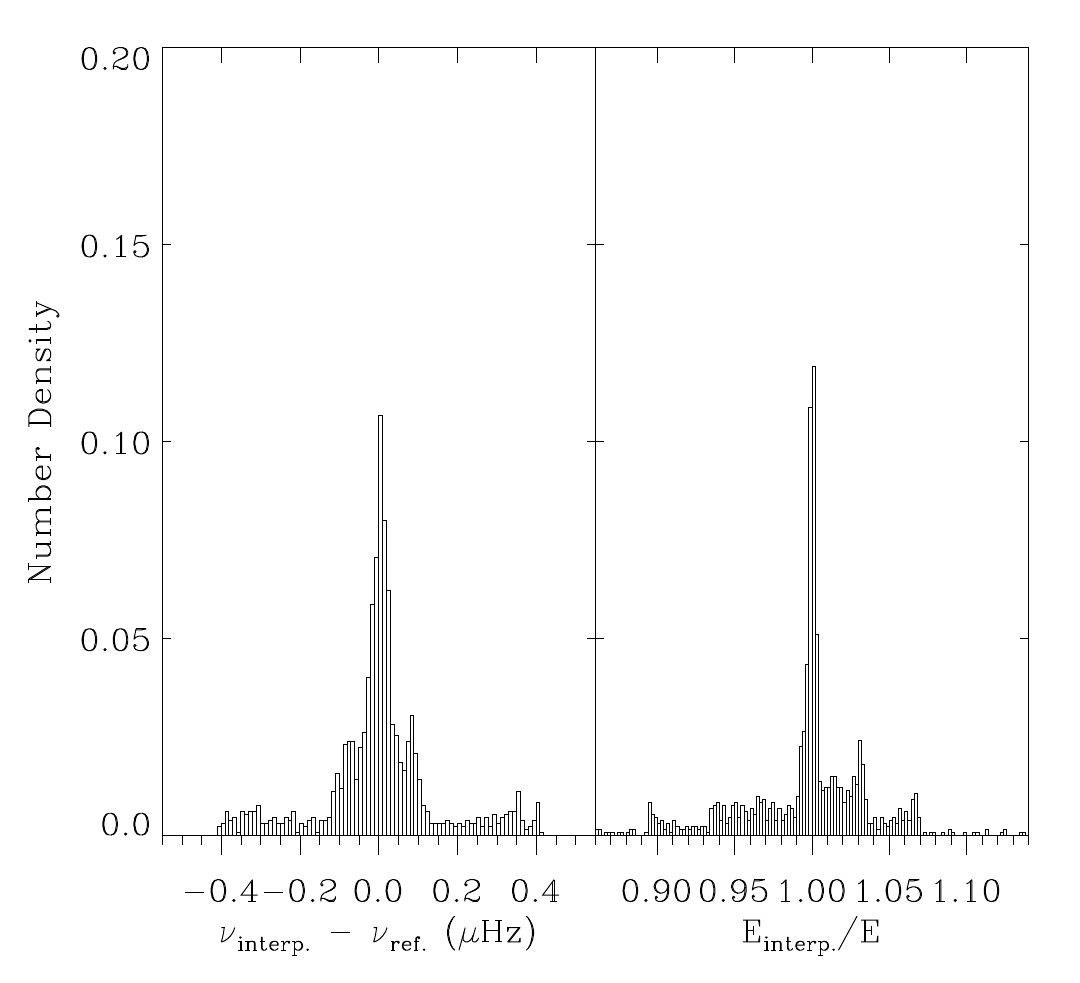}
    \caption{Same as Figure \ref{fig:inter_l1} but for three $\ell = 2$ mixed modes.} 
    \label{fig:inter_l2}
\end{figure}

We used the same method to test the interpolation for $\ell = 2$ mixed modes. Interpolations for three $\ell = 2$ modes are illustrated in Figure \ref{fig:inter_l2}. As shown in the top panel, frequencies and inertias of $\ell = 2$ mixed modes undergo sharper changes than $\ell = 1$ modes during the avoided crossing. The interpolation works quite well for p-dominated modes but not for some g-like modes. This can also be seen in the histograms of differences (lower panel). The histograms have Gaussian-like profiles at the centre, which correspond to p-dominated and well-fitted g-like modes, but also show a spread due to the poorly-fitted g-like modes.
We examined the differences at different evolutionary stages, as we did for the $\ell = 1$ modes.
For $\sim$85\% of modes in our test models, frequency differences were less than 0.2$\mu$Hz, while those of poor-fitted modes range from 0.2 to 0.5 $\mu$Hz. 
Interpolated mode inertias agreed very well (within 2\%) for p-dominated modes but the differences were up to $\sim$10\% when the modes are g-like. However, these g-dominated $\ell = 2$ mixed modes are not detectable from the {\em Kepler} data because their mode inertia are large (hence oscillation amplitudes are low). The most bumped $\ell = 2$ modes that were detected in our subgiant sample deviate by about half of the small separation ($\delta \nu_{02} = \nu _{\ell = 0} -\nu_{\ell = 2}$) from the acoustic ridge (the location of the acoustic mode had there been no mode mixing). For the modes within these ranges, the interpolation differences were mostly less than 5\%. 
%
%
It should also be noted that this difference in mode inertia does not significantly affect the fit. The mode inertia is used in Eq. \ref{eq:ball_sc} for correcting the surface term, and the correction ($\delta \nu$) is very small for g-like modes. For instance, the correction for the bumped $\ell = 2$ mode of KIC 11137075 at 1132$\mu$Hz is $\sim$1$\mu$Hz. A 5\% change in its mode inertia changes this frequency by $\sim$0.05$\mu$Hz.

\section{Fitting methods}\label{sec:fitting}

\subsection{The likelihood function}\label{sec:MLE}
\begin{figure}
	\includegraphics[width=\columnwidth]{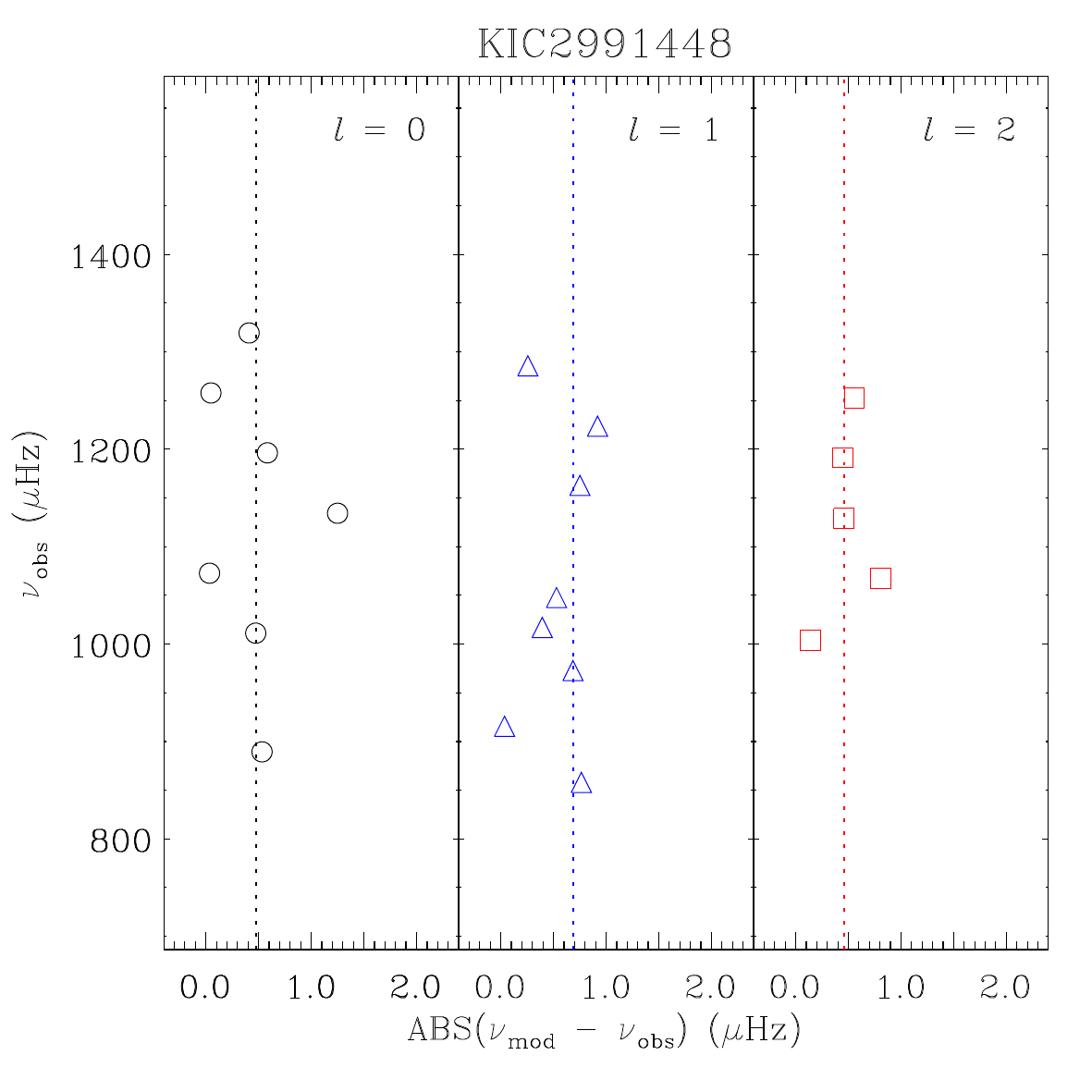}
    \caption{Systematic frequency offsets of theoretical models for KIC 2991448. Open circles, triangles, and squares are the absolute frequency differences between the best model and the observation. Dotted lines indicate the median values of the offsets.}
    \label{fig:systematics_uncertainty}
\end{figure}

We used the maximum likelihood estimation (MLE) to determine stellar
parameters. The likelihood function was defined as
\begin{equation}\label{likelihood}
\mathcal{L} {\rm{ = }} {e}^{-\frac{1}{2} (\chi_{\rm spec}^{2} + \chi_{\rm seis}^{2})},
\end{equation}
where $\chi_{\rm spec}^{2}$ and $\chi_{\rm seis}^{2}$ were calculated from spectroscopic and asteroseismic constraints. 
We calculated $\chi_{\rm spec}^{2}$ as 
\begin{equation}
\chi_{\rm spec}^{2}{\rm{ = }} \sum\limits_{i = 1}^n
{\frac{{{{({x_i} - {u_i})}^2}}}{{\sigma _{i}^2}}}, 
\end{equation}
where ${x_i}$ and ${u_i}$ indicate theoretical and
observed parameters, and ${\sigma _{i}}$ are the uncertainties of the observations.
Observed effective temperatures and metallicities were adopted as the spectroscopic constraints.
The radial modes are purely acoustic; the coupling between p- and g-modes is generally strong for $\ell = 1$ but weak for $\ell = 2$ modes. Modes for different $\ell$ therefore probe different regions of a star.
For this reason, we divided $\chi_{\rm seis}^{2}$ into three parts:
\begin{equation}\label{eq:chi2_seis1}
\chi_{\rm seis}^{2}{\rm{ = }}\chi_{\rm seis,\ell = 0}^{2} + \chi_{\rm seis,\ell = 1}^{2} +  \chi_{\rm seis,\ell = 2}^{2},
\end{equation}
where
\begin{equation}\label{eq:chi2_seis}
\chi_{\rm seis,\ell}^{2}{\rm{ = }} \sum\limits_{i = 1}^n
{w_i\frac{{{{({x_i} - {u_i})}^2}}}{{\sigma_{i}^2 + \sigma_{\rm{sys,} \ell}^2}}}
\end{equation} 
Here ${x_i}$, ${u_i}$, and ${\sigma _{i}}$ represent theoretical frequencies, observed frequencies and their uncertainties.
We introduced a model systematic uncertainty $\sigma_{\rm{sys}, \ell}$, which reflects the discrete spacing of the grid points and any missing physics of the model (e.g., poor modelling of the stellar surface). In Eq. \ref{eq:chi2_seis}, $w_i$ is a weighting factor which is described as
\begin{equation}\label{eq:weights}
w_i{\rm{ = }} {\rm exp}\Big(-\Big(\frac{({\nu_i} - {\nu_{\rm max}})^{2}}{2d^2}\Big)^{2}\Big)/\Big(\sum\limits_{i = 1}^n{w_i}\Big).
\end{equation}
For each star, $\sigma_{\rm{sys}, \ell}$ was determined from the frequency differences between the best seismic model and the observations. It should be noted that $\sigma_{\rm{sys}, \ell}$ depends on spherical degree $\ell$, because the sensitivities to the model input physics depend on $\ell$. The method to determine $\sigma_{\rm{sys}, \ell}$ is illustrated in Figure \ref{fig:systematics_uncertainty}. 
For each star, we first set up a 3-$\sigma$ box with the observed $T_{\rm eff}$ and [Fe/H] and found the best seismic model in the box using Eq. \ref{eq:chi2_seis1} and \ref{eq:chi2_seis}, but without $\sigma_{\rm{sys}, \ell}$. We then calculated the absolute frequency differences between the best model and the observations. The median values of the differences for each $\ell$ (dotted lines) were taken as the first guess of $\sigma_{\rm{sys}, \ell}$.  
We repeated the above process with $\sigma_{\rm{sys}, \ell}$ obtained in the previous iteration until the best model converged. The final $\sigma_{\rm{sys}, \ell}$ was used in the fitting procedure. The median value of $\sigma_{\rm{sys}, \ell}$ of all stars is 0.54 $\mu$Hz for $\ell$ = 0 modes and $\sim$0.85$\mu$Hz for $\ell =$ 1 and 2 mixed modes, which are generally large compared with observed uncertainty (the median observed uncertainty is 0.19$\mu$Hz, Figure 6 in Paper I).
Because of this, the model systematic uncertainty $\sigma_{\rm{sys}, \ell}$ becomes the major part of the term ${\sigma_{i}^2 + \sigma_{\rm{sys,} \ell}^2}$ in Eq.~\ref{eq:chi2_seis} for most modes. However, when the term ${\sigma_{i}^2 + \sigma_{\rm{sys,} \ell}^2}$ is similar for all observed modes, poorly-measured modes are similarly weighted as good ones, and this causes an issue in our fitting. Because poorly measured modes were generally poorly fitted. Their $\chi^{2}$ values are relatively large and also spread in big dynamical ranges compared with good modes. They hence play leading roles in valuating the average seismic $\chi^{2}$ and also significantly impact the probability distribution. Apparently, it is not appropriate because the fitting should be mainly determined by well-measured modes.
For this reason, we implemented a factor $w_{i}$ in the likelihood function, aiming to down-weight those poorly measured modes. To choose the weights, we inspected the observed mode frequencies and found that their squared uncertainty ($\sigma_{i}^2$) are generally small around $\nu_{\rm max}$ but obviously increase when closed to both edges. We hence applied a super-Gaussian window centred on $\nu_{\rm max}$ to calculate $w_{i}$ as described by Eq. \ref{eq:weights}.
The full-width at half-maximum $d$ is 0.75 of the frequency difference between the lowest and the highest radial modes, and we used a power of 2.0.This super-Gaussian profile arranges similar weights for mode frequencies from $\sim$0.9 to $\sim$1.1$\nu_{\rm max}$, starts to down-weight modes outside the centre region, and gives about half of the weight at $\nu_{\rm max}$ for the lowest and highest frequencies.
Detailed explanations of the reason to implement the weighting factor and a comparison between fitting results with and without $w_{i}$ are given in Appendix \ref{appC}.
The model systematic uncertainty $\sigma_{\rm sys}$ also gave a quick examination of the goodness of the fit. Large $\sigma_{\rm sys}$ indicated that even the best model was not good. After visual inspections of the \'echelle diagram, we set a threshold on $\sigma_{\rm sys, \ell}$ of 1.5$\mu$Hz for $\ell$ = 0 and 2, and 2$\mu$Hz for $\ell = 1$ modes. If the best model had a $\sigma_{\rm sys, \ell}$ above the threshold, the fit was determined to have failed.
%

For stars with good fits, we determined stellar parameters with the likelihood obtained above. We fitted the 1-D probability distributions with Gaussian functions to estimate the central values and their uncertainties.

%
%

\subsection{Summary of fitting procedure}
Here we summarise the fitting procedure described in Sections \ref{sec:interpolation} and \ref{sec:MLE}. 
For each star, we considered a cube with the observed $T_{\rm eff}$ ($\pm 3 \sigma$), [M/H]($\pm 3 \sigma$), and $\Delta \nu$ ($\pm$15\%). 
For each evolutionary track going through the cube, we then interpolated stellar parameters, mode frequencies, and mode inertias as a function of the age with a number of the models before and after the best model. 
%
We found that interpolating with 7 models (the best-fitting model plus 3 previous and 3 afterwards) was best for balancing the accuracy and the efficiency. %
We interpolated 1000 models (linear for stellar parameters and $\ell = 0$ modes, and cubic spline for $\ell$ = 1 and 2 modes) between every time step and hence obtained 6000 interpolated models for the 7 original models. We did not use the first 1000 and the last 1000 models to avoid edge effects and hence ended up with 4000 interpolated models for each evolutionary track. If $n$ tracks went through the observation cube, we would have $n$ $\times$ 4000 interpolated models.
Lastly, we adopted the methods mentioned in Section \ref{sec:MLE} to determine the stellar parameters of the star with all interpolated models. 

\section{Analysing dependencies}\label{sec:dependence}
With the fitting procedure mentioned above, we studied the dependencies of the inferred stellar parameters (mass, age, effective gravity, radius, and helium-core mass) on model inputs (helium fraction, metallicity, and mixing-length parameter) for the 36 stars with the model grids listed in Table \ref{tab:grid1}. Results are presented and discussed below. 

\subsection{Helium abundance}\label{subsec:helium}

\begin{figure*}
	\includegraphics[width=1.8\columnwidth]{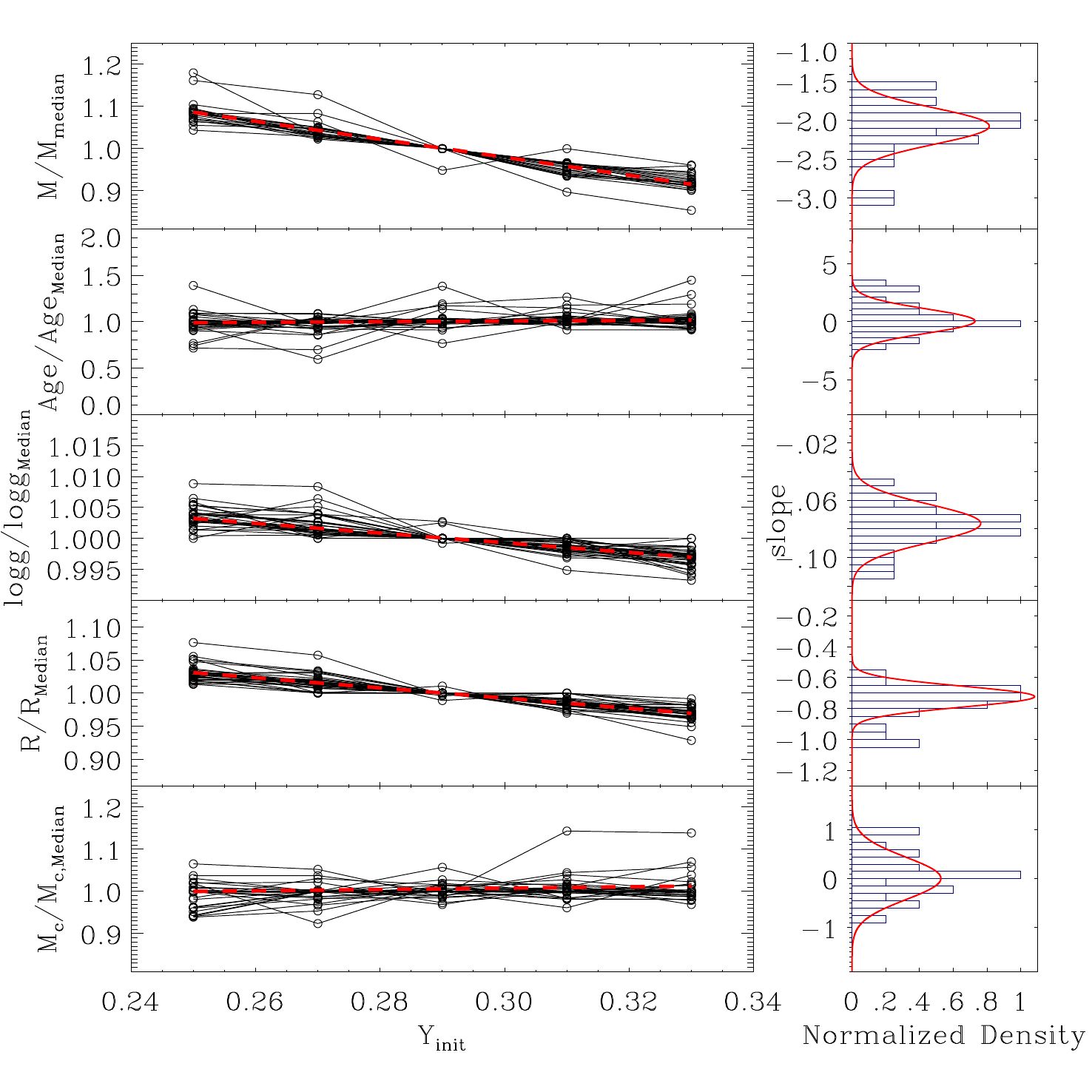}
    \caption{Left: dependencies of inferred mass, age, surface gravity, radius, and helium-core mass on the helium fraction given by `Grid-Y'. Inferred parameters of 36 stars with different initial helium abundance are represented by open circles, and each solid line joins the results of the same star. Inferred parameters of each star were fitted with linear functions as function of $Y_{\rm init}$, and the red dashed lines indicate the average slopes of all stars. Right: histograms of slopes from linear fits obtained in the left panels. Red solid lines indicate the Gaussian functions which fit the distributions.}
    \label{fig:y_effect}
\end{figure*}

We used models in `Grid-Y' to estimate dependencies on the helium fraction.
We estimated the five parameters for each star for different input helium fractions by fitting models to observations. Results are presented in Figure \ref{fig:y_effect}.
As illustrated, we normalised our inferred results with the median value of all 36 results for each parameter, lined up results of the same star, and fitted them with a linear function. 
We then plotted the histogram of the slopes (as shown on the right side) and fitted the density distribution with the Gaussian function to quantify the dependency. 
The dependence of each stellar parameter $i$ on the helium fraction was then described as
\begin{equation}\label{eq:slope1}
p_{i} = c + {(S_{i,Y} \pm \delta S_{i,Y})} {Y_{\rm{init}}},
\end{equation}
where $p_{i}$ are the normalised stellar parameters, and $S_{i, Y}$ and $\delta S_{i,Y}$ are the centres and the standard deviations of the Gaussian profiles.
The fractional change of a stellar parameter ($\Delta p_{i}$) for a given change in the helium abundance ($\Delta Y_{\rm{init}}$) can be written as 
\begin{equation}\label{eq:slope2}
\Delta p_{i} = {(S_{i,Y} \pm \delta S_{i,Y})} {\Delta Y_{\rm{init}}}.
\end{equation}
Based on the results in Figure~\ref{fig:y_effect}, we derived $S_{i, Y}$ and $\delta S_{i,Y}$ as summarised in Table \ref{tab:model_dependence1}. 
It can be seen that the inferred mass and radius are obviously helium-dependent, while inferred age, surface gravity, and the helium-core mass are not sensitive to the helium abundance.  

For most of stars, their helium abundances cannot be observed.  
To estimate the initial helium fraction of each star, we relied on the law of Galactic elements enrichment, which is described as
\begin{equation}\label{eq:Y_Z}
Y_{\rm{init}} = Y_0 + \frac{\Delta Y}{\Delta Z}Z_{\rm{init}}.
\end{equation}
The accepted value for the primordial helium abundance ($Y_0$) of the Galaxy ranges from 0.23 to 0.25
\citep[e.g.][]{2016A&A...594A..13P}, while the ratio ${\Delta Y}/{\Delta Z}$ is a loosely constrained parameter ranging from 1 to 3.
The initial helium abundance estimated for the Sun is 0.27 \citep{2009ARA&A..47..481A} and the study of {\em Kepler} LEGACY samples suggested a range from $\sim$0.25 to $\sim$0.30 \citep{2017ApJ...835..173S}. 
Following from the $5\%$ uncertainty in the helium abundance, the systematic uncertainties in estimates are $\sim$10$\%$ for masses and $\sim$3.5$\%$ for radii. 

\subsection{Heavy element abundance}\label{subsec:metals}
\begin{figure*}
	\includegraphics[width=1.8\columnwidth]{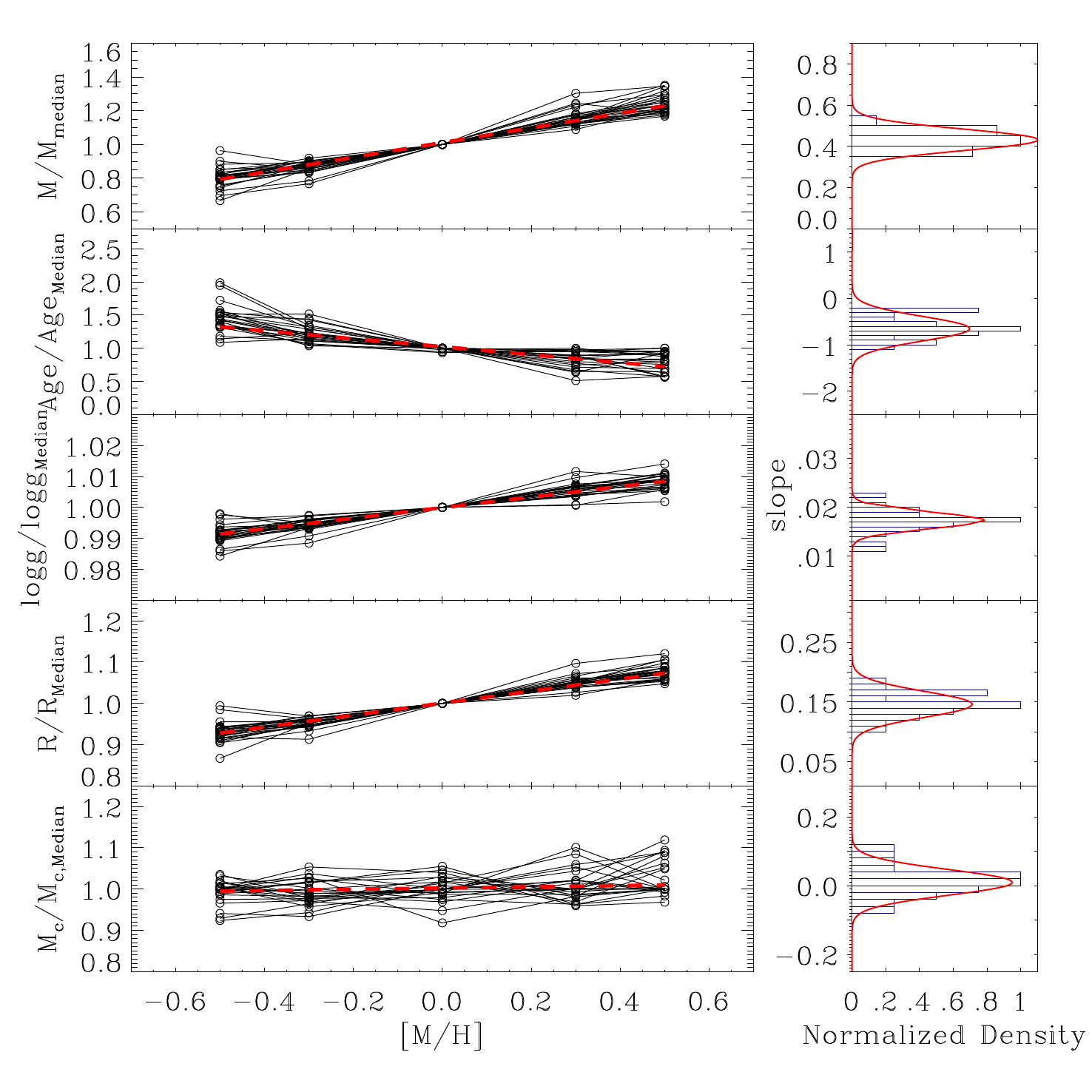}
    \caption{Same as Figure \ref{fig:y_effect} but for the input metallicity.}
    \label{fig:z_effect}
\end{figure*}

With the same approach, we defined $S_{i, Z}$ and $\delta S_{i, Z}$ and
used models in `Grid-Z' to investigate dependencies of inferred parameters on the input metallicity. 
The results are shown in Figure~\ref{fig:z_effect} and we summarise the derived $S_{i, Z}$ and $\delta S_{i, Z}$ in Table \ref{tab:model_dependence1}. 
It can be seen that our estimates of mass, age, and radius have strong correlations with the metallicity.
Given a change of 0.1 dex in [M/H], average systematic offsets are $4.3\%$ for the mass, $6.5\%$ for the age, and 1.5$\%$ for the radius, but less than 0.5\% for the surface gravity and the helium-core mass.   

Although [Fe/H] for most stars can be measured with good precision, there are still two issues. One is the $\alpha$-enhancement. As recent data from spectroscopic surveys showed \citep[e.g.][Fig.22]{BuderGalahDR2}, dwarfs and subgiants with [Fe/H] from -0.5 to +0.5 dex have a large scatter in [$\alpha$/Fe]. Taking carbon and oxygen as examples, the values of [C/Fe] and [O/Fe] at the solar [Fe/H] are spread by up to $\pm \sim$0.5 dex.        
Hence, good measurements of the $\alpha$-elements are useful to avoid the systematical uncertainties. For instance, \citet{2015MNRAS.447..680G} found a $\sim$10$\%$ change in the inferred age of star KIC 7976303 after applying its alpha-enhancement ([$\alpha$/Fe] = 0.4 dex) in their detailed modelling. The other issue is the solar mixture. Taking two of the most popular values, namely $(Z/X)_{\odot}$ = 0.0231 \citep{1998SSRv...85..161G} and $(Z/X)_{\odot}$ = 0.0181 \citep{2009ARA&A..47..481A}, the difference corresponds to a change of $\sim$0.1 dex in [M/H] for solar-metallicity stars, which could obviously affect parameter determinations.

\subsection{The mixing-length parameter}\label{subsec:alpha}

\begin{figure*}
	\includegraphics[width=1.8\columnwidth]{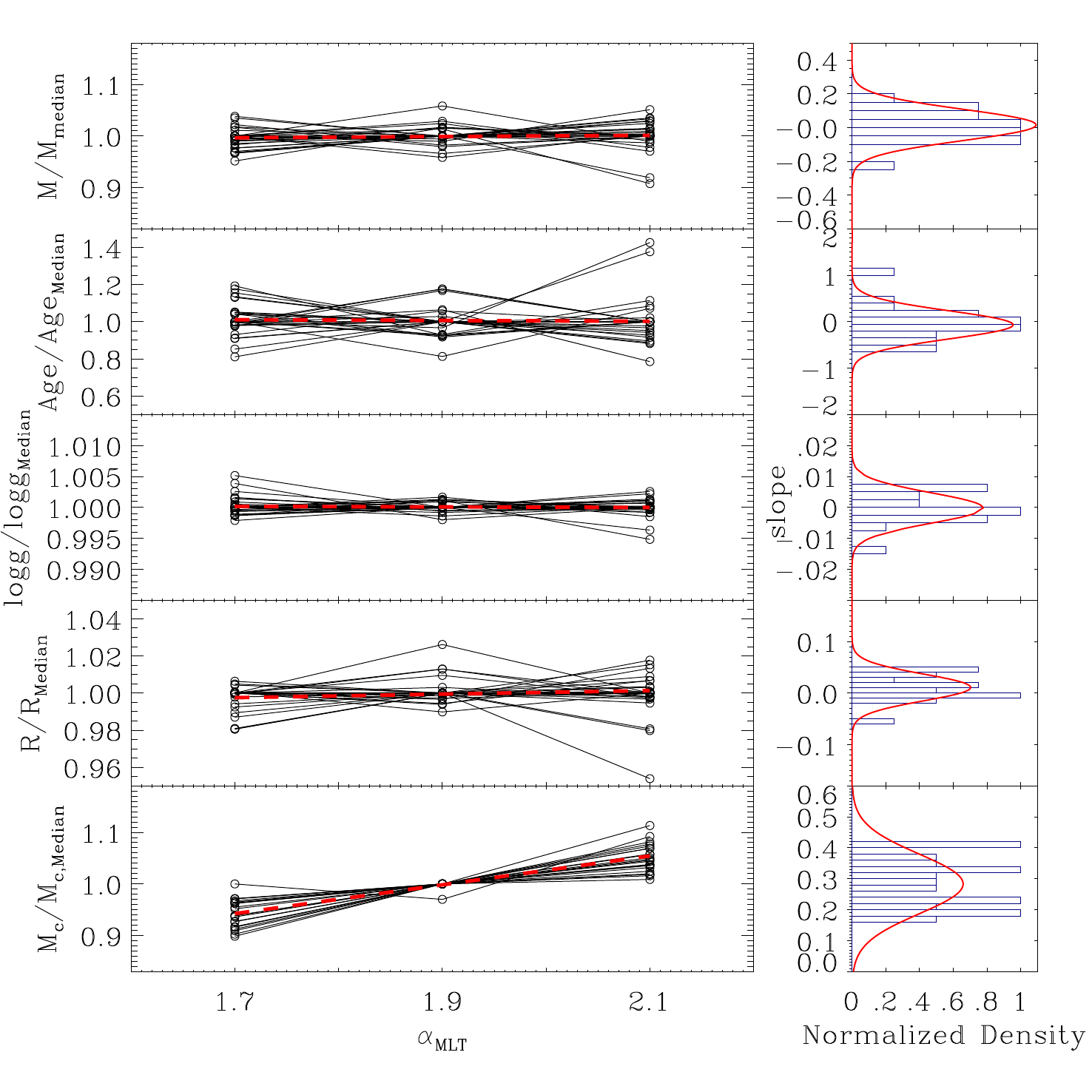}
    \caption{Same as Figure \ref{fig:y_effect} but for the input mixing-length parameter.}
    \label{fig:a_effect}
\end{figure*}

To estimate the dependencies of inferred parameters on the mixing-length parameter,
we used models in `Grid-$\alpha$' and derived $S_{i, \alpha}$ and $\delta S_{i,\alpha}$.
The results are presented in Figure~\ref{fig:a_effect} and summarised in Table \ref{tab:model_dependence1}. 
We found that the mixing-length parameter does not significantly affect determinations of mass, age, surface gravity, and radius. This agrees with previous studies \citep{2003A&A...399..243F, 2010Ap&SS.328...73P, LiTmuher}. However, estimates of helium-core masses show an obvious correlation with the mixing-length parameter and they vary by $2.8\%$ when $\alpha _{\rm MLT}$ is changed by 0.1. 

The mixing-length parameter is adjusted in theoretical models and needs to be calibrated. However, the calibrated value given by the Sun does not necessarily suit other types of stars.
Hydrodynamic simulations \citep[e.g.][]{2015A&A...573A..89M} are able to estimate the mixing length for a given set of surface properties. 
However, some conflicts between model calibrations and simulations have been recently found in red giants \citep{Tayar2017mlt_feh,Li2018sixredgiants,Ball2018redgiants}, for which stellar modelling required relatively large mixing-length parameters to fit observations. 
%
%
However, for subgiants, we found that mass and age determinations are not sensitive to the input mixing-length parameter. 

%


\begin{table}
	\centering
	\caption{Dependencies of the inferred stellar parameters on the input parameters. $p_{i}$ indicates a normalised inferred parameter ${i}$; $S_{i, N}$ and $\delta S_{i,N}$ define the slope and its uncertainty between the inferred parameter $i$ and input parameter $N$ determined by 36 subgiants; and $\Delta p_{i}$ refers as the fractional change of an inferred parameter $i$ when given changes of inputs.}   
	\label{tab:model_dependence1}
	\begin{tabular}{cccc} 
		\hline
		$p_i$ & $S_{i,Y}$ ($\delta S_{i,Y}$) & $S_{i,Z}$ ($\delta S_{i,Z}$) & $S_{i,\alpha}$ ($\delta S_{i,\alpha}$)\\
		\hline
		$p_{\rm M}$  & -2.1(0.2) & 0.43(0.05)& 0.0(0.1)\\
		$p_{\rm t}$ & -0.1(1.1) &  -0.65(0.26) & -0.1(0.3)  \\
		$p_{\rm{\log g}}$ & -0.08(0.01) & 0.02(0.01) & 0.00(0.01)  \\
        $p_{\rm R}$ & -0.72(0.07) & 0.15(0.02) & 0.01(0.03) \\
        $p_{\rm{M_{core}}}$ & 0.01(0.43)& 0.01(0.04) & 0.28(0.10) \\
        \hline
		$\Delta p_{i}$ & $\Delta Y_{\rm{init}}$ = 0.05 &$\Delta$[M/H] = 0.1 [dex] & $\Delta \alpha _{\rm{MLT}}$ = 0.1\\
		\hline
		$\Delta M$ [\%]      & -10.5 &  +4.3 & 0.0\\
		$\Delta t$  [\%]     & 0.5 &  -6.5 &  -1.0\\
		$\Delta \log g$  [\%]& -0.4 & +0.2 &  0.0\\
        $\Delta R$  [\%]     & -3.5 &  +1.5 & +0.1\\
        $\Delta M_{\rm core}^{\rm a}$  [\%]    & 0.05&  -0.1 & +2.8\\
        \hline
     \multicolumn{4}{p{.99\columnwidth}}{a. the helium core is defined as the centre regions where the mass fraction of hydrogen is below 0.01}  
	\end{tabular}
\end{table}

\section{Ages of the {\em Kepler} subgiants}\label{sec:results}

As presented above, changing the input helium abundance and the mixing-length parameter does not systematically change age determinations. The metallicity does matter and so it needs to be well measured. These results indicate that seismic ages of subgiants are less model-dependent than those of other types of stars. 
The weak dependencies on the helium fraction and the mixing-length parameter also mean that modelling subgiant ages can be more efficient because one may reduce two of the dimensions in the grid computation. The advantages make subgiants a very important population for determining ages.

\subsection{Grid-age}\label{sec:final_grid}

To determine ages of the 36 {\em Kepler} subgiants, we computed another grid with only two independent parameters, namely, mass and metallicity (`Grid-age'). Because metallicity is a crucial input parameter for modelling stellar ages, we calculated models with 22 values in a range from -0.55 to 0.50 dex. The model inputs are listed in Table \ref{tab:final_grid}. 
We determined the input helium fraction following the law of Galactic element enrichment (Eq. \ref{eq:Y_Z}) 
and adopted the primordial helium abundance given by \citet{2016A&A...594A..13P} ($Y_0$ = 0.249).
They used the Planck power spectra, Planck lensing, and some external
data such as baryonic acoustic oscillations. 
The ratio ${\Delta Y}/{\Delta Z}$ was derived by the initial
abundances of helium and heavy elements of the Sun. 
We used the $Y_{\rm{\odot,init}}$ and $Z_{\rm{\odot,init}}$ 
of the calibrated solar model given by
\citet{2011ApJS..192....3P}, which are 0.2744 and 0.0191 (different
from the present-day abundances of 0.243 and 0.0170), 
and hence the ratio ${\Delta Y}/{\Delta Z}$ is 1.33.
We took the solar-calibrated mixing-length parameter (1.9) 
as given by \citet{2011ApJS..192....3P} for the grid computation.
%
\begin{table}
	\centering
	\caption{The input physics of the `Grid-age'.}
	\label{tab:final_grid}
	\begin{tabular}{ccc} 
		\hline
		Input Parameter & Range & Increment\\
        \hline
		Mass [$M_{\odot}$]  & 0.80 -- 1.90 &  0.01\\
        $\rm{[M/H]}$ [dex] & -0.55 -- +0.50 & 0.05 \\
		$Y_{\rm{init}}$ & $Y_{\rm{init}}^{\rm a}$ = $Y_0$ + $\frac{\Delta Y}{\Delta Z}$$Z_{\rm{init}}$ & -\\
        $\alpha_{\rm{MLT}}$  & 1.9 &  -\\
        $f_{\rm{ov}}$/$f_0$ & 0.018/0.009 & - \\
        \hline
     \multicolumn{3}{p{.4\textwidth}}{$^{\rm a}Y_0$ = 0.249 and $\frac{\Delta Y}{\Delta Z}$ = 1.33 was adopted.}  
	\end{tabular}
\end{table}

\subsection{Model fitting}

\begin{figure}
    \includegraphics[width=\columnwidth]{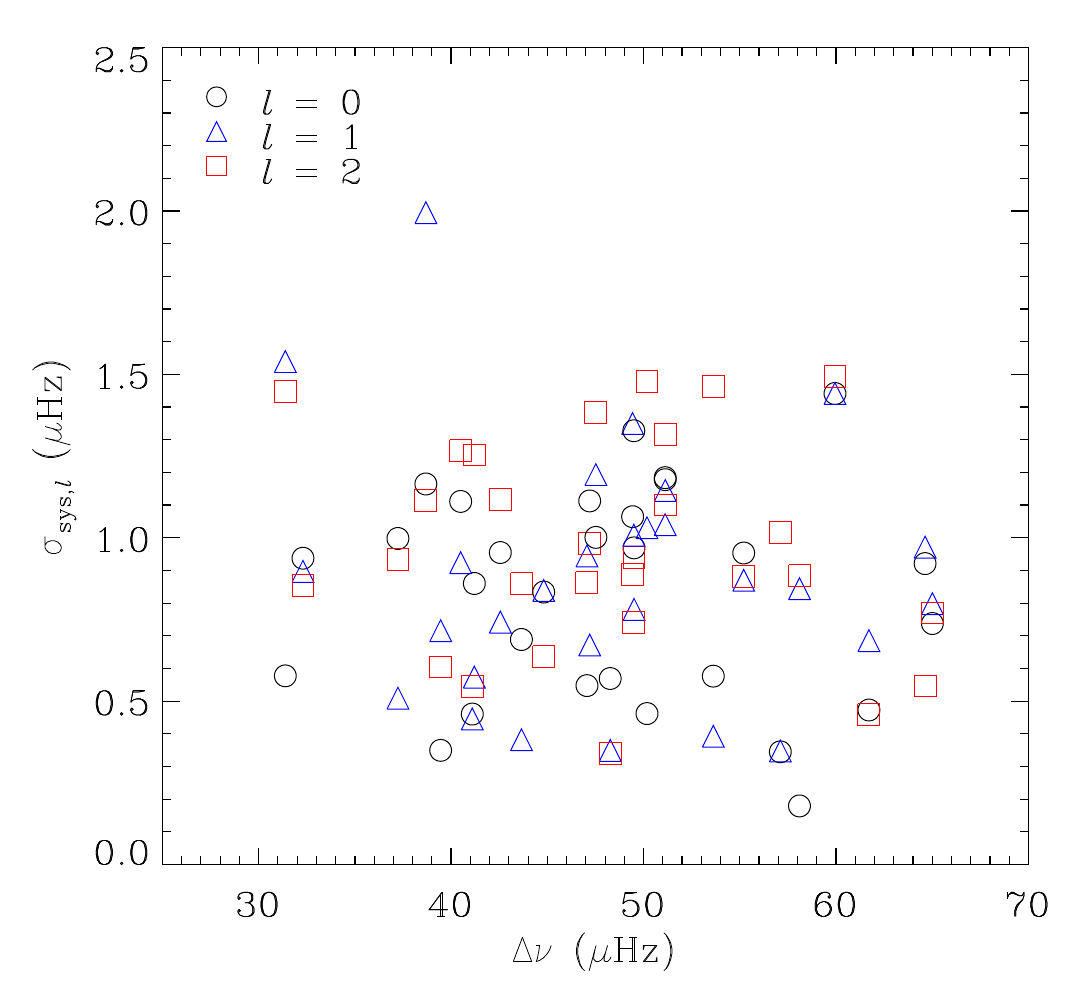}    
    \caption{Median differences between oscillation frequencies for $\ell$ = 0, 1, and 2 modes of the best-fitting models and observations of 31 {\em Kepler} stars.}
    \label{fig:systematics}
\end{figure}

We calculated likelihoods for the models in `Grid-age' based on the equations introduced in Section \ref{fitting}.
We firstly examined the goodness of fits by inspecting the differences between observations and models.
We removed five stars because their $\sigma _{\rm sys,\ell}$ exceeded the threshold (1.5$\mu$Hz for $\ell$ = 0 and 2 modes, and 2.0$\mu$Hz for $\ell$ = 1 modes). 
Models of the other 31 stars, as shown in Figure~\ref{fig:systematics}, generally fit the observations very well.  
We selected six subgiants from the early- to the late-subgiant phases and present their best-fitting models on the \'echelle diagram in Figure \ref{fig:echelle}. \'Echelle diagrams of all 31 subgiants are given in the Appendix \ref{appA}.
It can be seen that the radial and quadrupole modes generally show good matches but some of the bumped dipole modes have relatively large differences between models and observations. This indicates that the envelopes (probed by p- and p-dominated modes) are well modelled but the cores (probed by g-dominated modes) are not. 
These mismatches of the dipole modes are similar to those found in our previous study of the subgiant $\mu$ Her \citep{LiTmuher}. 
Improving the fits of dipole modes requires further studies of the stellar core, such as involving different overshooting schemes, the extra mixing caused by rotation, etc.
We also note that the surface terms of dipole modes at relatively high frequency in some of the stars are over-correcting (e.g., the dipole mode at 1100$\mu$Hz of KIC 10273246) and hence cause a larger offset than seen at lower frequencies. The issue mostly arises when dipole modes exceed the observed frequency ranges of radial and quadrupole modes. The surface correction seems not to work well for dipole modes at high-frequency edge and this cause mismatches. However, this over correction does not affect the model fitting much because the modes far from $\nu_{\rm max}$ had small weights in the likelihood functions.

\begin{figure*}
	\includegraphics[width=\columnwidth]{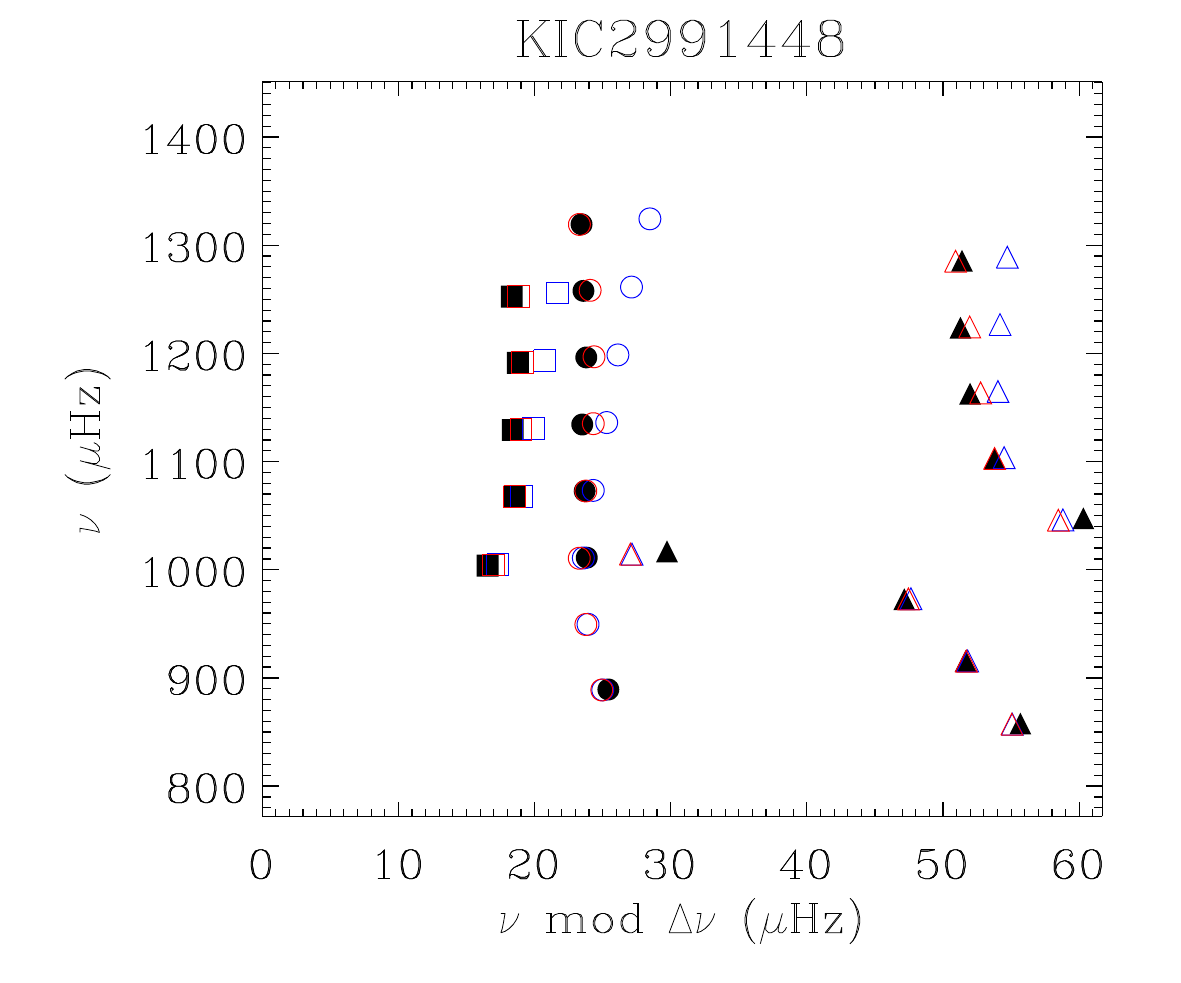}
    \includegraphics[width=\columnwidth]{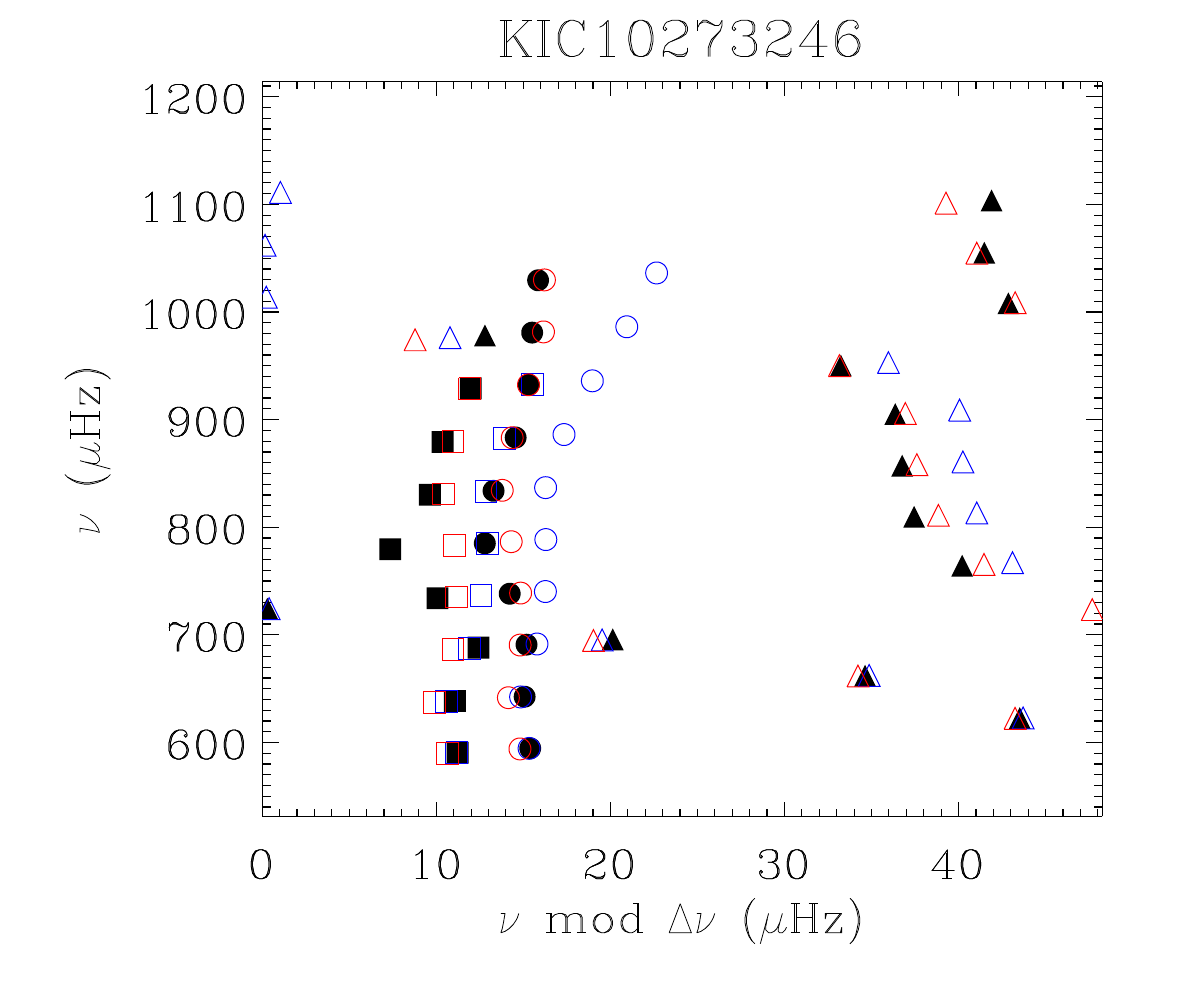}  
    \includegraphics[width=\columnwidth]{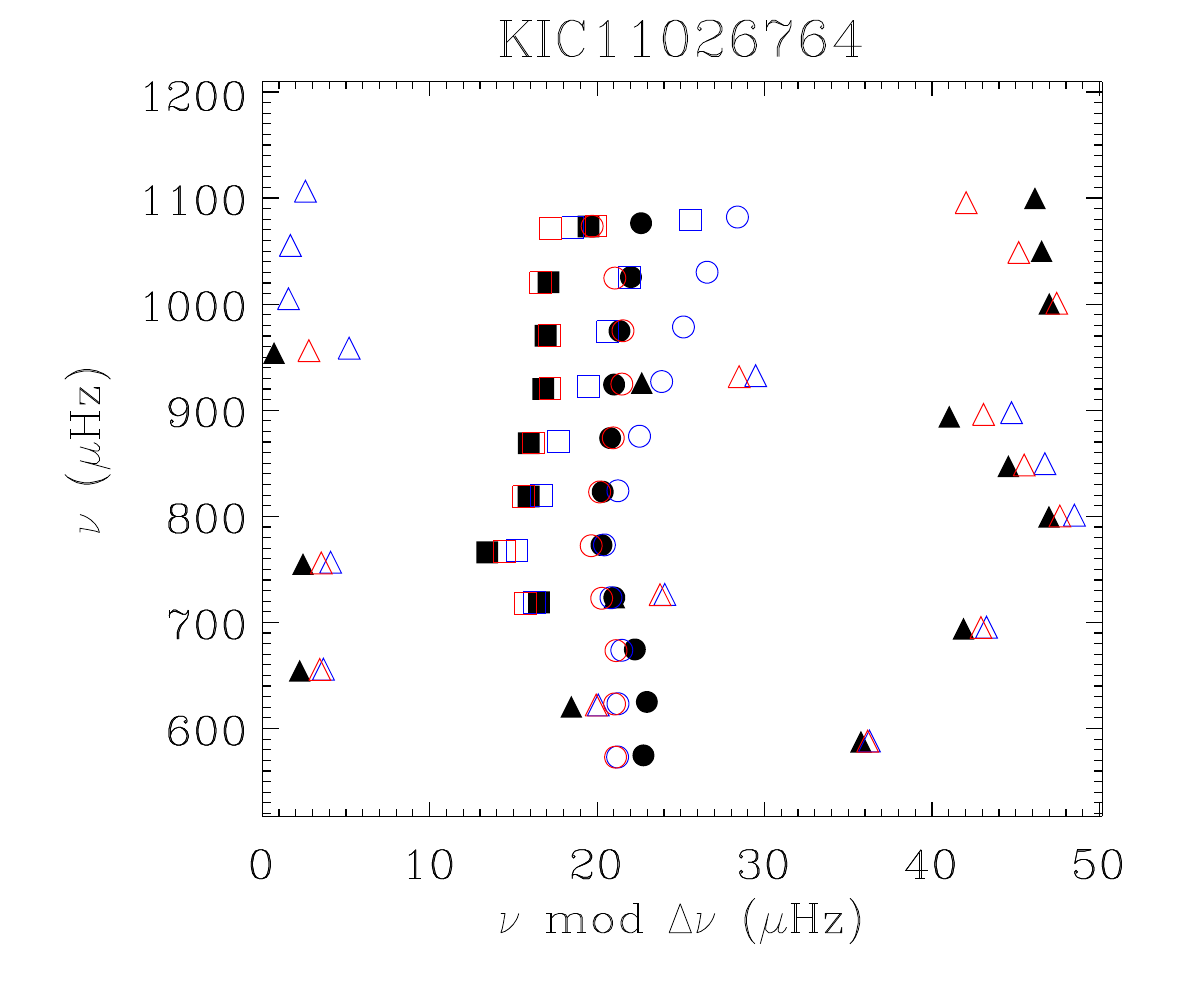}    
    \includegraphics[width=\columnwidth]{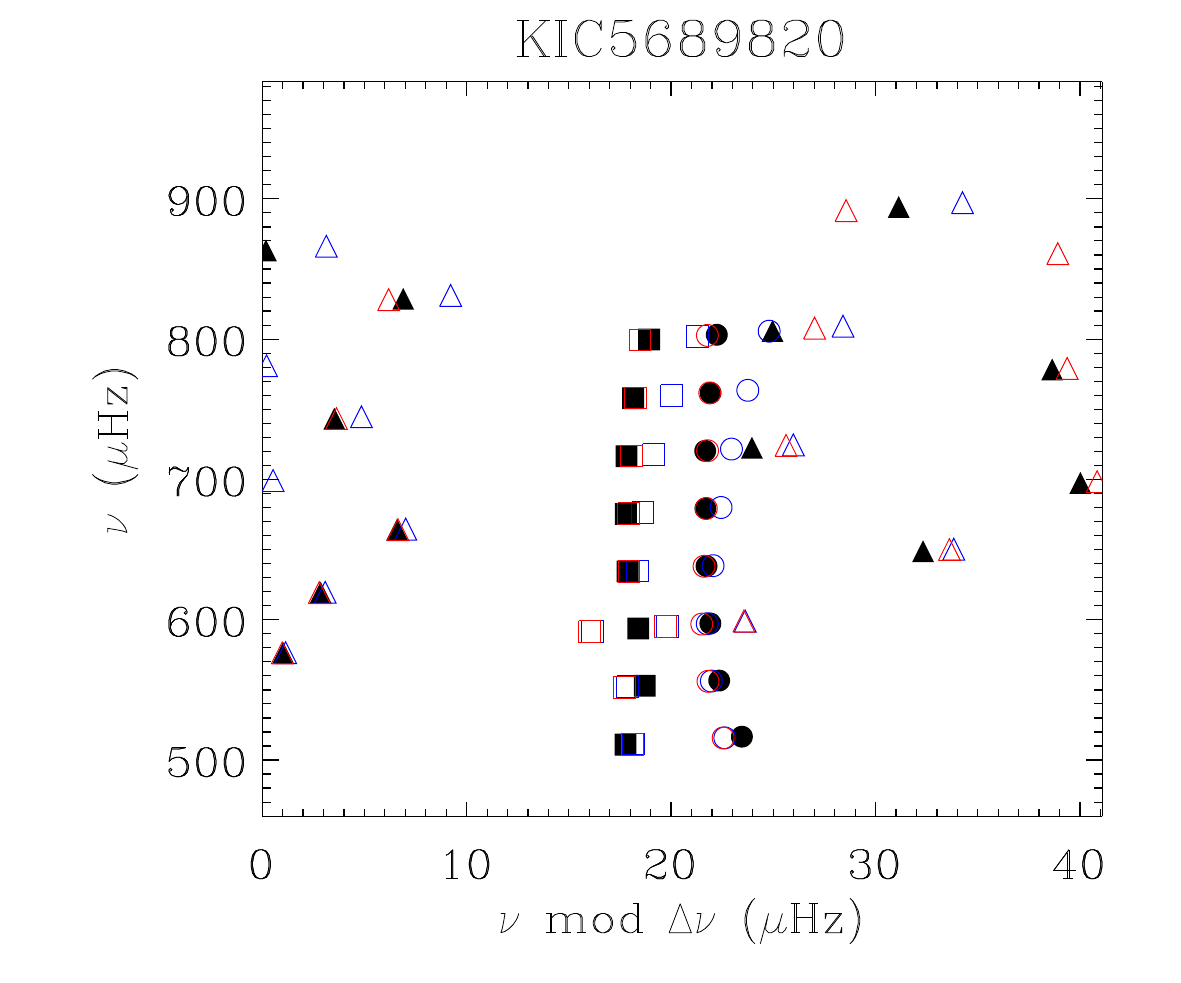}    
    \includegraphics[width=\columnwidth]{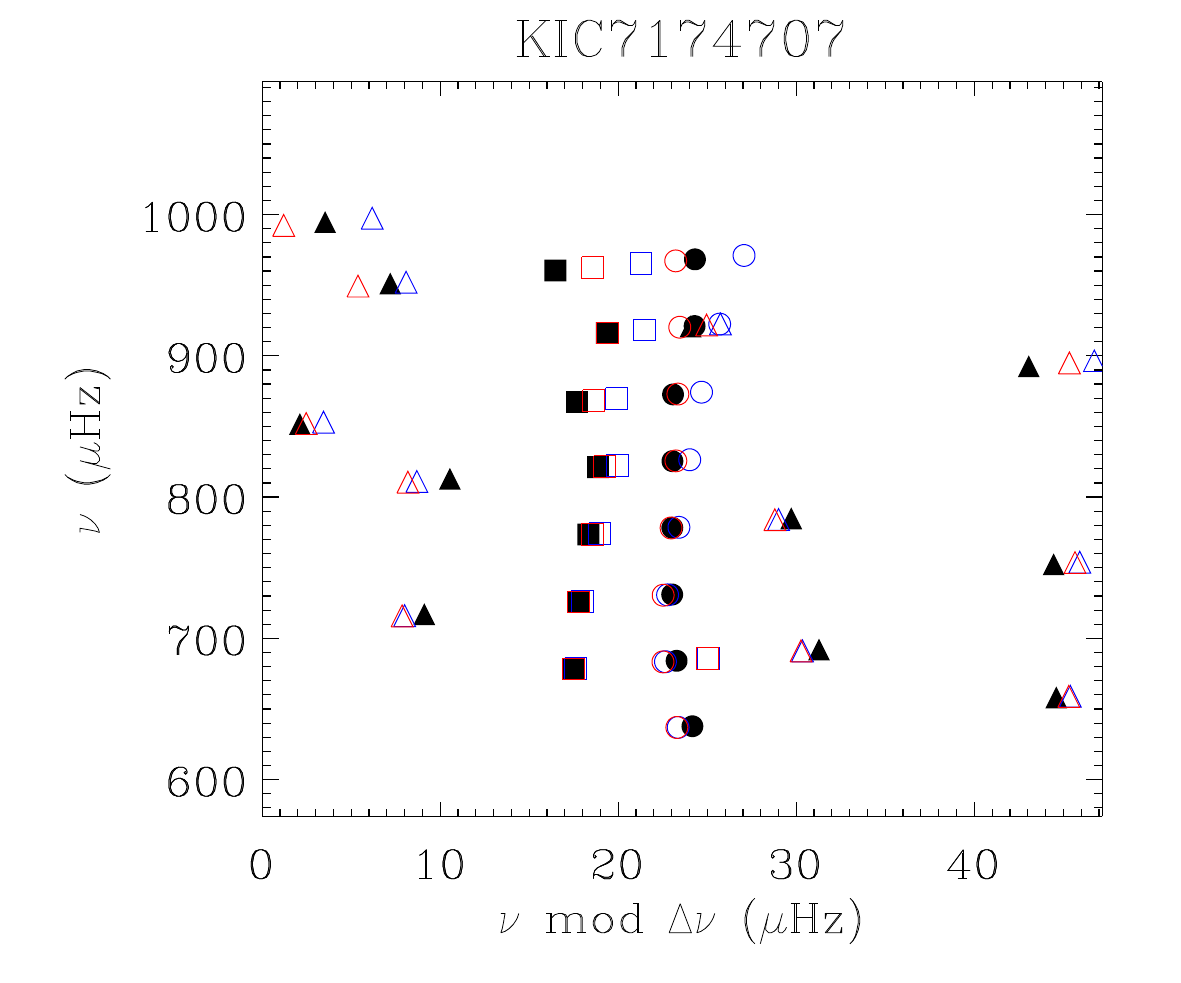}
    \includegraphics[width=\columnwidth]{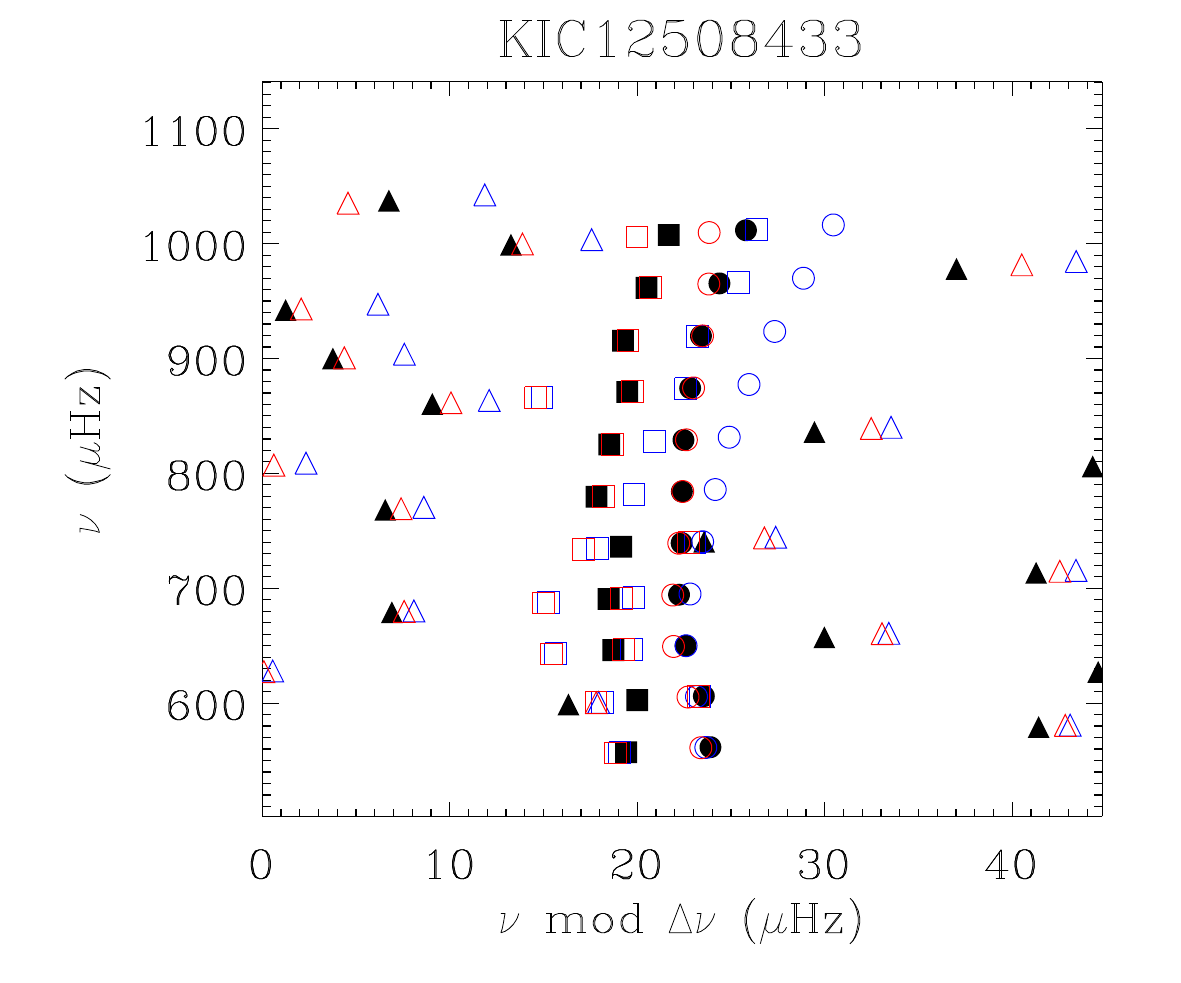}
    \caption{\'Echelle diagram of 6 examples given by the `Grid-age'. Graphs from left top to bottom right correspond to subgiants from early to late evolutionary phases (numbers of the avoided crossing). Filled symbols represent observed oscillation modes. Open blue and red symbols are theoretical modes before and after the surface correction. Circles, triangles, and squares indicate $\ell$ = 0, 1, and 2 modes.}
    \label{fig:echelle}
\end{figure*}

  
\subsection{Estimating stellar ages and other parameters}

\begin{figure*}
	\includegraphics[width=0.90\textwidth]{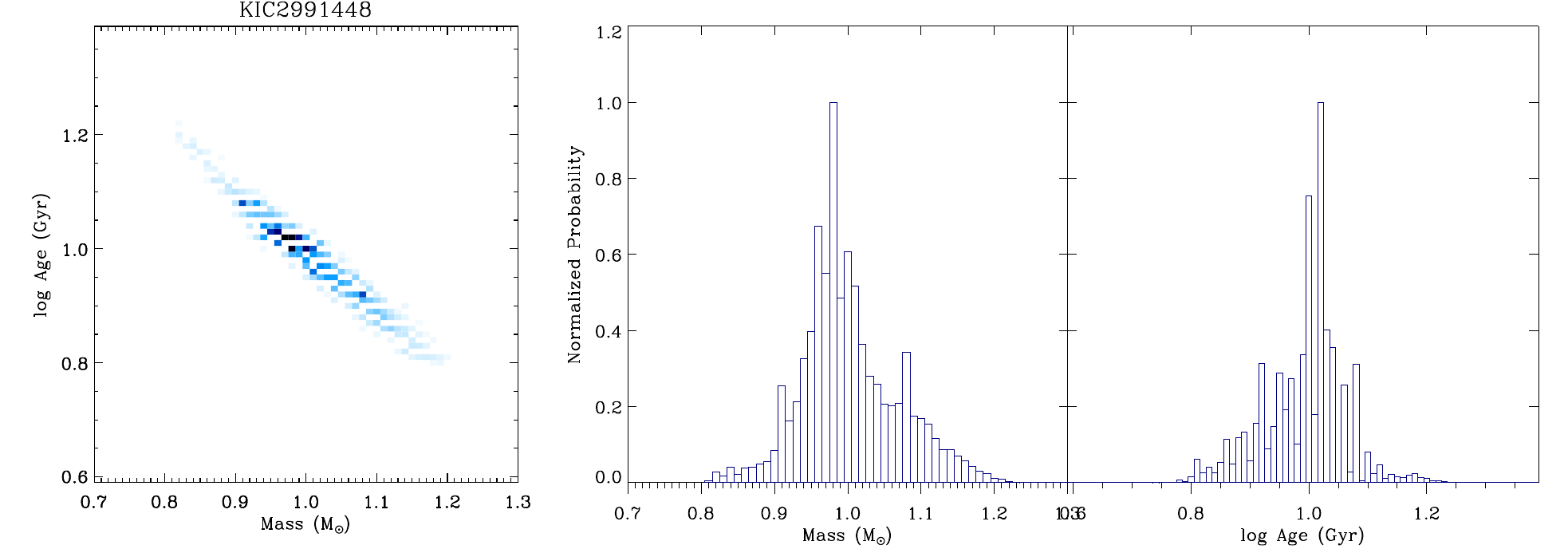}
    \includegraphics[width=0.90\textwidth]{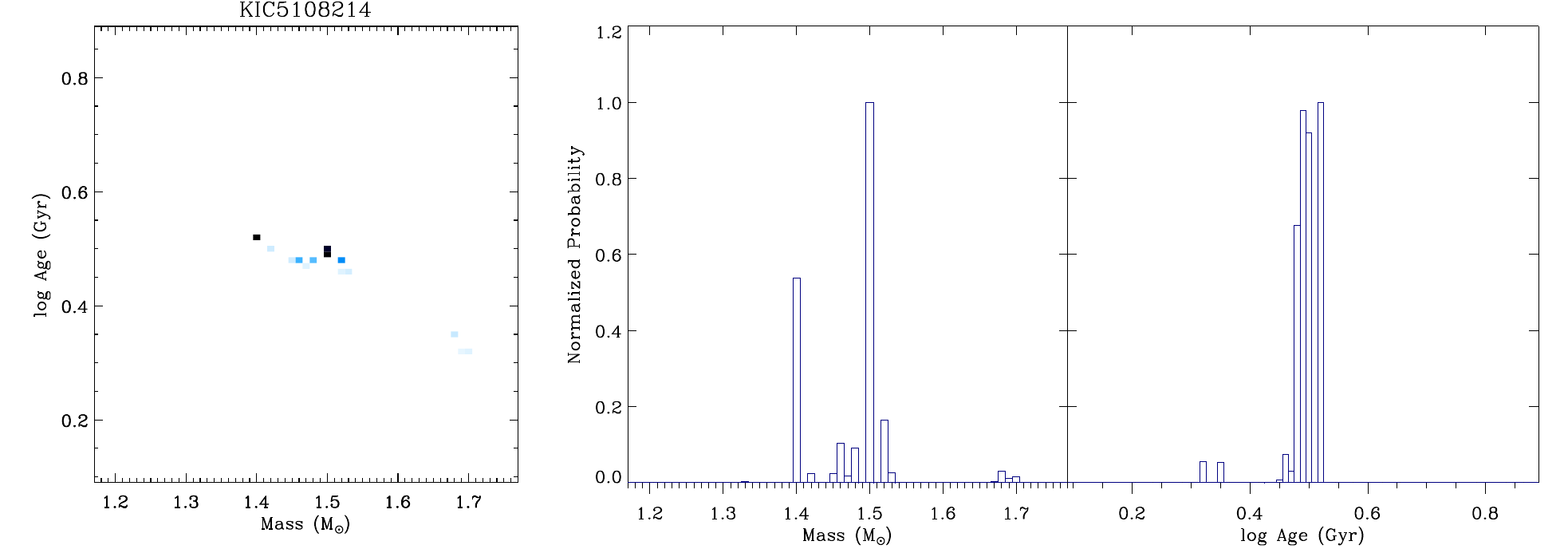}
    \includegraphics[width=0.90\textwidth]{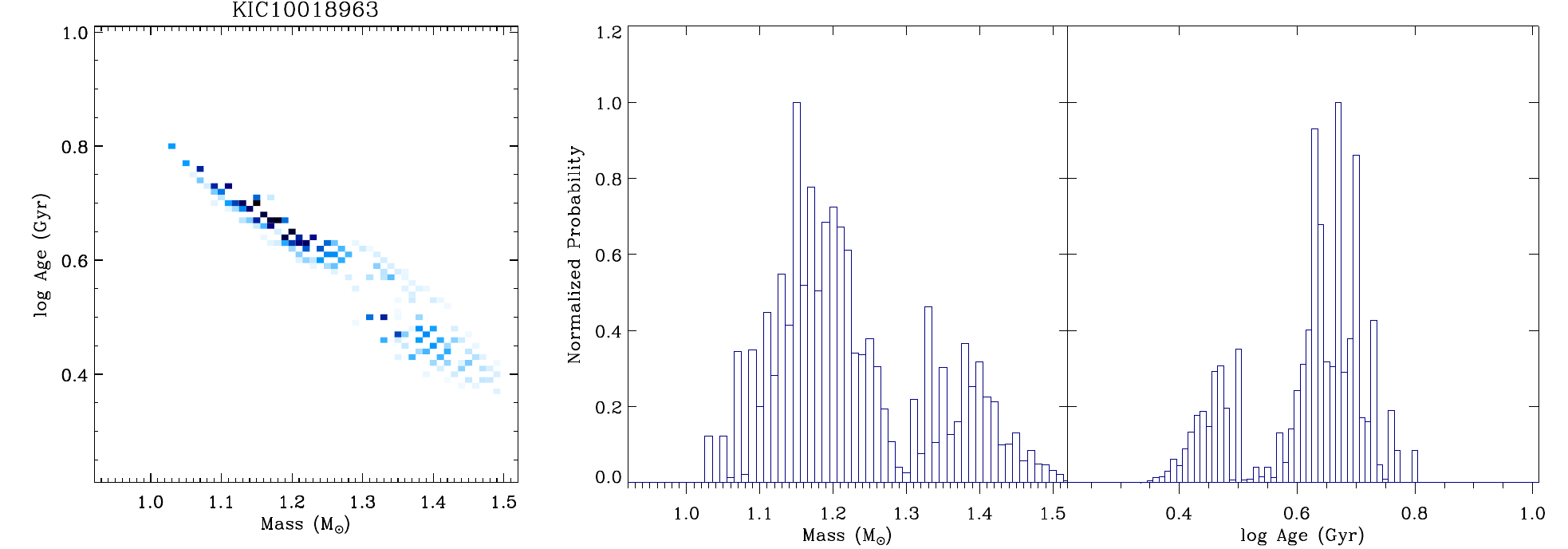}
    \caption{Three typical examples of the probability distributions of mass and age. Top: a relatively low-mass star (<1.3M$_{\odot}$) with sufficient grid sampling. Middle: a relatively massive stars (>1.3M$_{\odot}$) without sufficient grid sampling. Bottom: star close to the turn-off point or the base of red giant branch with broad and possible bimodal distributions. Probability distributions for all stars are presented in Appendix \ref{appB}.
    }
    \label{fig:estimation}
\end{figure*}

We used the marginal probability distributions to estimate masses, ages, surface gravities, radii, luminosities, and helium-core masses for the 31 {\em Kepler} subgiants. We present three typical examples of the probability distributions of mass and age in Figure~\ref{fig:estimation} and summarise as follows.
\begin{itemize}
     \item The probability distributions of 21 stars are continuous with clear centres, as in the example presented in the top panel of Figure~\ref{fig:estimation}. These stars all have relatively lower masses ($<1.3M_{\odot}$).
     \item Our grid sampling seems insufficient for six relatively massive stars with masses larger than $\sim$1.3M$_{\odot}$ and hence their distributions are not continuous as can be seen in the middle panel. This indicates that modelling these stars requires a denser grid.
     \item We also note that four stars show extraordinarily broad distributions, as shown in the bottom panel of Figure~\ref{fig:estimation}. Among them, KIC 5689620 and KIC 7174707 are the two coolest stars in our sample ($T_{\rm eff}$ = 5037K and 5168K) and close to the bottom of red giant branch. Their good-fitting models are at both late-subgiant and early-red-giant phases and hence make the distributions broad. The two phases in principle would give two different ages for each of the two stars, although they may be so close that this just adds to the uncertainty. 
     The other two stars, KIC 6766513 and KIC 10018963, are hot stars ($T_{\rm eff}$ = 6227K and 6117K) closed to the turn-off point. Moreover, KIC 10018963 presents a bimodal distribution with two maximums at $\sim$1.2M$_{\odot}$ and $\sim$1.4M$_{\odot}$ (and also two ages). These broad distributions indicate that larger uncertainties of mass and age are expected in subgiants close to the turning points on HR diagram. 
\end{itemize}
%
%
%

As mentioned in Section \ref{sec:MLE}, we fitted the probability distribution with a Gaussian function to estimate each stellar parameter.
For KIC 10018963, whose mass and age distributions are bimodal, we estimated two masses and ages.
We summarise our inferred stellar parameters for 31 stars in Table~\ref{tab:results}. Because the helium fraction and the mixing-length parameter were not adjusted in our `Grid-age', the uncertainties of the stellar parameters were mainly determined by the metallicity uncertainty. 
Average uncertainties of our estimates are $\sim15\%$ for the age and $\sim0.2\%$ for the surface gravity, corresponding to the typical observed uncertainties for the matellicity (0.15--0.20 dex), which agree with the dependencies we found in Section \ref{sec:dependence}. 
The average precision of other three inferred parameters are $\sim 4\%$ for the mass, $\sim 1.3\%$ for the radius, and $\sim 2.5\%$ for the helium-core masses. However, the estimates of mass and radius correlate with the inputs of helium fraction, and those of helium-core mass depend on the mixing-length parameters, hence our results have systematic uncertainties of a few percent.
%
%
%

\begin{table*}
	\centering
	\caption{Theoretical Stellar parameters of 31 {\em Kepler} subgiants given by `Grid-age'.}
	\label{tab:results}
	\begin{tabular}{ccccccc} 
		\hline
		KIC &Mass & Age  & $\log g$ & Radius & Luminosity & He-core mass \\
     		& [M$_{\odot}$] & [Gyr]&  [dex] & [R$_{\odot}$]& [L$_{\odot}$] &  [$M_{\odot}$]  \\   
            \hline
     2991448& 0.99$\pm$0.04&10.18$\pm$ 1.18&3.974$\pm$0.010& 1.70$\pm$ 0.04& 2.60$\pm$ 0.20& 0.094$\pm$ 0.004\\
   4346201& 1.33$\pm$0.02& 3.03$\pm$ 0.07&3.974$\pm$0.004& 1.98$\pm$ 0.03& 5.20$\pm$ 0.20& 0.098$\pm$ 0.007\\
   5108214& 1.50$\pm$0.01& 3.15$\pm$ 0.17&3.800$\pm$0.003& 2.55$\pm$ 0.03& 7.00$\pm$ 0.35& 0.128$\pm$ 0.005\\
   5607242& 1.23$\pm$0.09& 4.44$\pm$ 0.93&3.764$\pm$0.011& 2.38$\pm$ 0.06& 4.60$\pm$ 0.35& 0.139$\pm$ 0.003\\
   5689820& 1.19$\pm$0.12& 6.45$\pm$ 2.21&3.778$\pm$0.014& 2.33$\pm$ 0.08& 3.10$\pm$ 0.35& 0.143$\pm$ 0.005\\
   5955122& 1.12$\pm$0.05& 5.81$\pm$ 0.65&3.862$\pm$0.007& 2.04$\pm$ 0.04& 4.50$\pm$ 0.30& 0.107$\pm$ 0.004\\
   6064910& 1.51$\pm$0.02& 2.21$\pm$ 0.12&3.856$\pm$0.004& 2.42$\pm$ 0.02& 8.90$\pm$ 0.50& 0.110$\pm$ 0.002\\
   6370489& 1.12$\pm$0.05& 5.13$\pm$ 0.66&3.896$\pm$0.009& 1.98$\pm$ 0.03& 5.10$\pm$ 0.35& 0.096$\pm$ 0.003\\
   6442183& 0.96$\pm$0.07&10.57$\pm$ 1.95&4.002$\pm$0.011& 1.62$\pm$ 0.04& 2.50$\pm$ 0.20& 0.085$\pm$ 0.007\\
   6693861& 1.01$\pm$0.05& 7.57$\pm$ 0.97&3.828$\pm$0.009& 2.04$\pm$ 0.05& 3.70$\pm$ 0.30& 0.119$\pm$ 0.004\\
   6766513& 1.33$\pm$0.10& 2.62$\pm$ 0.15&3.926$\pm$0.016& 2.11$\pm$ 0.05& 6.40$\pm$ 0.35& 0.101$\pm$ 0.002\\
   7174707& 1.08$\pm$0.10& 8.40$\pm$ 2.12&3.840$\pm$0.012& 2.07$\pm$ 0.06& 2.70$\pm$ 0.20& 0.128$\pm$ 0.003\\
   7199397& 1.39$\pm$0.03& 3.09$\pm$ 0.08&3.754$\pm$0.005& 2.57$\pm$ 0.02& 7.40$\pm$ 0.35& 0.135$\pm$ 0.003\\
   7747078& 1.11$\pm$0.06& 6.22$\pm$ 0.79&3.912$\pm$0.008& 1.92$\pm$ 0.03& 4.00$\pm$ 0.25& 0.098$\pm$ 0.004\\
   7976303& 1.10$\pm$0.05& 5.36$\pm$ 0.59&3.886$\pm$0.007& 1.97$\pm$ 0.03& 5.00$\pm$ 0.30& 0.096$\pm$ 0.003\\
   8524425& 0.97$\pm$0.04&10.40$\pm$ 1.14&3.952$\pm$0.009& 1.71$\pm$ 0.03& 2.60$\pm$ 0.20& 0.097$\pm$ 0.006\\
   8702606& 1.22$\pm$0.09& 4.33$\pm$ 0.77&3.752$\pm$0.010& 2.42$\pm$ 0.06& 4.80$\pm$ 0.45& 0.138$\pm$ 0.005\\
   8738809& 1.53$\pm$0.01& 2.51$\pm$ 0.17&3.916$\pm$0.004& 2.26$\pm$ 0.03& 6.20$\pm$ 0.25& 0.107$\pm$ 0.006\\
   9512063& 1.12$\pm$0.07& 5.81$\pm$ 0.89&3.866$\pm$0.009& 2.04$\pm$ 0.04& 4.30$\pm$ 0.35& 0.110$\pm$ 0.006\\
  10018963*& 1.18$\pm$0.07/1.40$\pm$0.04& 4.56$\pm$0.52/2.81$\pm$0.35&3.944$\pm$0.017& 1.94$\pm$ 0.05& 5.00$\pm$ 0.40& 0.090$\pm$ 0.004\\
  10147635& 1.47$\pm$0.06& 2.75$\pm$ 0.06&3.740$\pm$0.005& 2.69$\pm$ 0.05& 8.50$\pm$ 0.40& 0.133$\pm$ 0.005\\
  10273246& 1.49$\pm$0.08& 2.84$\pm$ 0.60&3.906$\pm$0.009& 2.25$\pm$ 0.03& 6.90$\pm$ 0.45& 0.107$\pm$ 0.005\\
  10920273& 0.94$\pm$0.05&12.20$\pm$ 1.52&3.926$\pm$0.008& 1.75$\pm$ 0.04& 2.30$\pm$ 0.20& 0.108$\pm$ 0.003\\
  10972873& 1.06$\pm$0.07& 7.97$\pm$ 1.28&3.954$\pm$0.010& 1.80$\pm$ 0.04& 3.10$\pm$ 0.20& 0.096$\pm$ 0.005\\
  11026764& 1.18$\pm$0.08& 5.69$\pm$ 0.68&3.884$\pm$0.009& 2.06$\pm$ 0.04& 3.80$\pm$ 0.25& 0.114$\pm$ 0.004\\
  11137075& 0.91$\pm$0.05&13.21$\pm$ 2.17&4.002$\pm$0.009& 1.58$\pm$ 0.04& 2.00$\pm$ 0.20& 0.093$\pm$ 0.006\\
  11193681& 1.36$\pm$0.07& 3.72$\pm$ 0.10&3.812$\pm$0.007& 2.40$\pm$ 0.04& 4.90$\pm$ 0.40& 0.134$\pm$ 0.003\\
  11395018& 1.23$\pm$0.07& 4.73$\pm$ 0.61&3.860$\pm$0.009& 2.16$\pm$ 0.04& 4.50$\pm$ 0.50& 0.117$\pm$ 0.004\\
  11414712& 1.18$\pm$0.08& 4.93$\pm$ 0.70&3.804$\pm$0.010& 2.25$\pm$ 0.05& 4.50$\pm$ 0.40& 0.127$\pm$ 0.004\\
  11771760& 1.39$\pm$0.06& 2.65$\pm$ 0.07&3.650$\pm$0.004& 2.95$\pm$ 0.03& 9.50$\pm$ 0.85& 0.153$\pm$ 0.003\\
  12508433& 1.27$\pm$0.07& 4.97$\pm$ 0.70&3.832$\pm$0.010& 2.25$\pm$ 0.05& 3.60$\pm$ 0.30& 0.135$\pm$ 0.003\\
\hline
    \multicolumn{7}{p{.7\textwidth}}{$^{*}$Two masses and ages were determined because of the bimodal probability distributions.}\\
	\end{tabular}
\end{table*}


\subsection{Comparison with previous studies}\label{sec:comparison}

We compared our inferred ages with previous results that were also based on detailed modelling.
We found 14 stars from our sample in the literature and list those results in Table~\ref{tab:comparison}. 
The major differences of the input physics between previous studies and ours are as follow: 
\begin{itemize}
\item \citet{2017A&A...600A.128B} modelled two stars in our sample (KIC 8702606 and 12508433) with adjustable helium abundance and mixing-length parameter, and they adopted an Eddington grey atmosphere as the boundary condition and included element diffusion. 
\item Ages given by \citet{2014A&A...564A..27D} for the same two stars (KIC 8702606 and 12508433) were from two models. First, \textsc{CESAM2K} models were calculated with adjustable helium fraction and mixing-length parameter. Differences in input physics include: the Eddington grey law for describing the atmosphere, the Canuto-Goldman-Mazzitelli (CGM) formalism for the convection, and no convective overshooting in the core. Second, \textsc{ASTEC} models were computed with free initial helium content and an Eddington grey atmosphere. 
\item \citet{2014ApJS..214...27M} used the \textsc{AMP} code to model six stars in our sample. They adopted the \citet{1993A&A...271..587G} (GN93) solar mixture and helium diffusion, but neglected convective overshooting.  
\item \citet{2015MNRAS.447..680G} studied KIC 7976303 with the \textsc{YREC} code and they used alpha-enhanced opacity and neglected core overshooting.
\item \citet{2015A&A...580A..44T} also used the \textsc{YREC} for modelling KIC 6442183 and 11137075 and they adopted GN93 solar mixture, an Eddington grey atmosphere, and element diffusion of helium and heavy elements. 
\item Recent studies of KIC 6766513 and KIC 10147635 \citep{2017RAA....17...44L} using the \textsc{YREC} included element diffusion of helium and heavy elements but not convective overshooting.
\item \citet{2013ApJ...763...49D} characterised KIC 10920273 and KIC 11395018 with five different pipelines and averaged results. The five modelling approaches are all slightly different from our modelling in terms of atmosphere, element diffusion, overshooting, etc.   
\end{itemize}
Although the input physics adopted in these studies is not quite the same as ours, it can be seen in Table~\ref{tab:comparison} that inferred ages for most stars show good agreement (within 2$\sigma$).
Large age differences always correlate with differences in input heavy elements.
We found only three previous results that deviate from our determinations by more than 2$\sigma$. For the cases of KIC 8524425 and KIC 8702606, \citet{2010ApJ...723.1583M} used a higher [M/H] and also a large value of the solar mixture, and hence estimated smaller ages than ours. For KIC 10920273, we obtained a significantly larger age than \citet{2013ApJ...763...49D} that is presumably due to our lower input metallicity.
The good match overall suggests that seismic ages of subgiants could be relatively robust with respect to different stellar code and input physics, provided the metallicity is fixed.

\begin{table*}
	\centering
	\caption{Comparisons with literature ages based on the detailed modelling.}
	\label{tab:comparison}
	\begin{tabular}{c|ccc|cccc} 
		\hline
            \multicolumn{1}{c|}{}&\multicolumn{3}{c|}{This work}&\multicolumn{4}{c}{Literature}\\
		KIC & Age & [M/H] & Solar Mixture& Age & [M/H] & Solar Mixture& Ref. \\
     		& [Gyr]& [dex]& &[Gyr] & [dex] && \\   
            \hline
5689820&6.45$\pm$2.21&0.22&GS98&7.35$\pm$0.10,5.6-6.9& 0.24,0.24 & GS98,GN93 & 1,2 \\
5955122& 5.81$\pm$ 0.65&-0.20&GS98&5.26$\pm$0.58& -0.17&GN93& 3 \\
6442183&10.57$\pm$ 1.95&-0.20&GS98&$8.65^{+1.12}_{-0.06}$&-0.11&GN93&5\\
6766513& 2.62$\pm$ 0.15&-0.17&GS98&$3.94^{+0.52}_{-0.30}$&-0.18&GS98&6\\
7747078&6.22$\pm$ 0.79&-0.24&GS98&6.26$\pm$0.92&-0.26&GN93&3\\
7976303&5.36$\pm$ 0.59&-0.50&GS98&4.78$\pm$0.58,4.88$\pm$0.08& -0.53,-0.53 & GS98,GN93&3,4\\
8524425&10.40$\pm$ 1.14&0.07&GS98&7.98$\pm$0.46&0.14&GN93&3\\
8702606&4.33$\pm$ 0.77&-0.15&GS98&3.84$\pm$0.03,3.8-4.1,2.23$\pm$0.13 & -0.09,-0.09,-0.09 & GS98,GN93,GN93&1,2,3 \\
10147635&2.75$\pm$ 0.06&-0.02&GS98&3.34$^{+0.54}_{-0.30}$&-0.08&GS98&6\\
10920273&12.20$\pm$1.52 &-0.16 &GS98&7.12$\pm$0.47&-0.04&-&7\\
11026764&5.69$\pm$ 0.68&0.04&GS98&5.00$\pm$0.53&0.05&GN93&3\\
11137075&13.21$\pm$2.17&-0.13&GS98&$10.36^{+0.01}_{-0.20}$&-0.06&GN93&5\\
11395018&4.73$\pm$0.61& 0.02&GS98&4.57$\pm$0.23&0.13&-&7\\
12508433&4.97$\pm$0.70&0.23&GS98&5.07$\pm$0.13,5.1-5.9&0.25,0.25 & GS98,GN93&1,2 \\
\hline \\
    \end{tabular} 
    \\
References:1 -- \cite{2017A&A...600A.128B} (\textsc{MESA} ); 2--\citet{2014A&A...564A..27D} (\textsc{ASTEC} and \textsc{CESAM2K}); 3--\citet{2014ApJS..214...27M} (\textsc{AMP}); 4 -- \citet{2015MNRAS.447..680G} (\textsc{YREC}); 5 -- \citet{2015A&A...580A..44T} (\textsc{YREC}); 6 -- \citet{2017RAA....17...44L} (\textsc{YREC});7 --\citet{2013ApJ...763...49D} (\textsc{multiple pipelines}).
\end{table*}

\section{The p--g diagram}
\label{sec:p-g-diagram}

In this section, we step back from the detailed modelling of individual frequencies to address a topic that was presented in Paper~I, namely the so-called p--g diagram.  As suggested by \cite{bedding2014asteroseismology}, this compares the observed frequencies of the avoided crossings with theoretical models.  The rationale is that much of the information from the dipole mixed modes is contained in the positions of these avoided crossings, which coincide with the underlying g-modes that would be present if the core could be isolated from the envelope \citep{aizenman++1977-avoided-crossing}.  The p--g diagram is made by plotting the frequencies of avoided crossings, which we denote by $\gamma$, against the large separation of the p~modes ($\Delta\nu$).  This diagram is a quick and convenient way to compare observations with theory and to make first estimates of mass and age.

\begin{figure*}
 \centering
\includegraphics[width=2.0\textwidth]{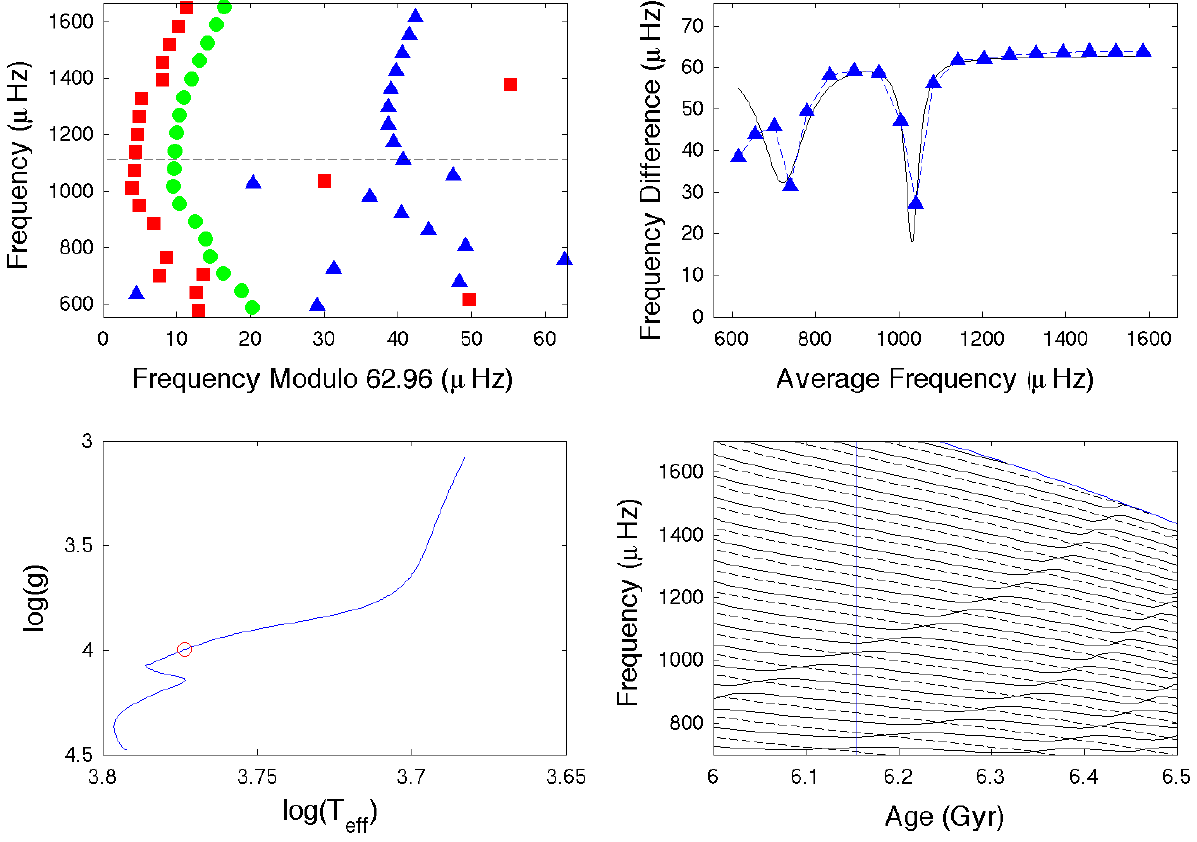}
\caption{Measuring the avoided crossing in the model of a 1.1-\Msun\ star with solar metallicity ($Z =
0.0142$). Upper left: \'Echelle diagram for this model. Green circles represent $\ell = 0$ modes, blue triangles represent $\ell = 1$ modes and red squares represent $\ell=2$ modes.  The horizontal dashed line indicates~$\nu_{\rm max}$. Upper right: Frequency differences between consecutive $\ell = 1$ modes, with Lorentzian fits to the minima. Lower left: Evolutionary track showing the surface gravity plotted against effective temperature, with the red circle indicating the model shown in the upper panels. Lower right: Evolution with time of frequencies for $\ell=0$ modes (dashed lines) and $\ell = 1$ modes (solid lines), with the acoustic cutoff frequency shown in the upper right as a blue line. The vertical line shows the age of the model in the upper panels.}
    \label{fig:pg-snapshot}
\end{figure*}

\begin{figure*}
\includegraphics[width=0.75\textwidth]{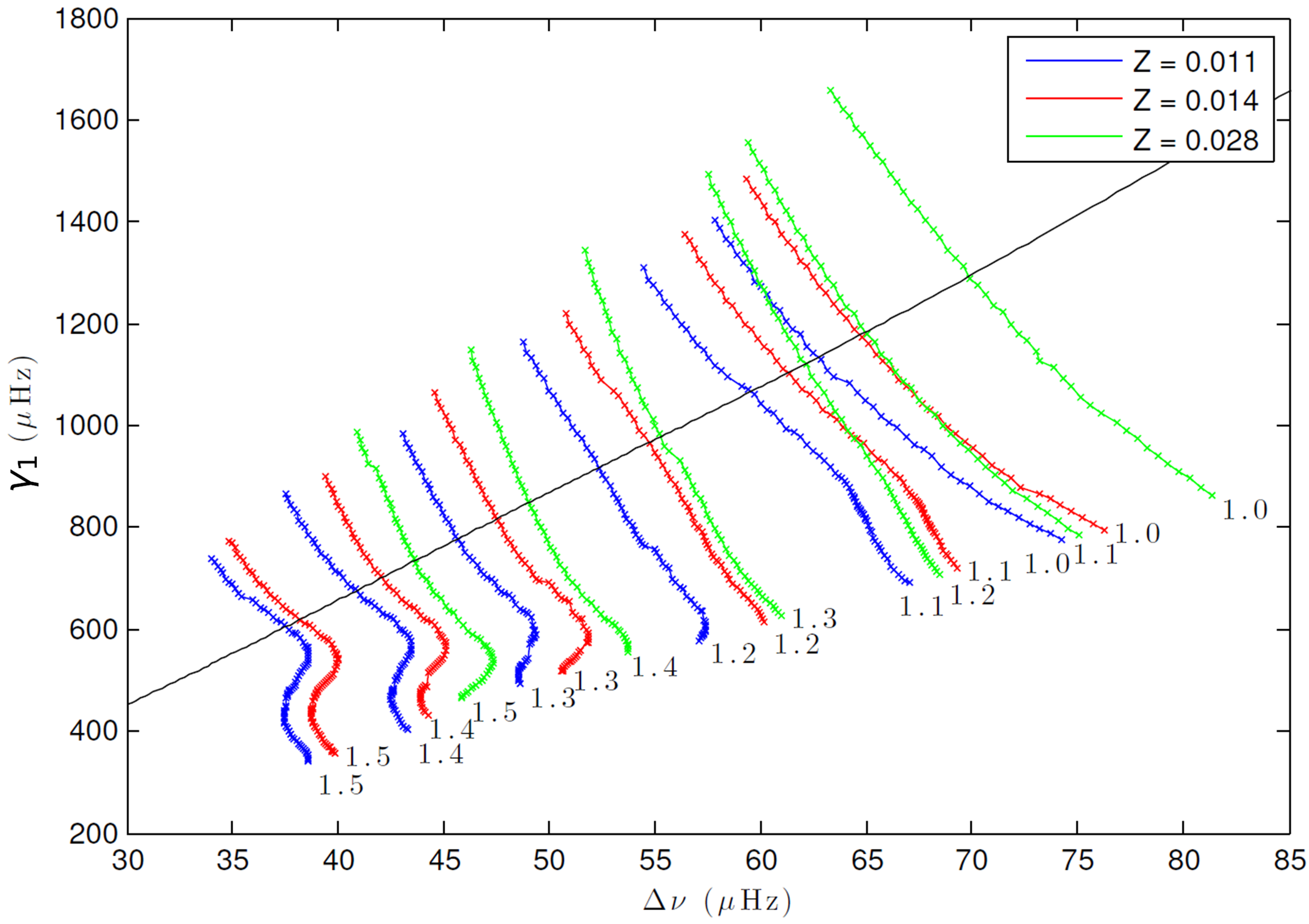}
\includegraphics[width=0.75\textwidth]{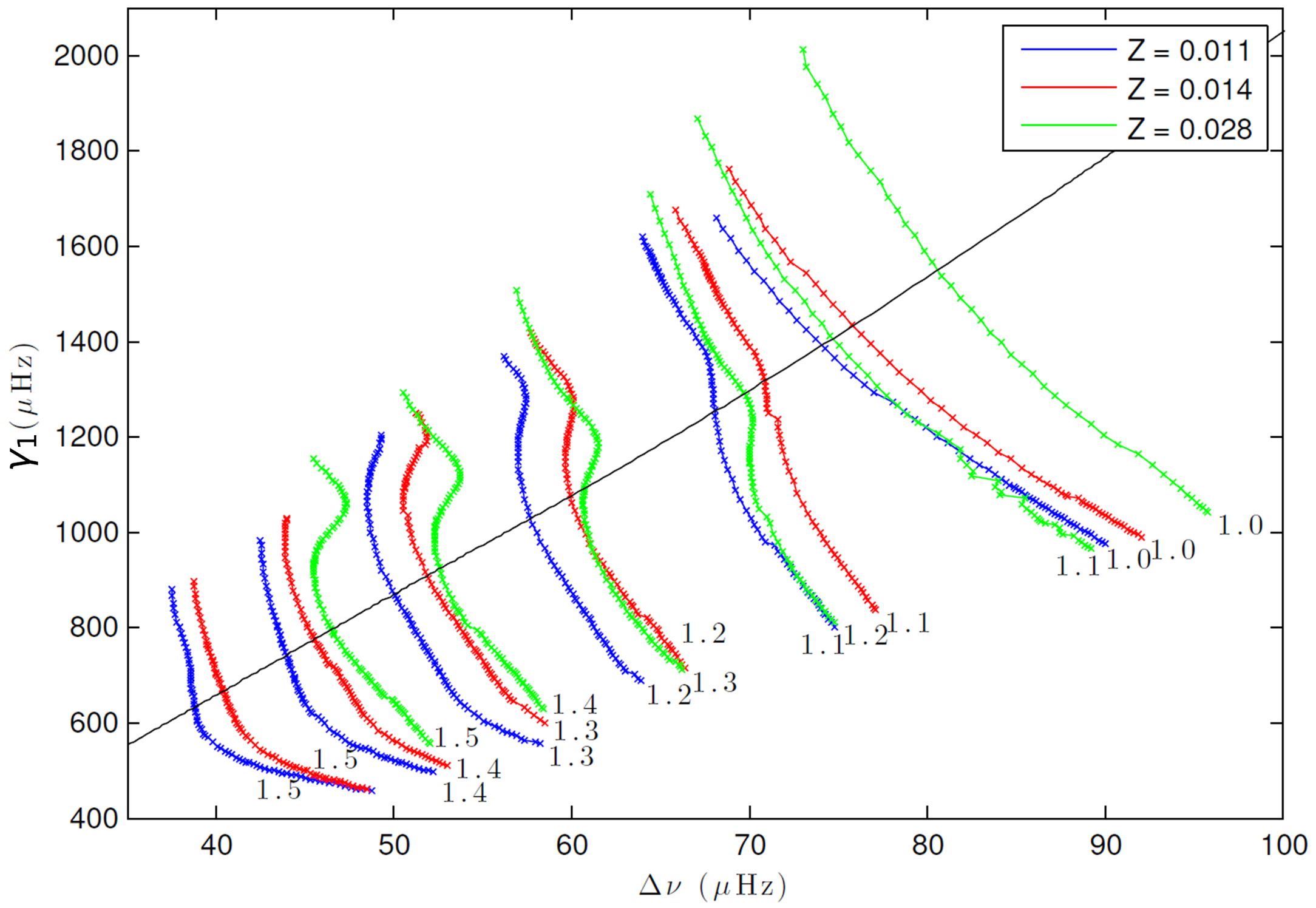}
\caption{Theoretical p-g diagram, showing the frequency of $\ell=1$(top) and $\ell = 2$ (bottom) avoided crossings versus the large separation.  The vertical axis, $\gamma_1$, is the frequency of the first avoided crossing ($n_g=1$).  We show models with masses ranging from 1.0 to 1.5\,\Msun\ in increments of 0.1\,\Msun\ (as indicated on each curve) and three different metallicities. Crosses indicate the individual models, and evolution is from bottom-right to top-left.  The diagonal black line is $\nu_{\rm max}$.}
    \label{fig:pg-1}
\end{figure*}

Here, we briefly describe our method of calculating the theoretical p--g diagram using the MESA models described above. Figure~\ref{fig:pg-snapshot} shows a 1.1-\Msun\ model in the early subgiant phase.  The lower-right panel shows the oscillation frequencies as a function of time for the $\ell=0$ and $\ell=1$ modes.  The latter show upward-sloping features that correspond to the underlying g modes, whose frequencies increase with time.  The avoided crossings occur each time the frequency of one of these g modes comes close to a p mode. We located the $\ell = 1$ avoided crossings by plotting the pairwise differences between consecutive orders (upper-right panel) and measured the positions of the avoided crossings by fitting Lorentzian profiles (solid curve).  Figure~\ref{fig:pg-1} shows the results for the dipole modes ($\ell=1$), where we show the avoided crossing with the highest frequency (i.e., the g~mode with radial order $n_g=1$). The models evolve towards the upper left, with the p-mode frequency spacing ($\Delta\nu$) decreasing with time and the g-mode frequency ($\gamma_1$) increasing.  The black diagonal line shows the approximate value of $\nu_{\rm max}$, so that modes that are close to this frequency should have the highest amplitudes.  Provided the metallicity is known, this diagram can be used to estimate the mass and age based on two simple asteroseismic parameters, $\gamma_1$ and~$\Delta\nu$, as shown in Paper~I (see also \citealt{campante++2011-subgiants-kic10273246-kic10920273}).  Note that this diagram applies to stars in the early to mid subgiant phase, where $\gamma_1$ falls within the observed oscillation envelope.  

\begin{figure}
\centerline{\includegraphics[width=1.0\linewidth]{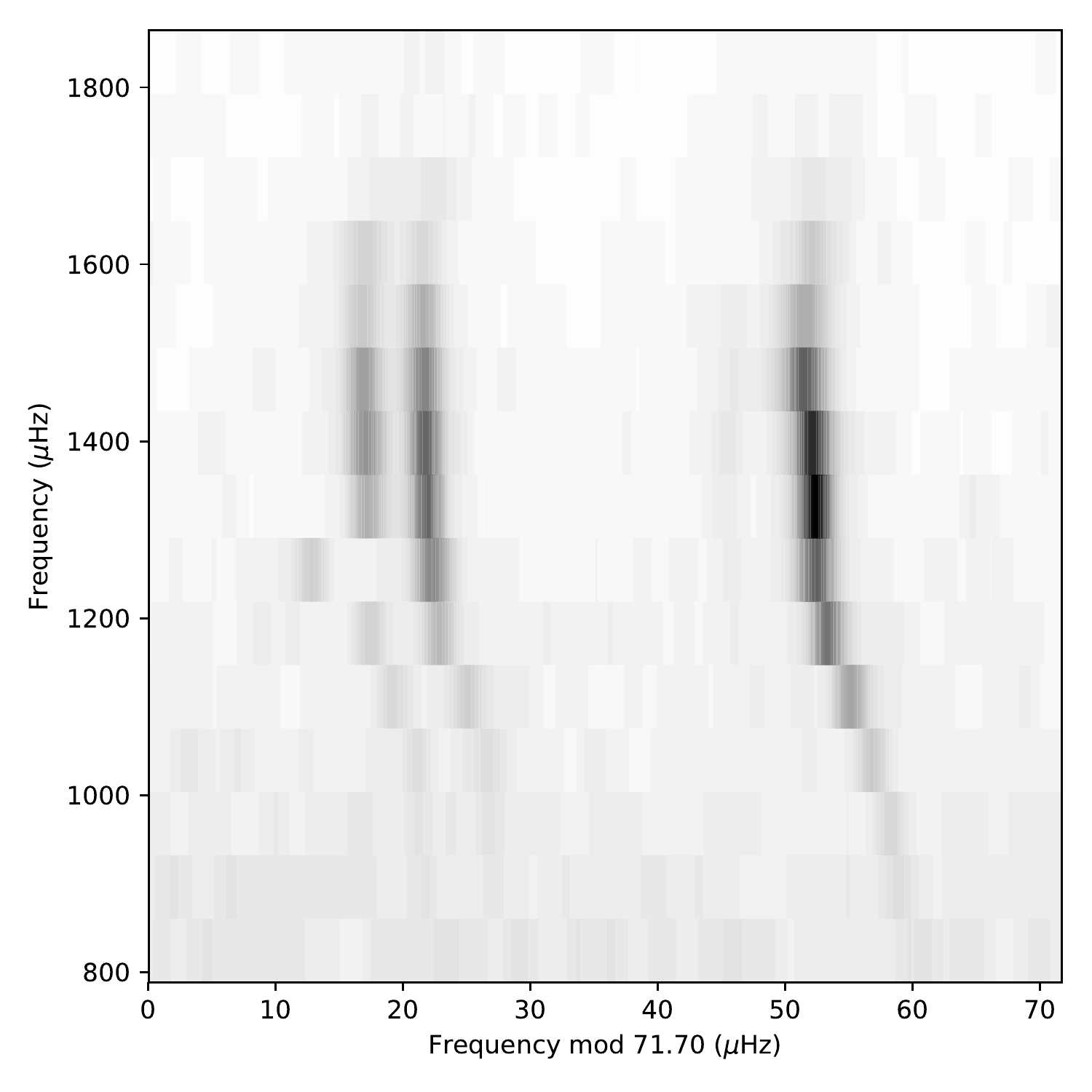}}
\caption{Power spectrum observed by {\em Kepler} of the late-main-sequence star KIC~11244118, shown in {\'e}chelle format with an avoided crossing of the $\ell=2$ modes at 1230\,$\mu$Hz.}
    \label{fig:kic-11244118}
\end{figure}

At the bottom of Figure~\ref{fig:pg-1} shows a similar diagram for the highest-frequency $\ell=2$ avoided crossing.  Note that these models are less evolved than those for $\ell=1$, reflecting the fact that $\gamma_1$ for $\ell=2$ modes falls within the observed oscillation envelope at an earlier stage in evolution.  This is why bumping of $\ell=2$ modes is sometimes seen on the main sequence, as in the example shown in Fig.~\ref{fig:kic-11244118} of KIC~11244118 \citep{2012A&A...543A..54A,2012ApJ...749..152M,2014ApJS..214...27M}.  This means that the p--g diagram for $\ell=1$ modes is applicable to early subgiants, while the diagram for $\ell=2$ modes is applicable to stars on the late main sequence.  One can also add the avoided crossing with the second-highest frequency (i.e., $\gamma_2$, the g~mode with radial order $n_g=2$), which would extend the application to more evolved subgiants.  For even more evolved stars on the red giant branch, the period spacing of the g modes has proved to be a valuable diagnostic \citep{2011Sci...332..205B,2011Natur.471..608B,2014A&A...572L...5M,2016A&A...588A..87V}.

\section{Conclusions}\label{sec:conclusions}
In this work, we modelled 36 {\em Kepler} subgiants by fitting their asteroseismic frequencies to study the age dependencies on three model input parameters. The findings and conclusions are as follows:  
\begin{itemize}
\item We found that seismic ages of subgiants do not change systematically with the input helium fraction or the mixing-length parameter. However, they strongly depend on the input metallicity.
\item The lack of dependencies allowed us to remove two free dimensions ($Y_{\rm init}$ and $\alpha _{\rm MLT}$) in the computation of model grids when modelling ages of subgiants, which significantly improved the efficiency.
\item The accuracy and the efficiency in deriving stellar ages with asteroseismology is a big advantage of subgiants.  
\item We estimated ages for 31 {\em Kepler} subgiants with an average uncertainty of $\sim$15\% (corresponding to a 0.15-0.20 dex uncertainty in the metallicity).   
\item Our ages agree well with previous results derived with different stellar codes and input physics. We conclude that seismic-determined ages of subgiants are much less model-dependent than for dwarfs and red giants.
\item For other stellar parameters, determinations of the surface gravity do not change systematically with any of the three input parameters; inferred masses and radii correlate with the helium fraction; and estimates of the helium-core mass relate to the mixing-length parameter.
\item We show that the so-called p--g diagram, which plots the evolution of the highest-frequency avoided crossing versus the p-mode large separation, can be used to estimate mass and age (provided the metallicity is known).
\end{itemize}

While the $\it Kepler$ mission provided short-cadence data for only tens of subgiants, the on-going Transiting Exoplanet Survey Satellite (TESS) mission \citep{2016SPIE.9904E..2BR}
and future projects like PLAnetary Transits and Oscillations of stars (PLATO) \citep{2014ExA....38..249R} will obtain asteroseismic data for thousands of subgiants. Thus, further works about subgiant ages with a bigger sample will be useful for studies of the galactic evolution. To improve the results in this work, a denser model grid or better interpolation schemes (especially for relatively massive stars) will be necessary for solving the issues with poorly sampled probability distributions. Moreover, including other effects such as diffusion and convective overshooting which impact the main-sequence lifetime could reveal additional systematic uncertainty on the age estimates. Further studies of those effects are hence required.

\section*{Acknowledgements}
We gratefully acknowledge the {\em Kepler} team, whose efforts made these results possible.
We gratefully acknowledge support from the Australian Research Council (grant DE~180101104), and from the Danish National Research Foundation (Grant DNRF106) through its funding for the Stellar Astrophysics Center (SAC).
This work is also supported by the Joint Research Fund in Astronomy (U1631236) under cooperative agreement between the National Natural Science Foundation of China (NSFC) and Chinese Academy of Sciences (CAS).
This work has also received funding from the European Research Council (ERC) under the European Union’s Horizon 2020 research and innovation programme (CartographY GA. 804752).
This research was partially conducted during the Exostar19 program at the Kavli Institute for Theoretical Physics at UC Santa Barbara, which was supported in part by the National Science Foundation under Grant No. NSF PHY-1748958.



\bibliographystyle{mnras}
\bibliography{aalib} 



\clearpage
\appendix
\onecolumn
\section{The implementation of the weighting factor}\label{appC}
\begin{figure}
\centering
 \includegraphics[width=0.6\columnwidth]{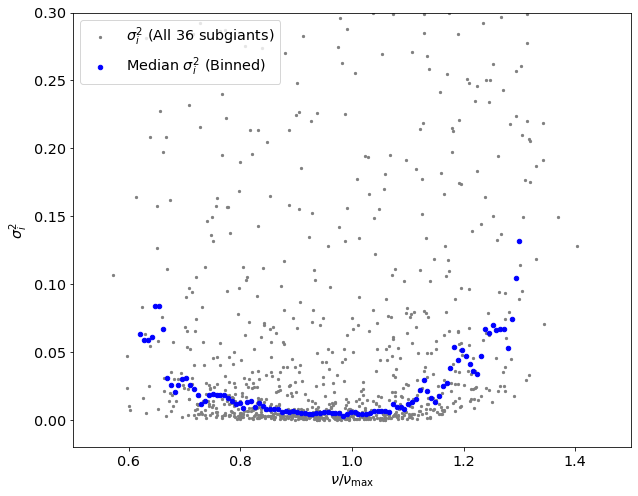}
    \caption{Squared observed uncertainties against scaled mode frequencies of 36 Kepler subgiants. Grey points indicate the observed mode frequencies; blues points are median values of binned observed uncertainties.}
    \label{fig:obs_errors}
\end{figure}

In this section, we explain the reason for implementing the weighting factor $w_{i}$ in Eq. \ref{eq:weights}. In Eq. \ref{eq:chi2_seis}, we applied a model systematic uncertainty $\sigma_{sys}$, which is $\sim$0.5$\mu$Hz for $\ell=0$ modes and $\sim$0.8$\mu$Hz for $\ell = 1$ and 2 modes. These values are much larger than most observed uncertainties. We show all observed frequency uncertainties and their binned median values as a function of $\nu_{\rm max}$ in Figure \ref{fig:obs_errors}. As it can be seen, most observed uncertainties are much smaller than the model systematic uncertainties. Because of this, $\sigma^{2}_{sys}$ becomes the dominant part of the term $\sigma^{2}_{i}$ + $\sigma^{2}_{sys}$ for most modes. Consequently, the weights of mode frequencies are mainly determined by $\sigma^{2}_{sys}$ and are hence relatively uniform. As a result, poorly measured modes (which are generally poorly fitted) impact the fitting more significantly than they should.

\begin{figure}
\centering
 \includegraphics[width=\columnwidth]{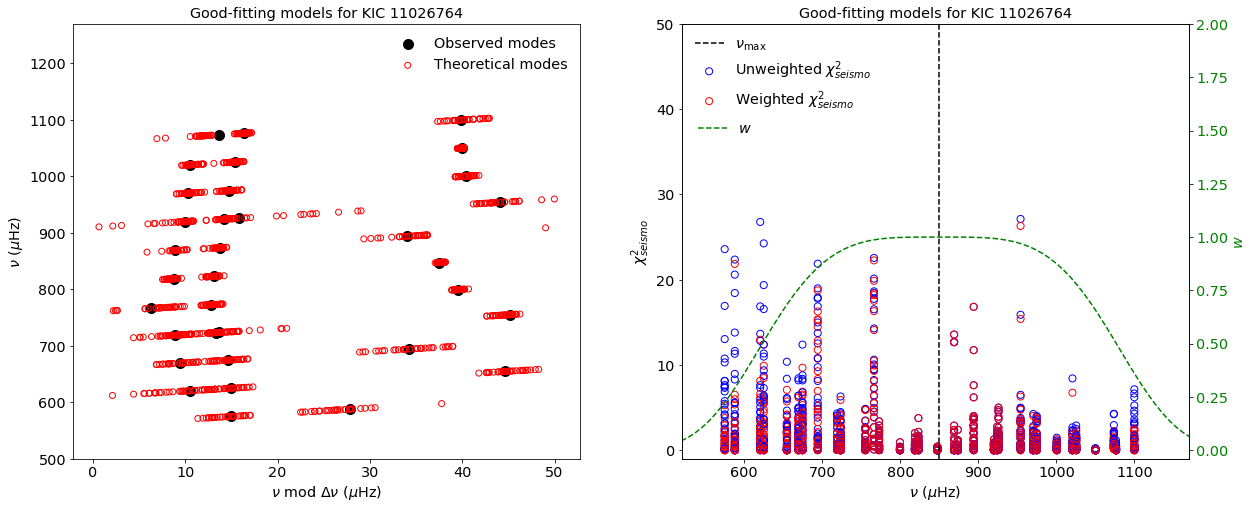}
    \caption{Good-fitting models for KIC 11026764 on the \'Echelle diagram (left) and seismic $\chi^{2}$ as a function of the mode frequency before and after weighting (right). The green dashed line in the right graph, using the right-hand ordinate, shows the weight function $w_i$.}
    \label{fig:wi_eff}
\end{figure}

In Figure \ref{fig:wi_eff}, we include an example to illustrate how poorly measured modes affect our fitting and how the weighting factor solves the issue. We first fitted models to observed frequencies with no weighting and plotted the top 30 models in the \'Echelle diagram. It can be seen that mode frequencies close to the upper and lower edges are poorly fitted (because they are poorly measured) compared with those around $\nu_{\rm max}$. With a normal $\chi^{2}$ function, poorly measured modes are down-weighted by their large uncertainty and hence their $\chi^{2}$ values are generally at the same level with other modes. However, because of the implementation of $\sigma_{sys}$, which makes the term $\sigma^{2}_{i}$ + $\sigma^{2}_{sys}$ relatively uniform for all modes, seismic $\chi^{2}$ values of poorly fitted modes are relatively large and also cover very wide ranges. We show seismic $\chi^{2}$ values of individual modes in the right panel. As it shown that seismic $\chi^{2}$ values close to both edges are up to 5 times larger than those around $\nu_{\rm max}$. Clearly, it is not sensible to let poorly measured modes lead the fitting. We hence introduced the weighting factor ($w_{i}$) to down-weight them. According to the uncertainty distribution in Figure \ref{fig:obs_errors}, observed uncertainties are generally small around $\nu_{\rm max}$, obviously rise at $\sim$0.8$\nu_{\rm max}$ and $\sim$1.2$\nu_{\rm max}$, and keep increasing with the distance away from $\nu_{\rm max}$. We aimed to adjust seismic $\chi^{2}$ values for all modes to the same level, and hence a super Gaussian is applied to calculate $w_{i}$ (indicated by the green dashed line). It can be seen that the weighting factor significantly reduces the large $\chi^{2}$ at both edges.

\begin{figure}
\centering
 \includegraphics[width=0.6\columnwidth]{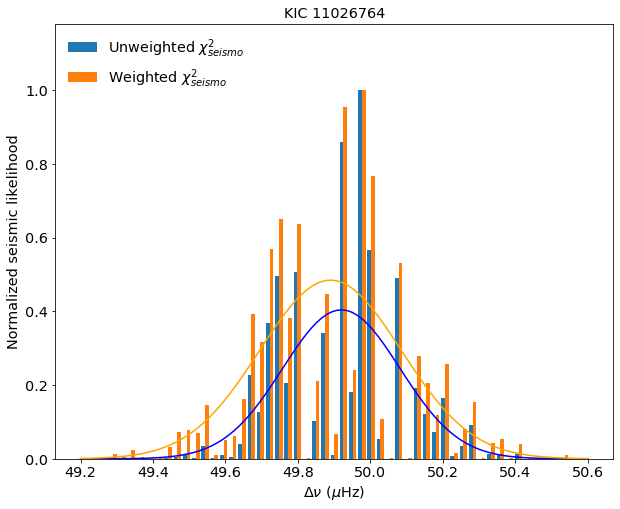}
    \caption{Probability distributions of theoretical $\Delta \nu$ for KIC 11026764 with and without the weighting factor.}
    \label{fig:prob_diff}
\end{figure}

We lastly demonstrate the effect of weighting factor on the probability distribution in Figure \ref{fig:prob_diff}. We plotted the likelihood as a function of $\Delta \nu$ because this describes how the likelihood changes when model frequencies move away from the best fit. The obvious difference between the distributions before and after weighting is the width. Without the weighting factor, the $\chi^{2}$ values of poorly measured modes substantially increase when the goodness of fitting decrease. Their influences on the average seismic likelihood also rise up in the meantime. This leads to a faster reduction in the likelihood. Thus, equal weighting for poorly measured modes could affect the estimate for uncertainty. For this reason, we find that it is more sensible to re-weight mode frequencies to ensure that the fitting is mainly determined by well-measured modes.

\clearpage
\section{\'Echelle diagrams of the best-fitting models of 31 {\em Kepler} subgiants based on `Grid-age' (sorted by KIC number)}\label{appA}

\begin{figure}
 \includegraphics[width=0.5\columnwidth]{KIC2991448-eps-converted-to.pdf}
   \includegraphics[width=0.5\columnwidth]{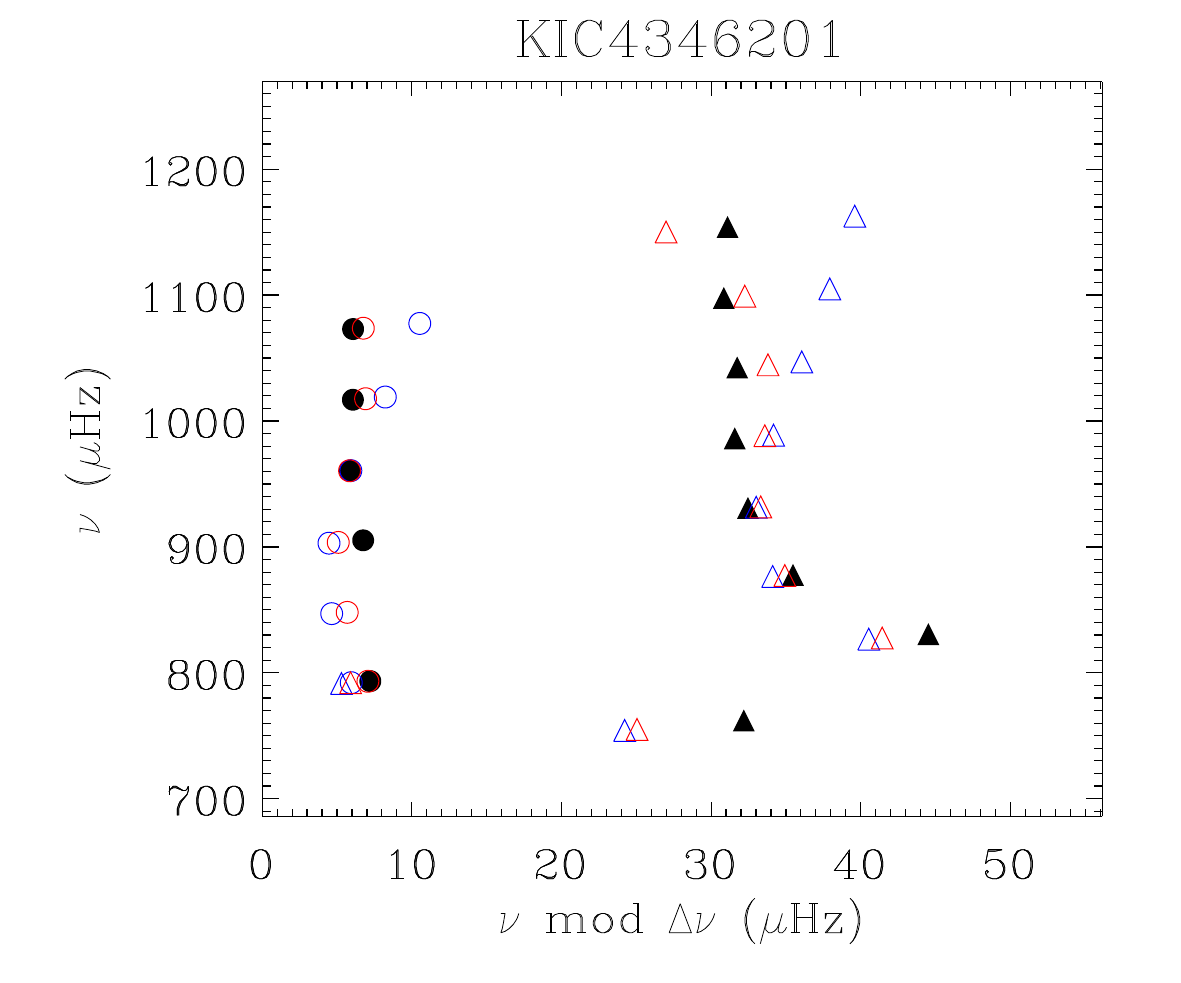}
 \includegraphics[width=0.5\columnwidth]{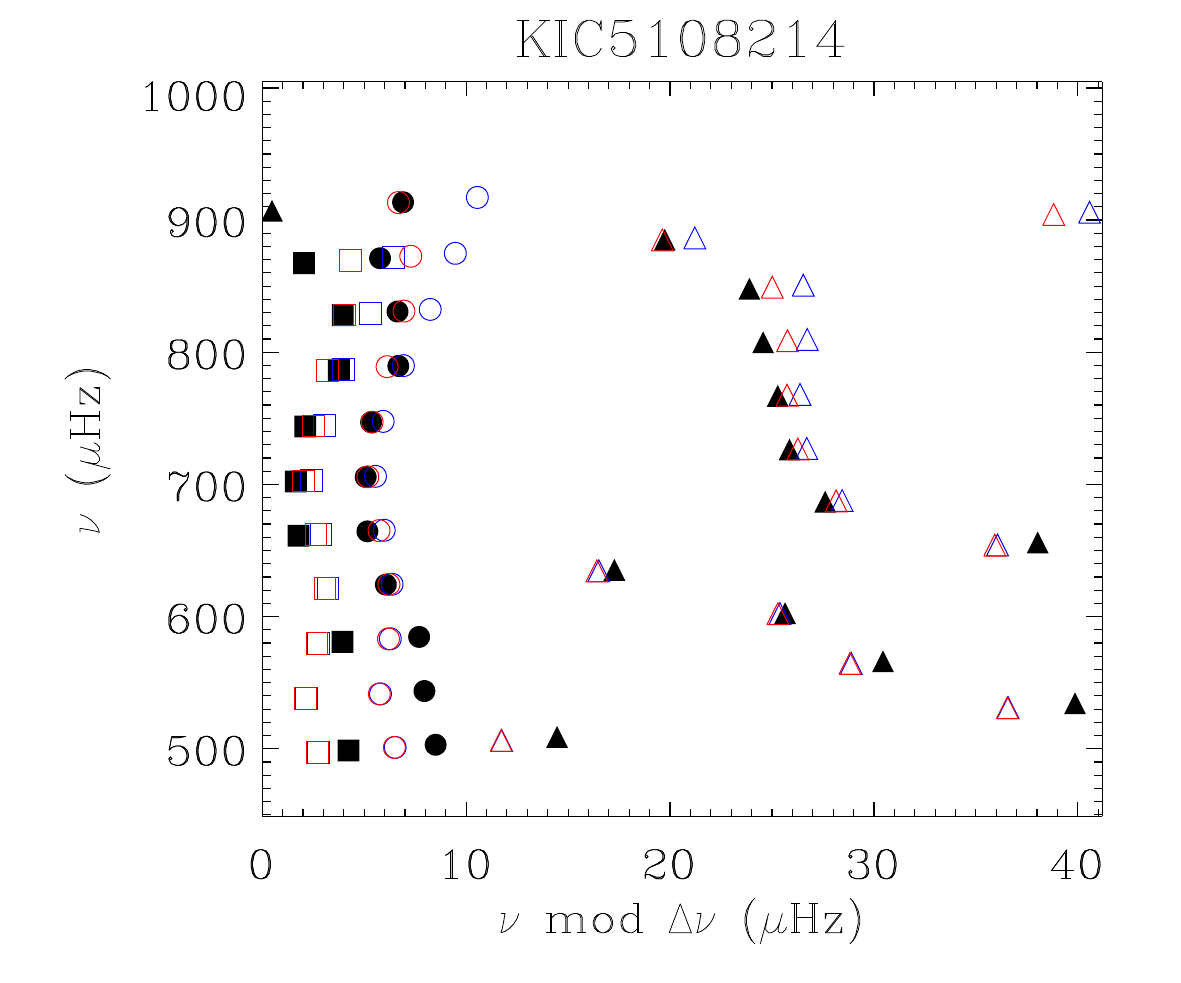}
  \includegraphics[width=0.5\columnwidth]{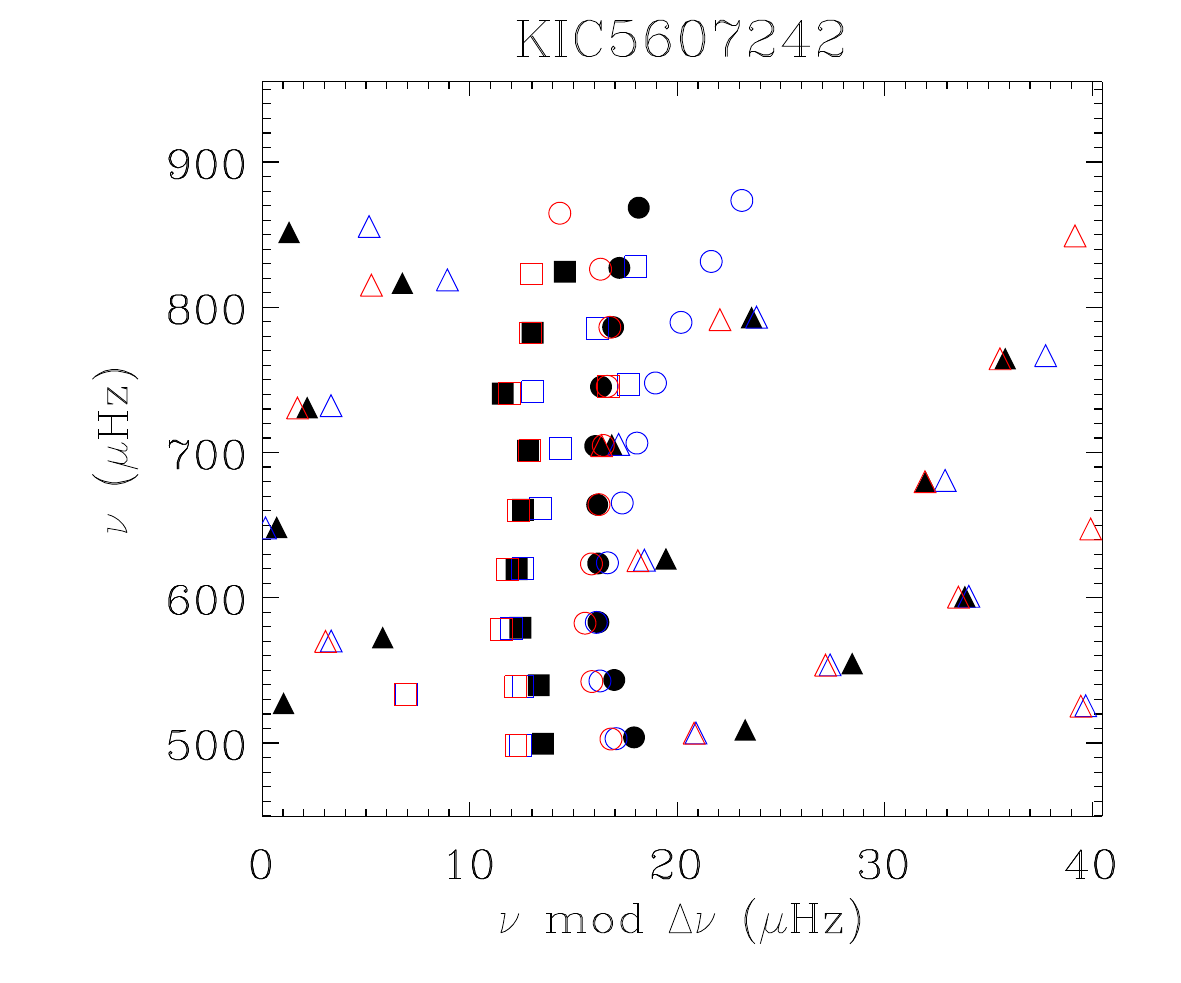}
  \includegraphics[width=0.5\columnwidth]{KIC5689820-eps-converted-to.pdf}
    \includegraphics[width=0.5\columnwidth]{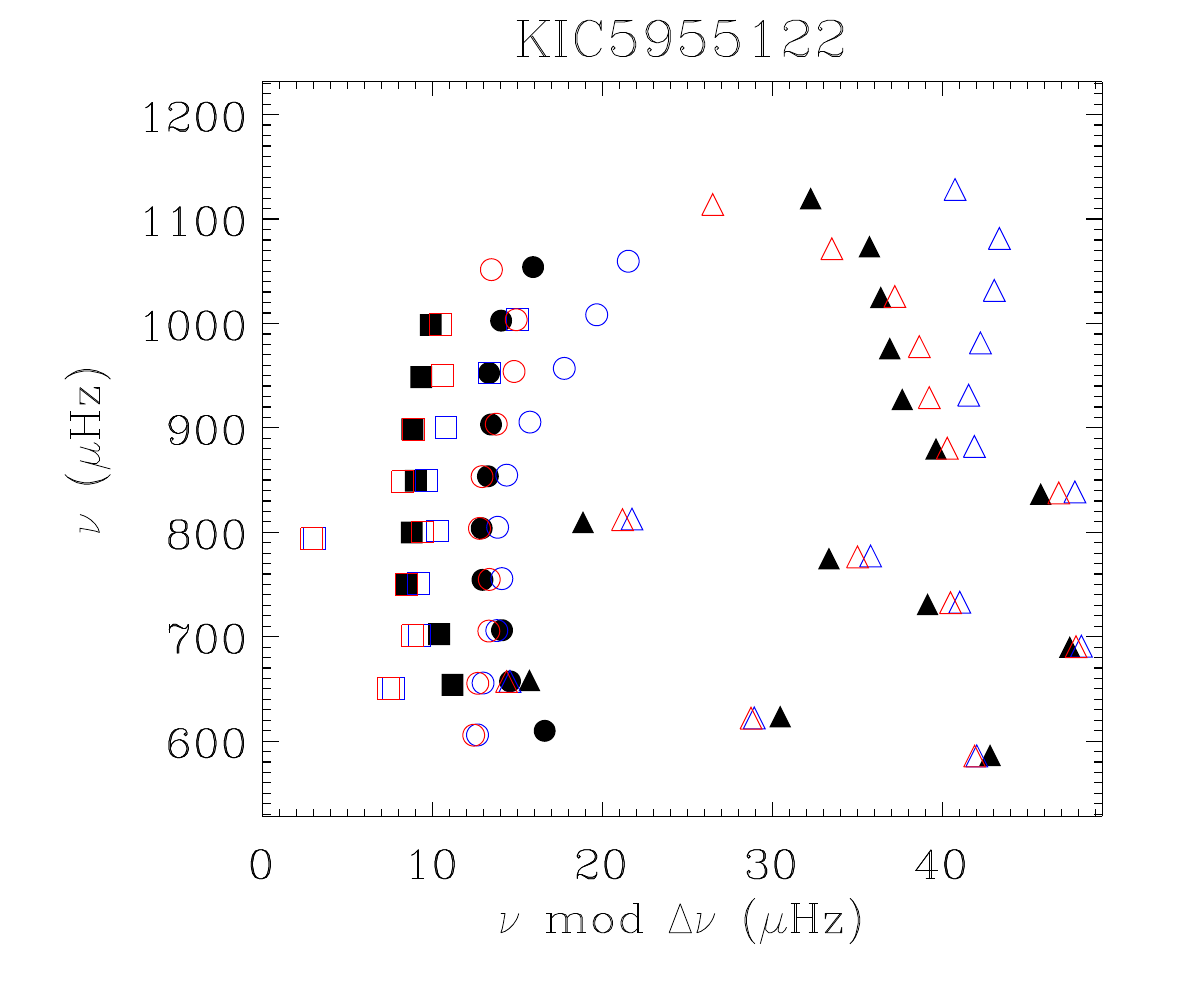}
    \label{fig:echelle_1}
\end{figure}  

\begin{figure}
  \includegraphics[width=0.5\columnwidth]{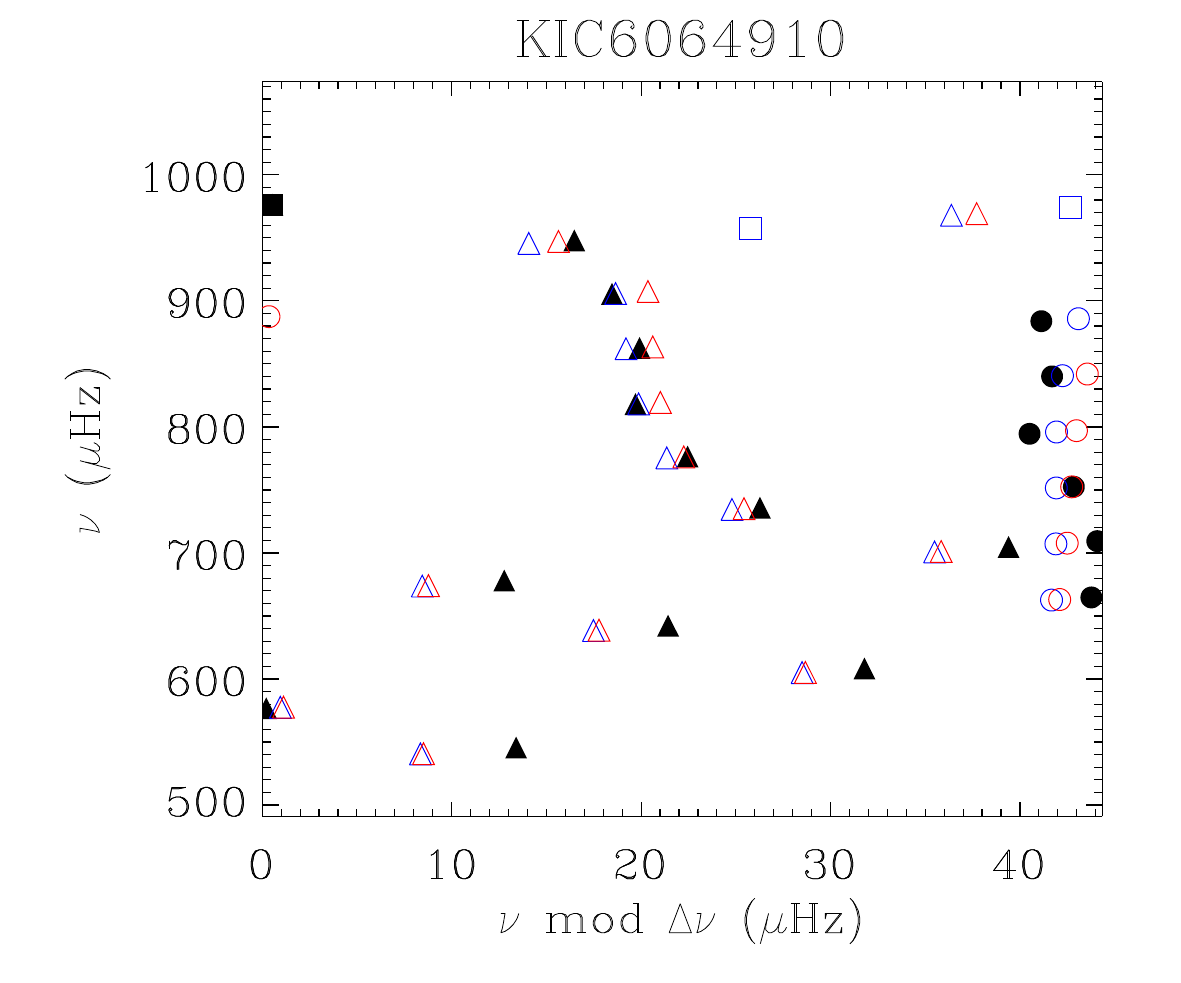}
  \includegraphics[width=0.5\columnwidth]{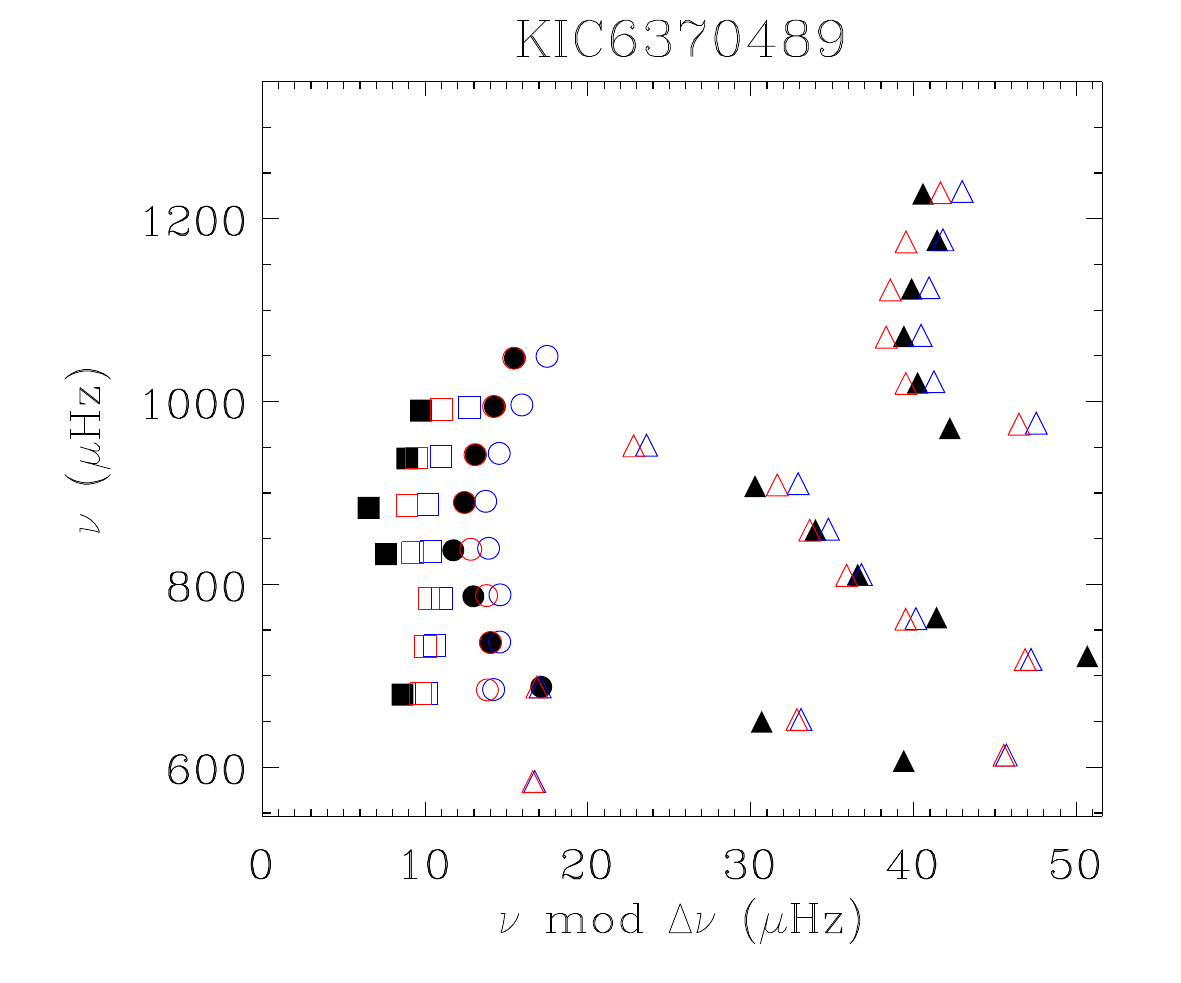}
  \includegraphics[width=0.5\columnwidth]{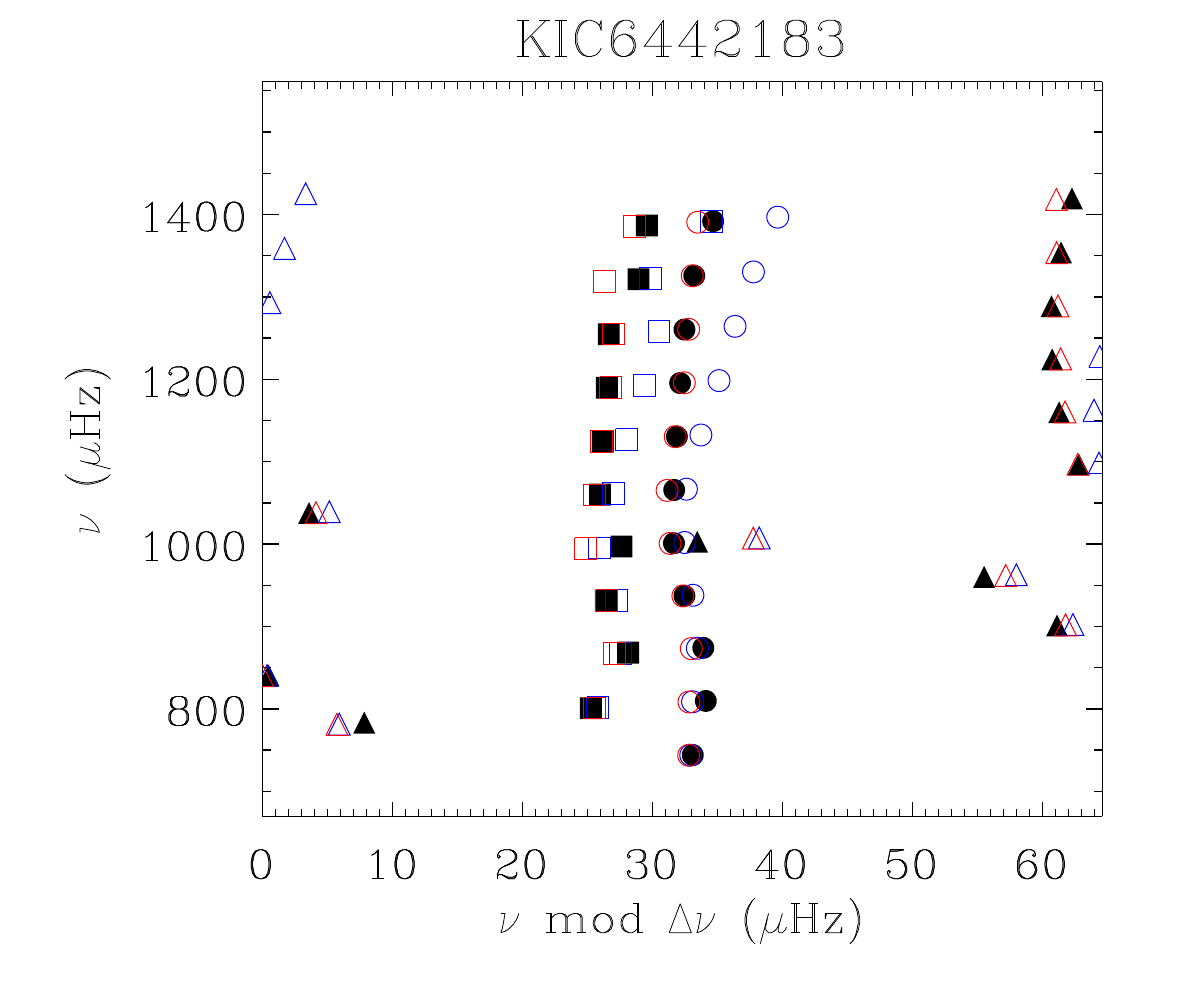}
  \includegraphics[width=0.5\columnwidth]{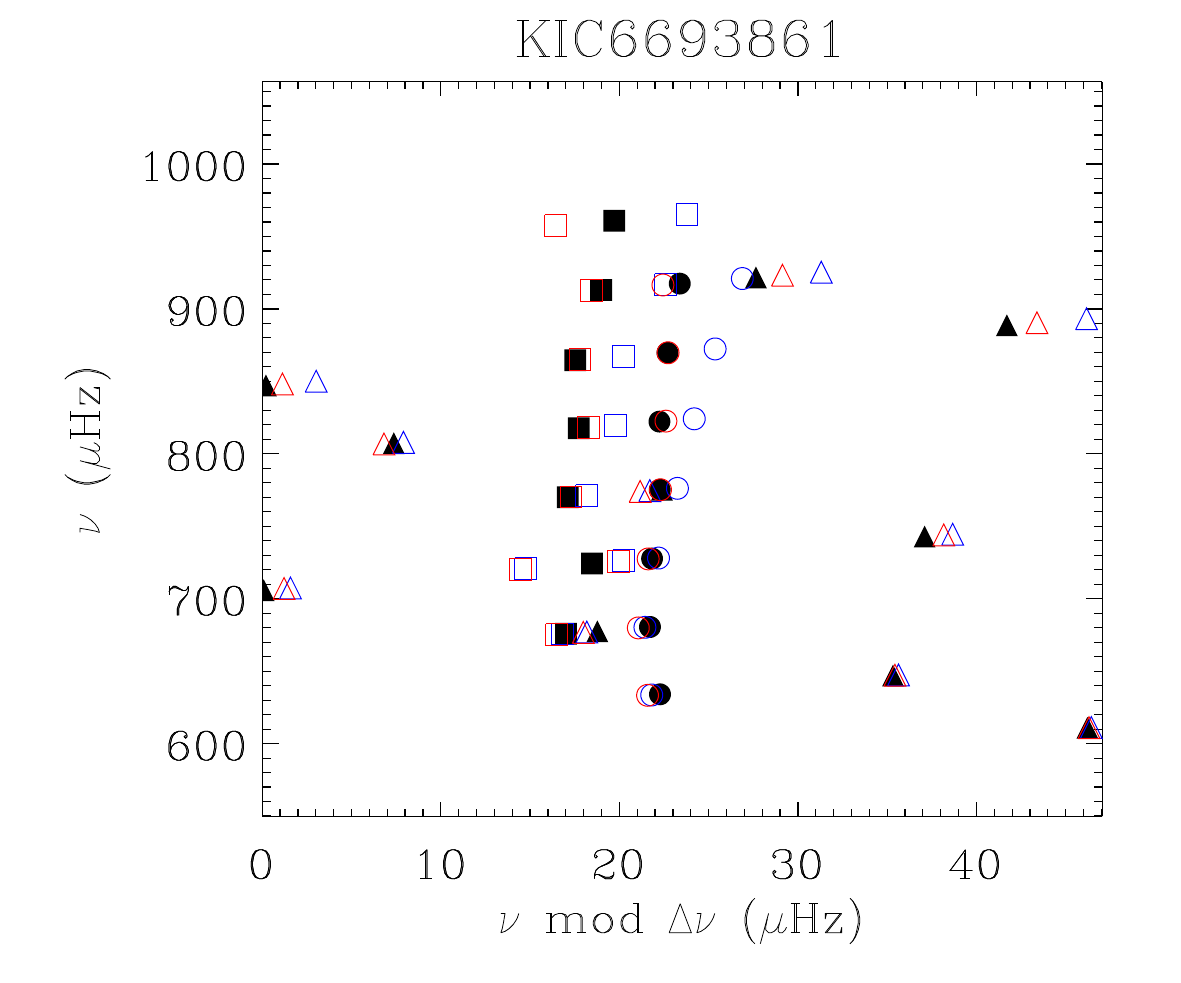}
  \includegraphics[width=0.5\columnwidth]{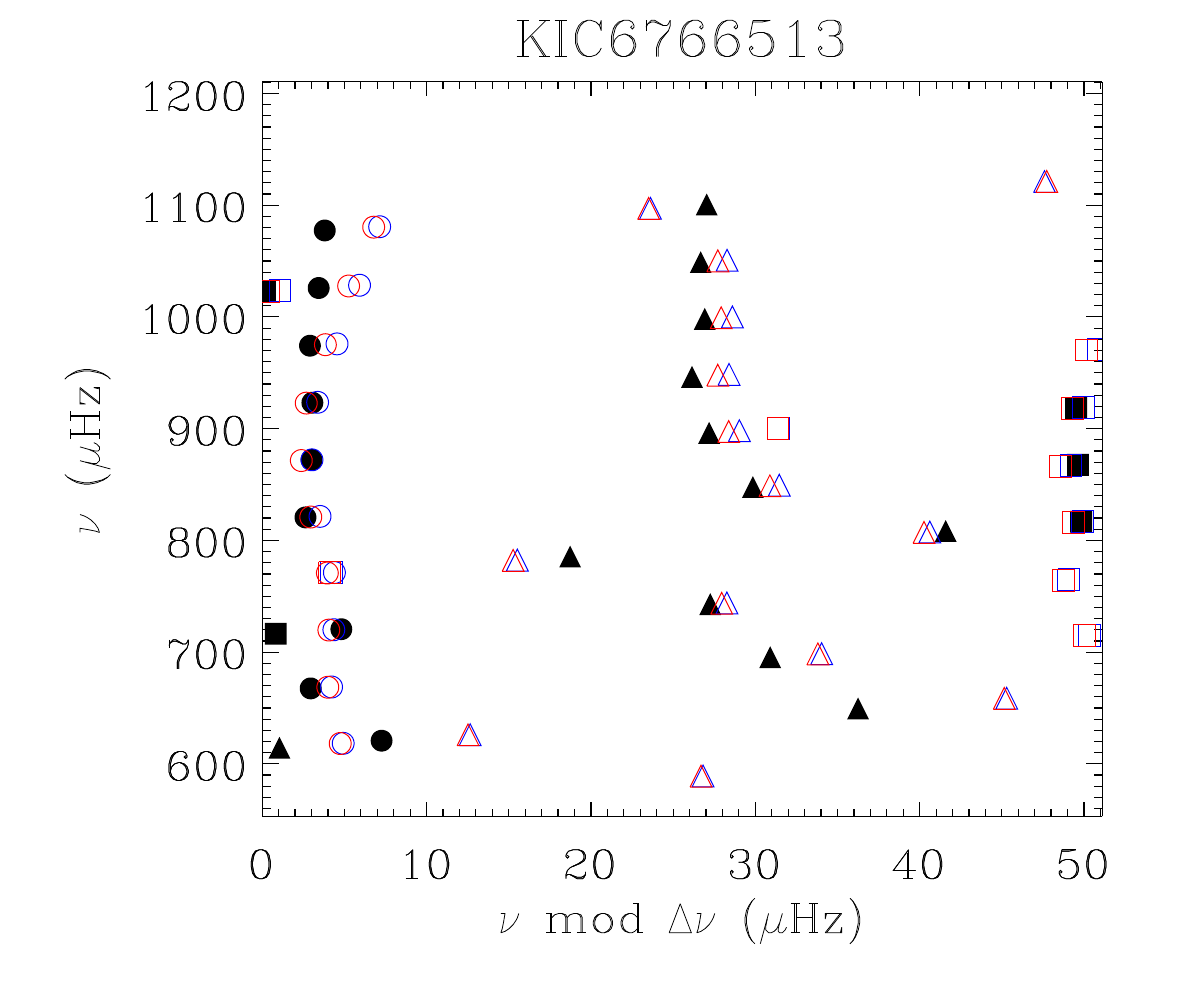}
  \includegraphics[width=0.5\columnwidth]{KIC7174707-eps-converted-to.pdf}
    \label{fig:echelle_2}
\end{figure}  
\begin{figure}
  \includegraphics[width=0.5\columnwidth]{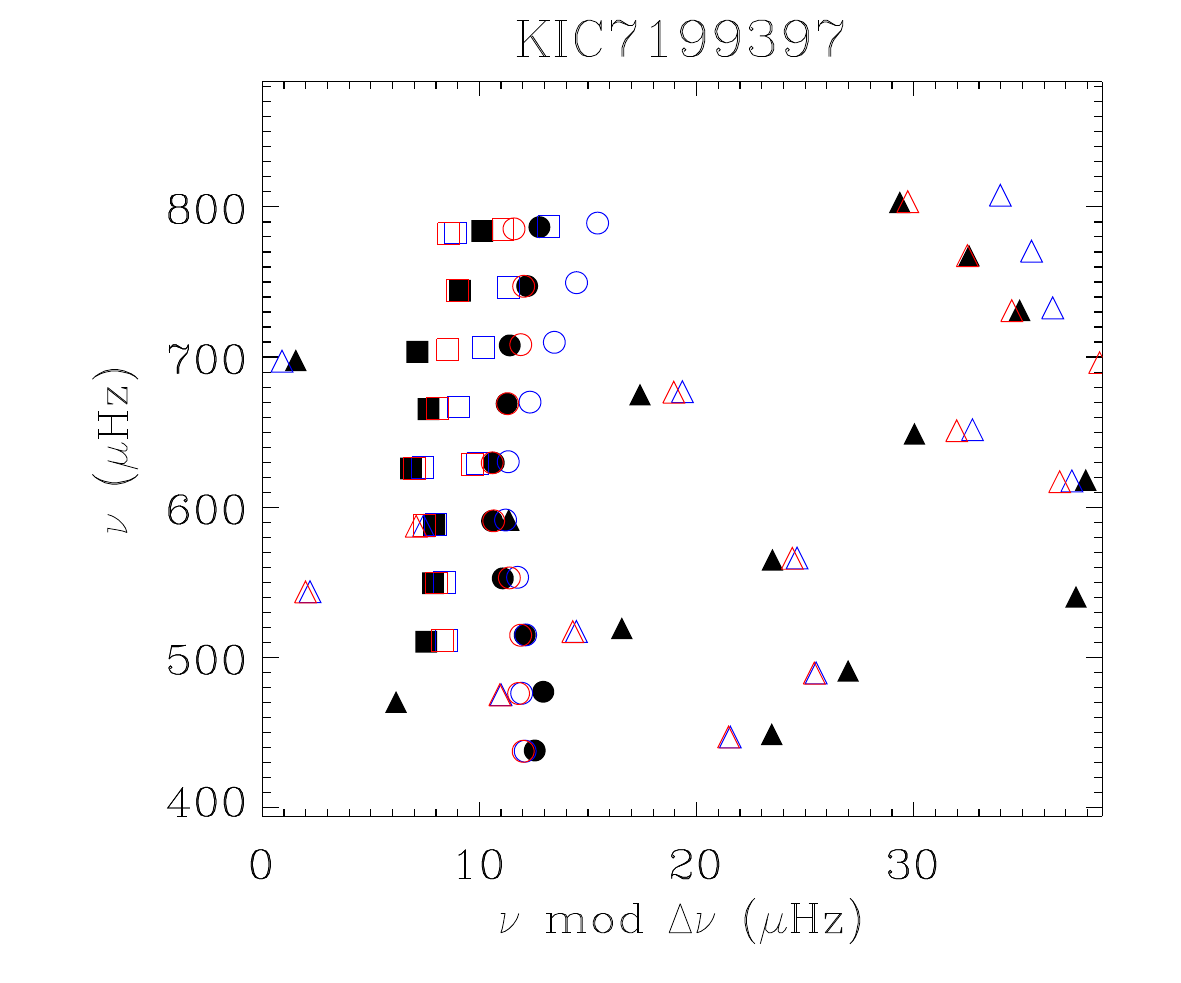}
  \includegraphics[width=0.5\columnwidth]{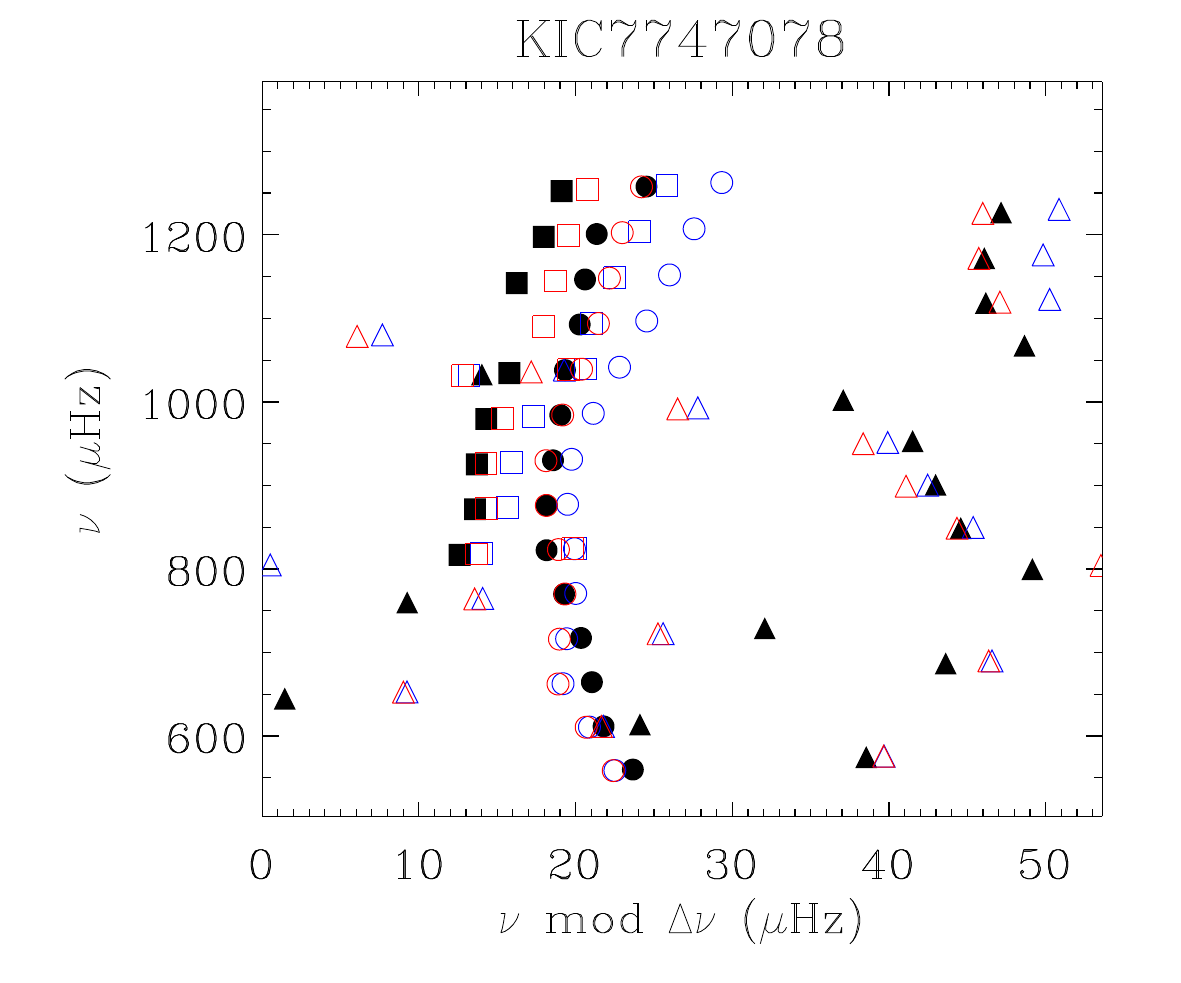}
  \includegraphics[width=0.5\columnwidth]{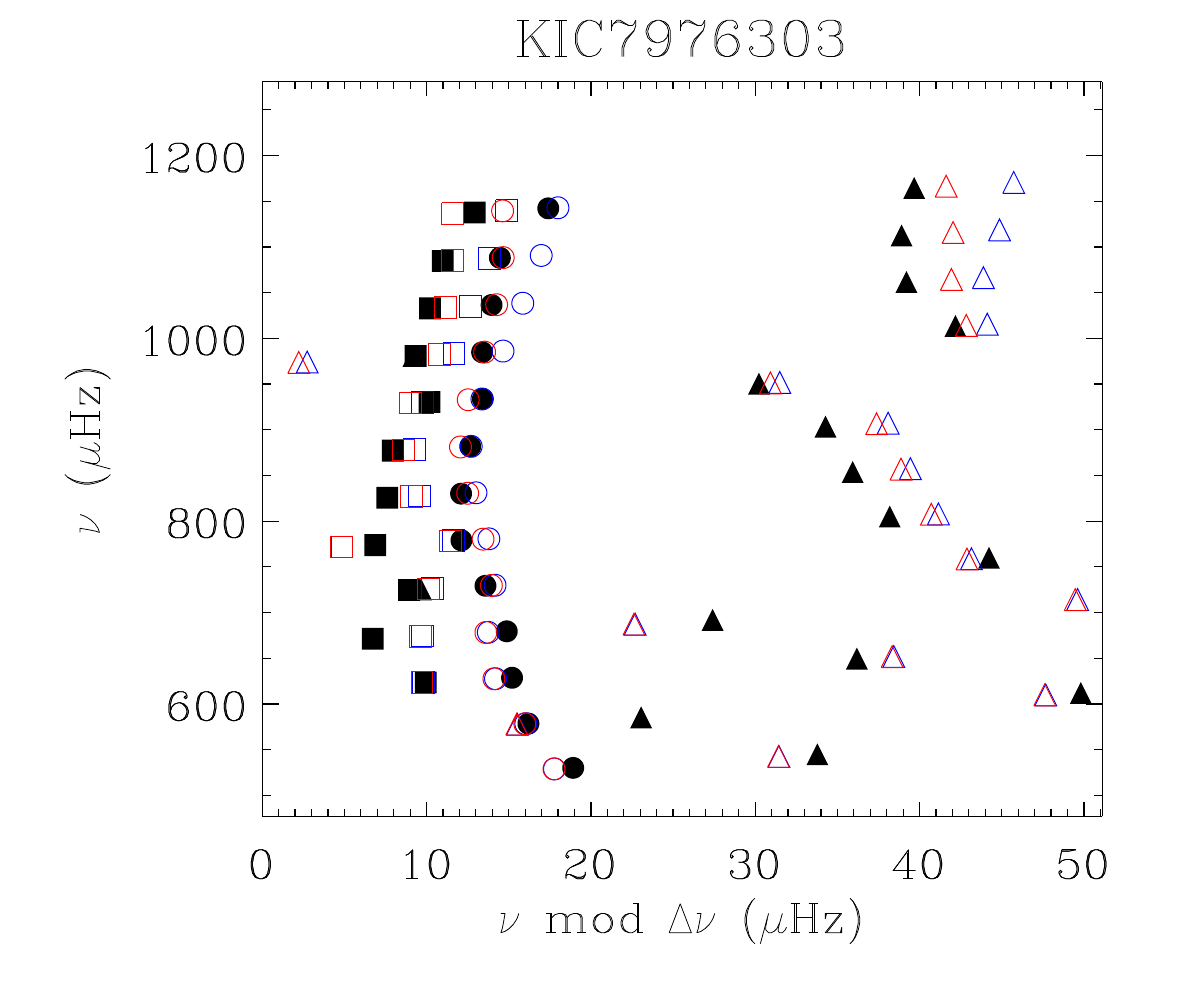}
  \includegraphics[width=0.5\columnwidth]{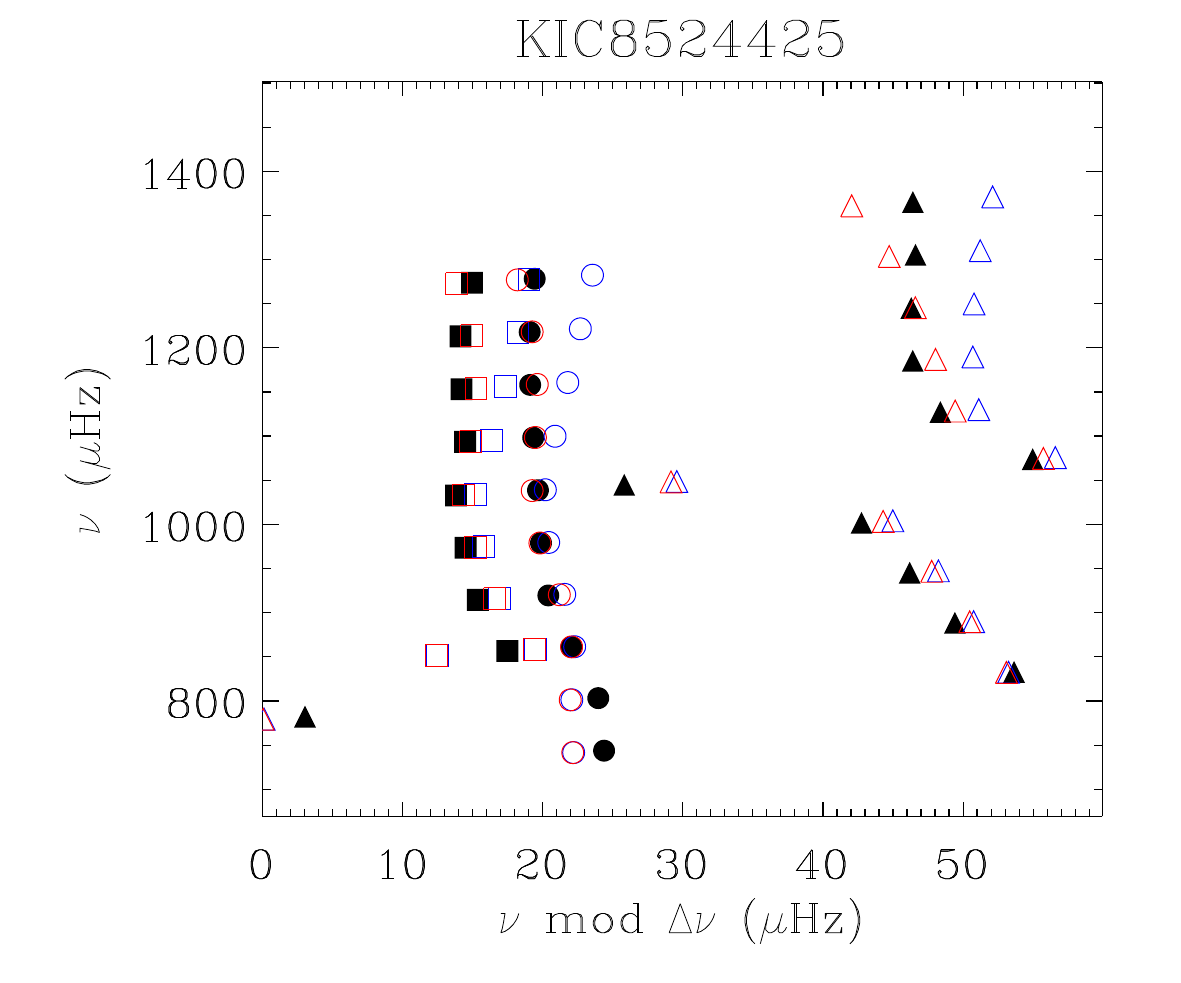}
  \includegraphics[width=0.5\columnwidth]{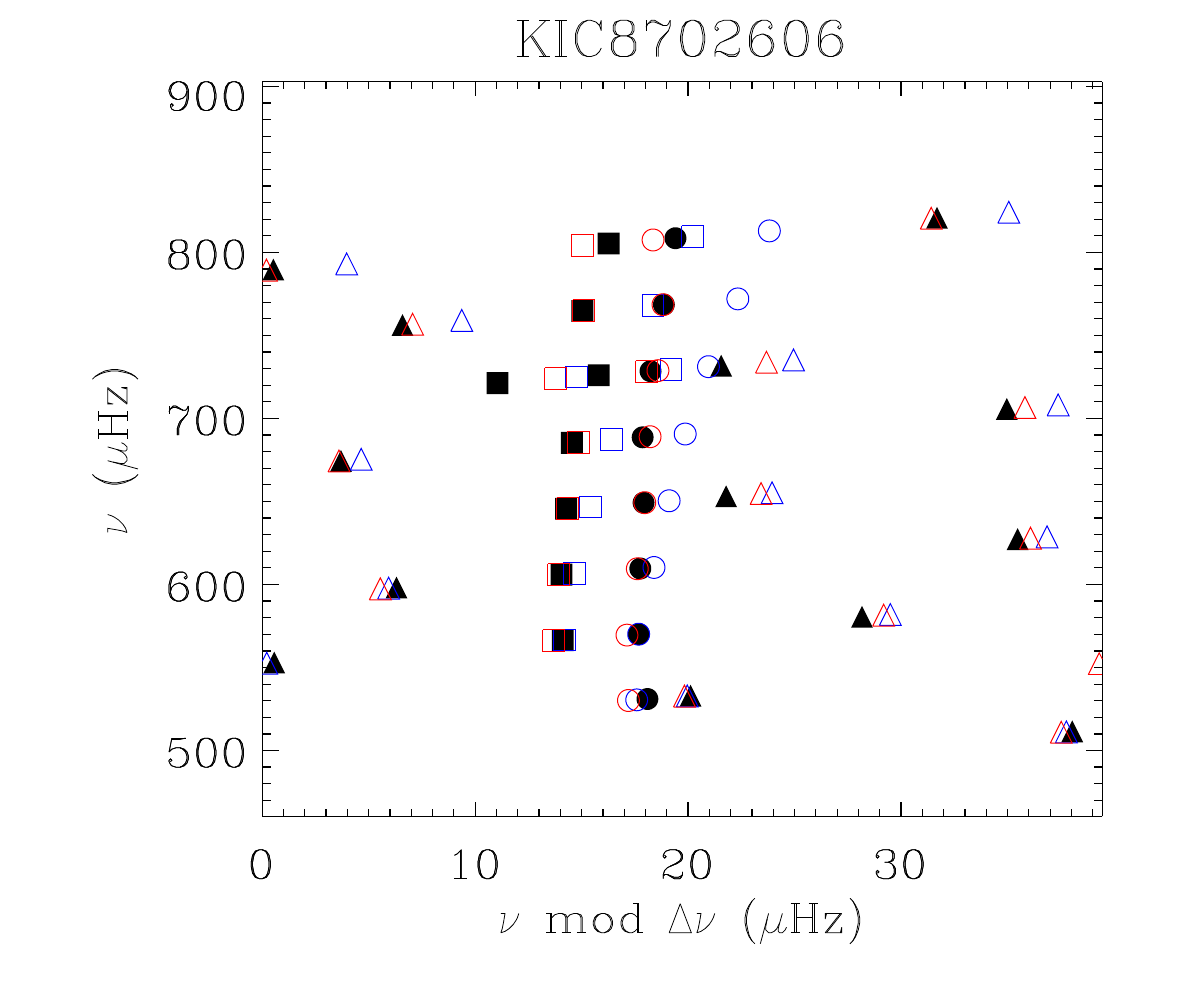}
  \includegraphics[width=0.5\columnwidth]{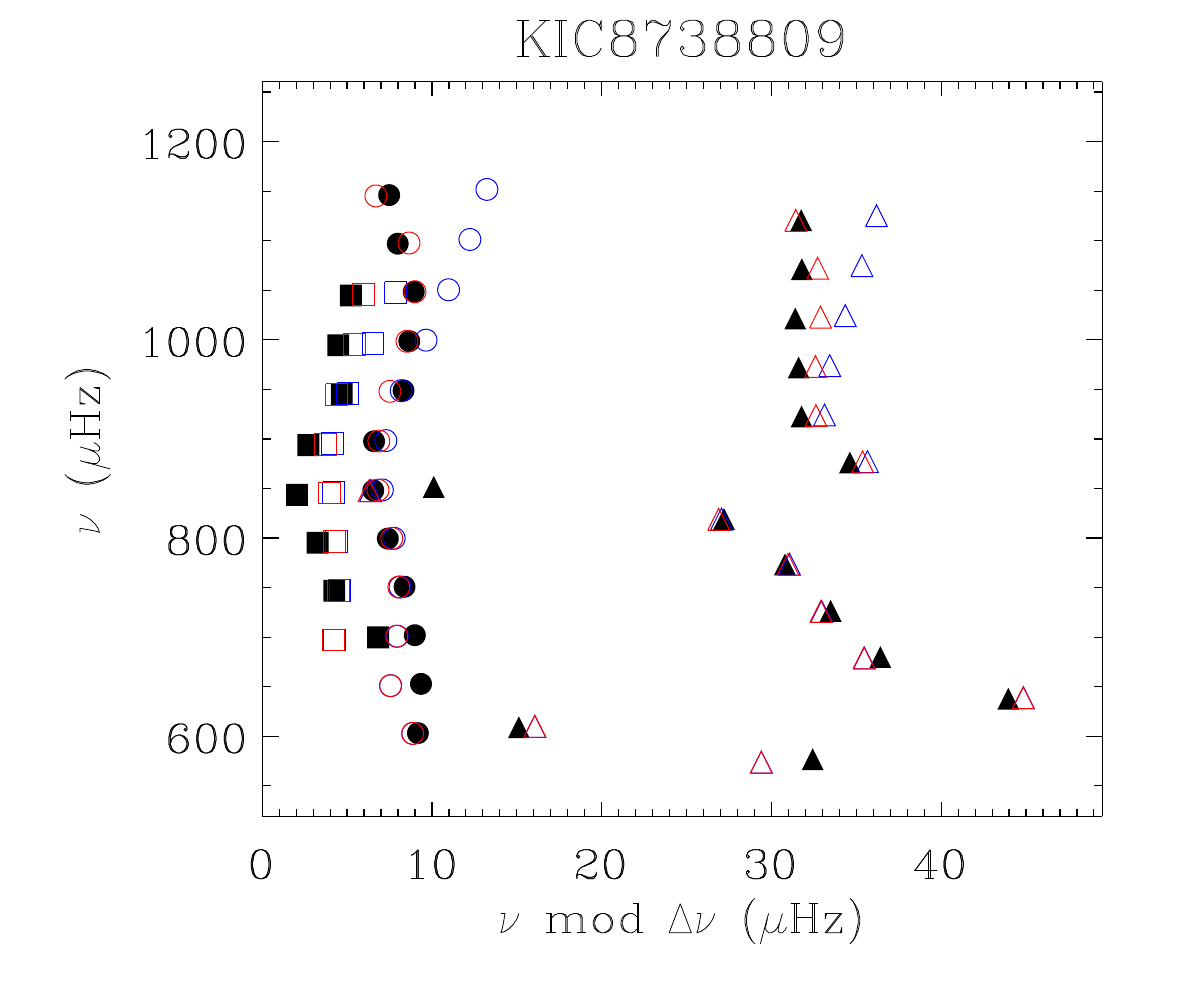}
    \label{fig:echelle_3}
\end{figure}  
 \begin{figure}
 \includegraphics[width=0.5\columnwidth]{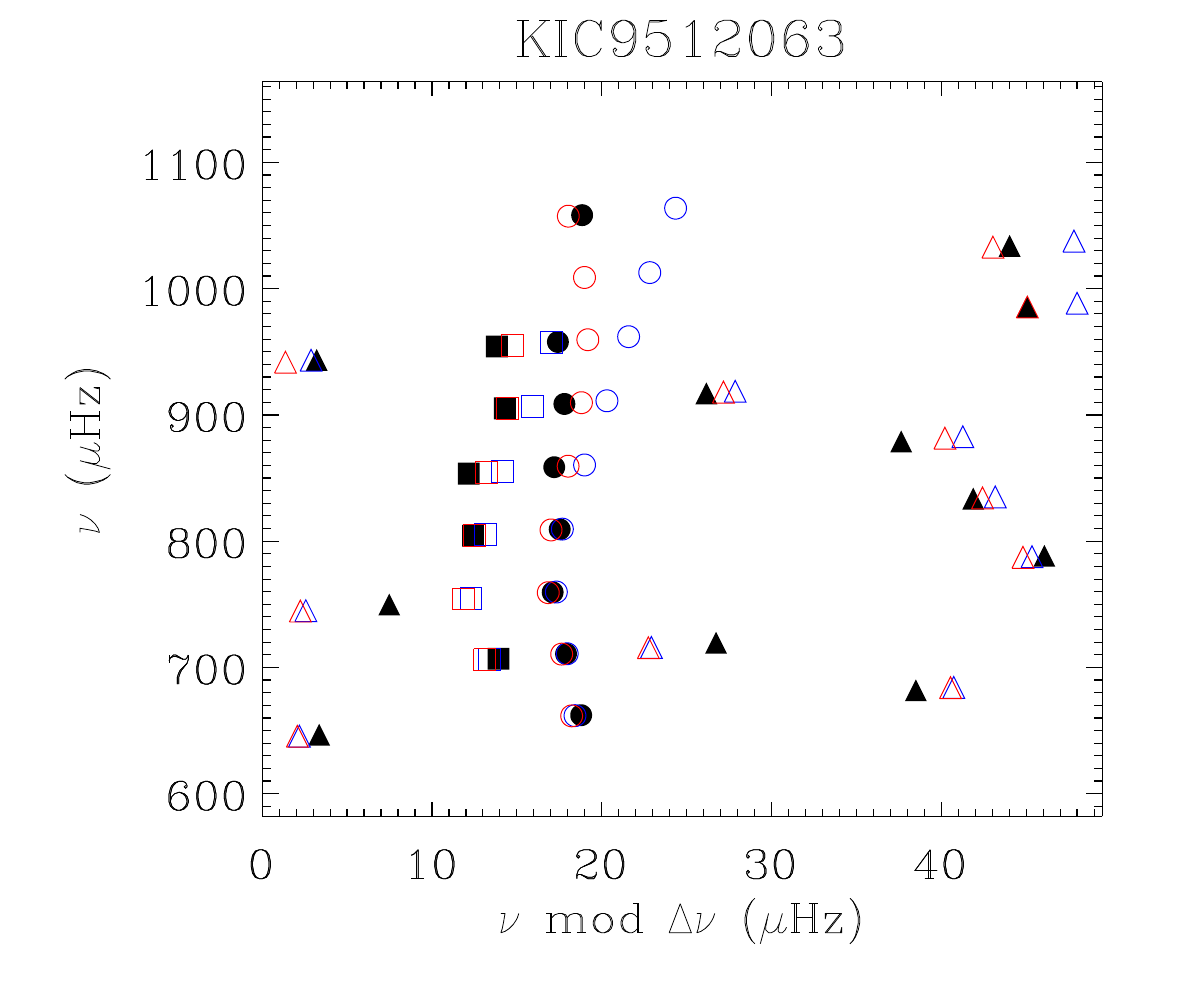}
 \includegraphics[width=0.5\columnwidth]{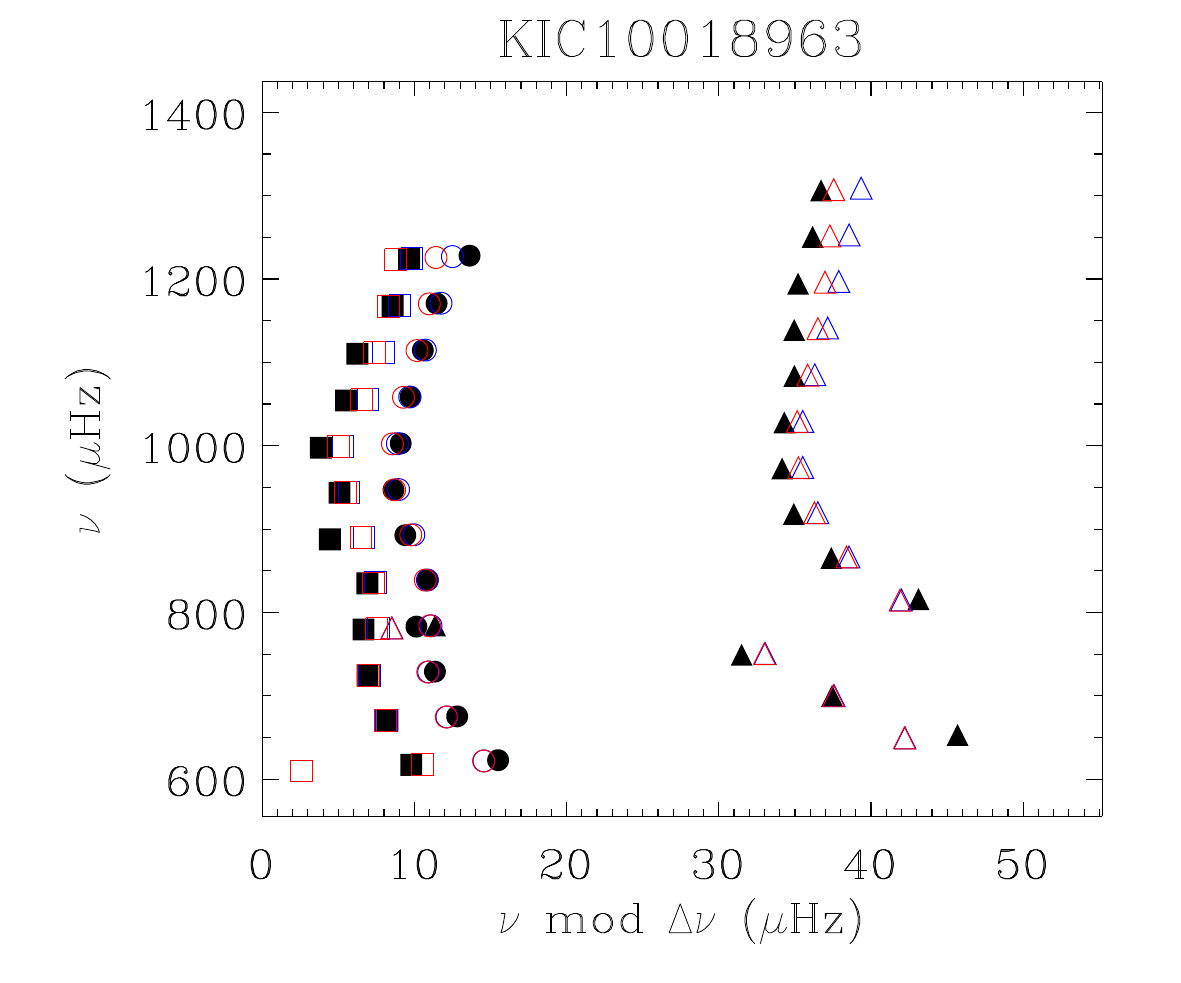}
 \includegraphics[width=0.5\columnwidth]{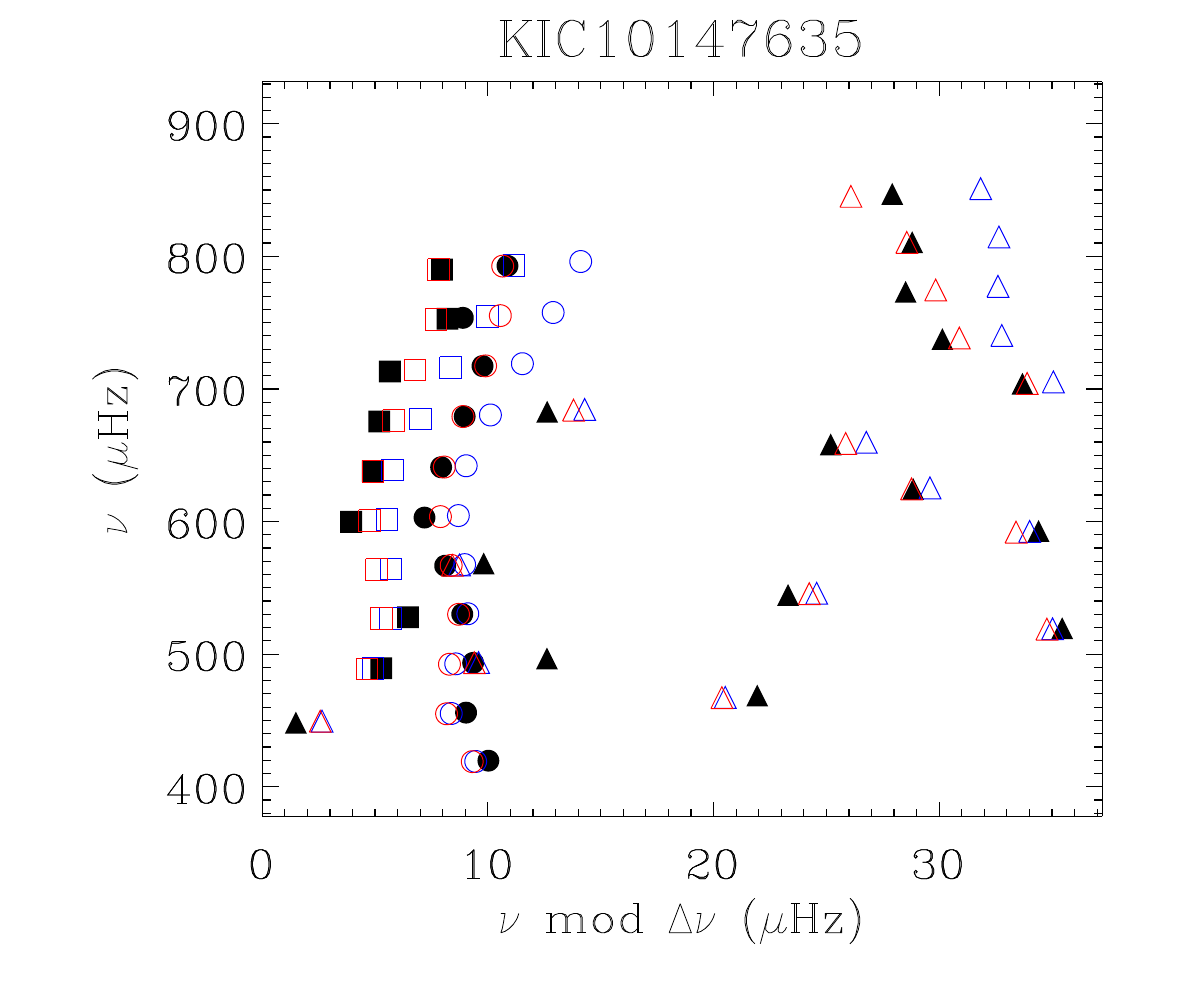}
 \includegraphics[width=0.5\columnwidth]{KIC10273246-eps-converted-to.pdf}
 \includegraphics[width=0.5\columnwidth]{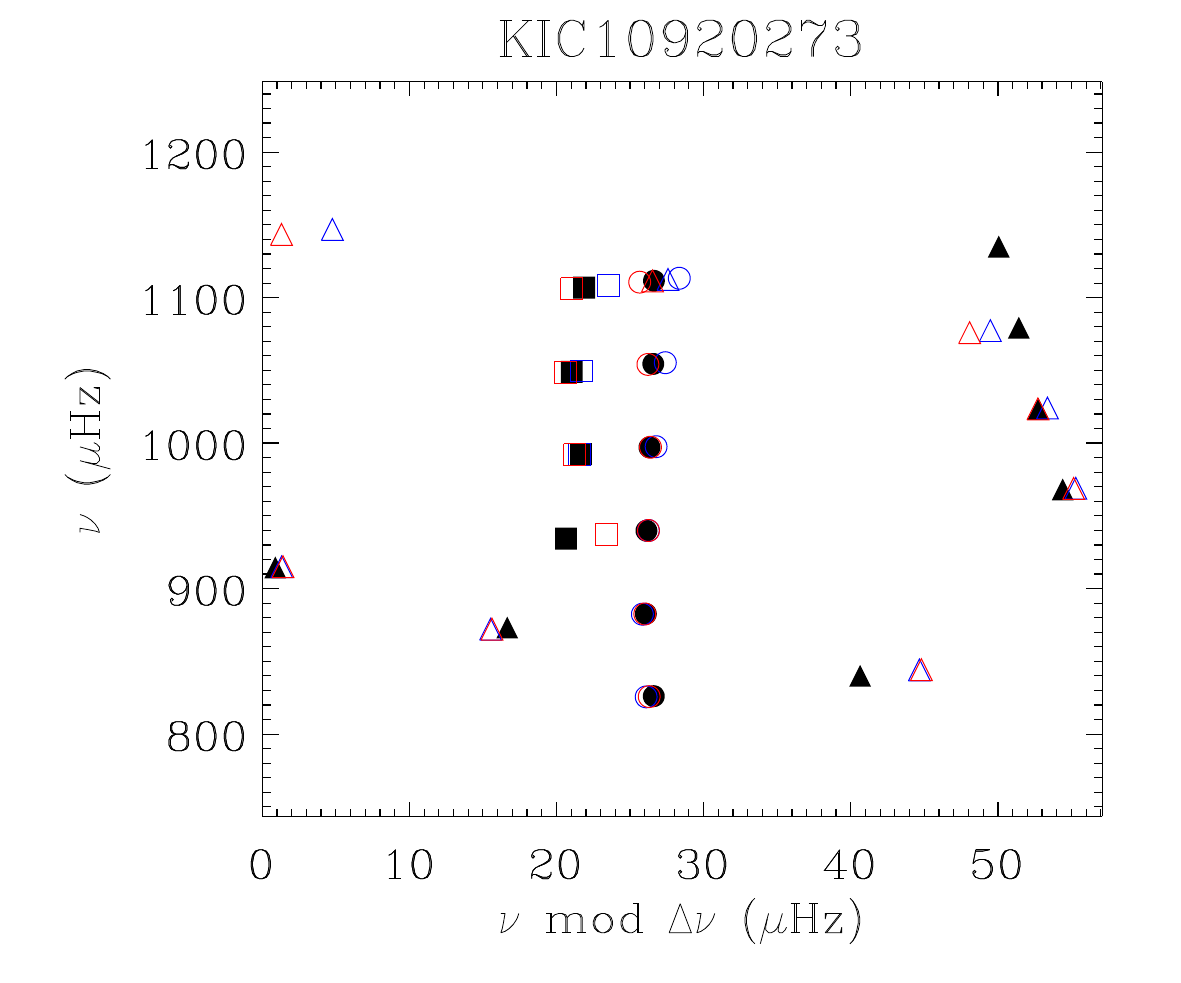}
   \includegraphics[width=0.5\columnwidth]{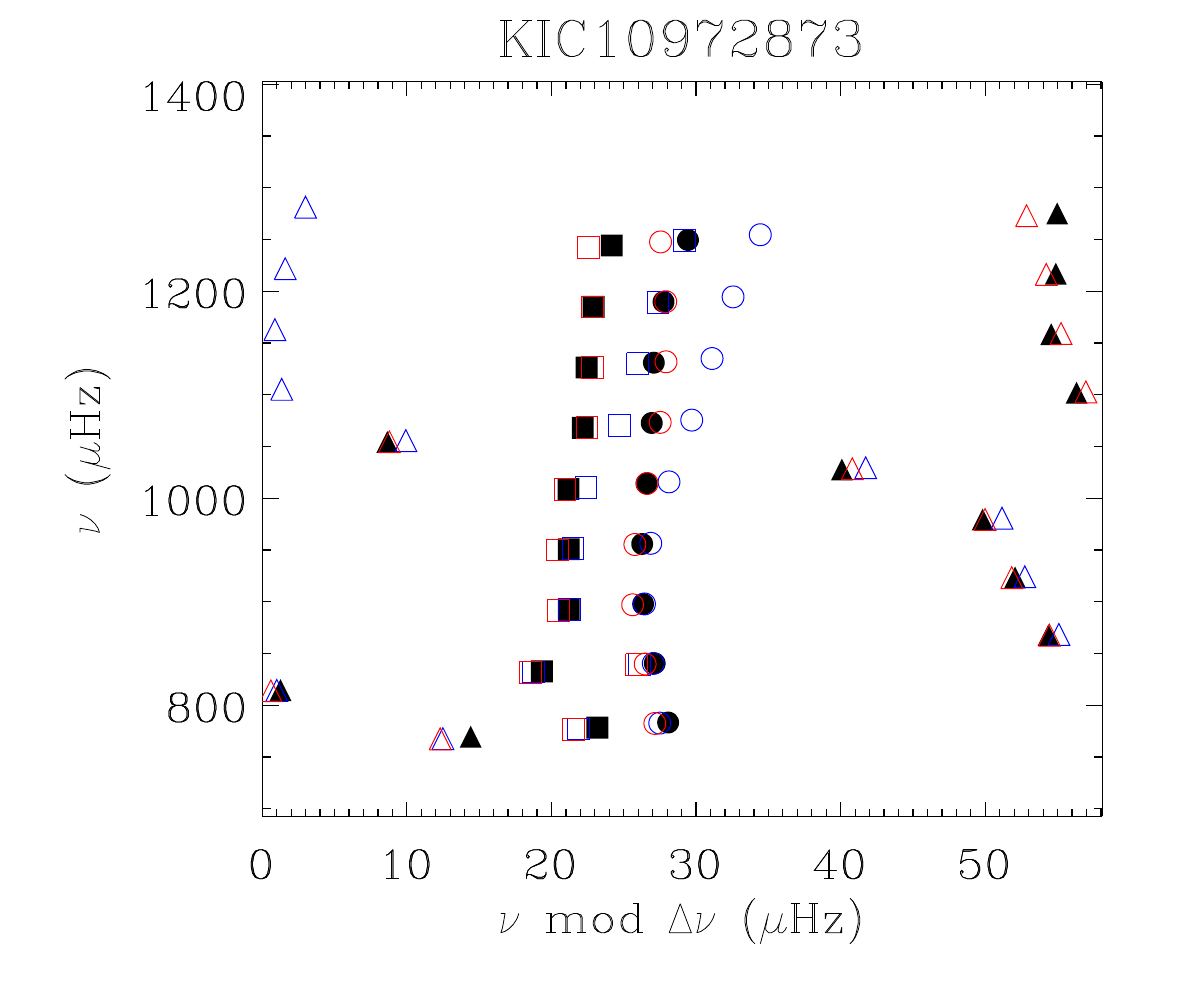}
    \label{fig:echelle_4}
\end{figure}  

 \begin{figure}
\includegraphics[width=0.5\columnwidth]{KIC11026764-eps-converted-to.pdf}
 \includegraphics[width=0.5\columnwidth]{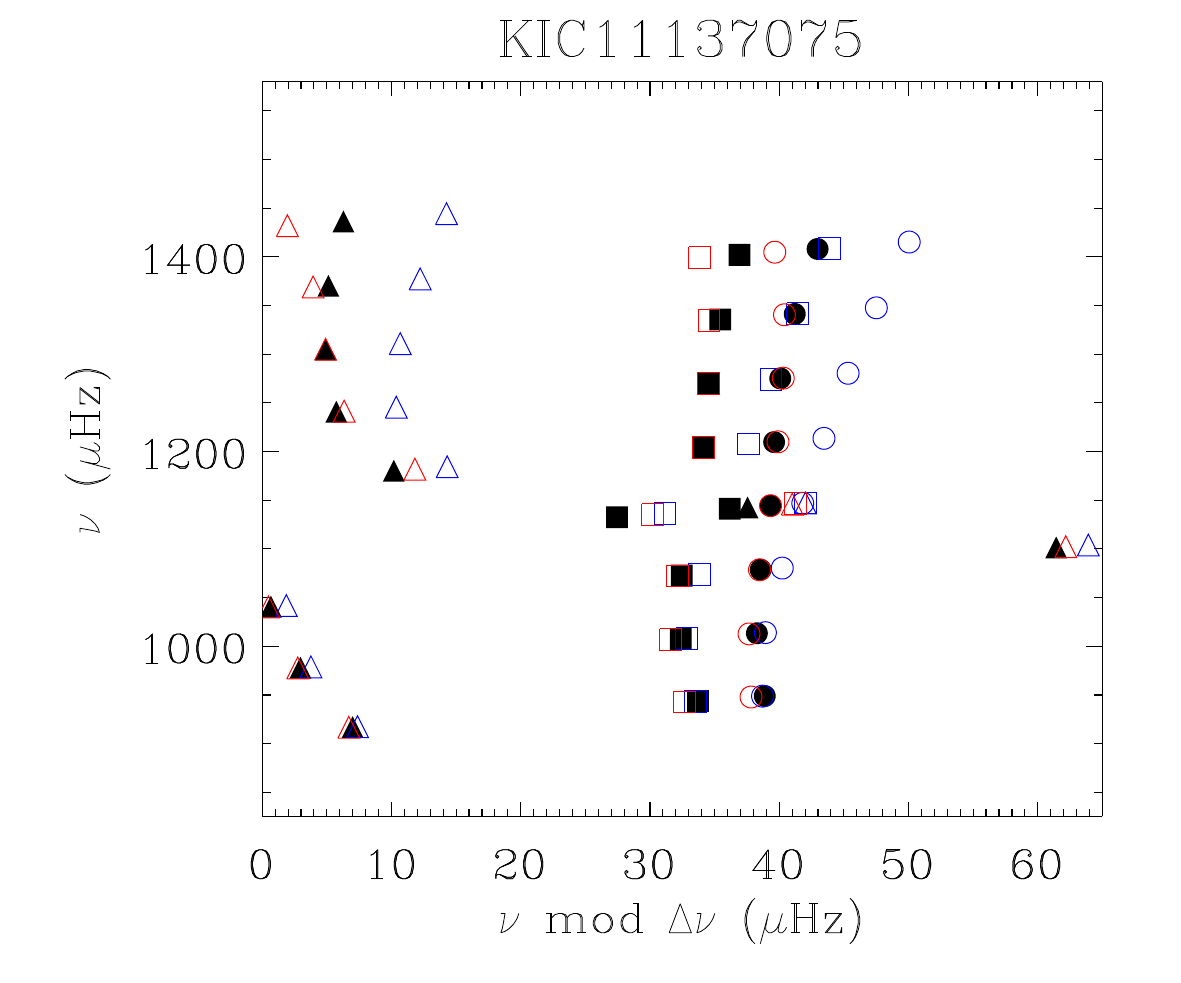} 
 \includegraphics[width=0.5\columnwidth]{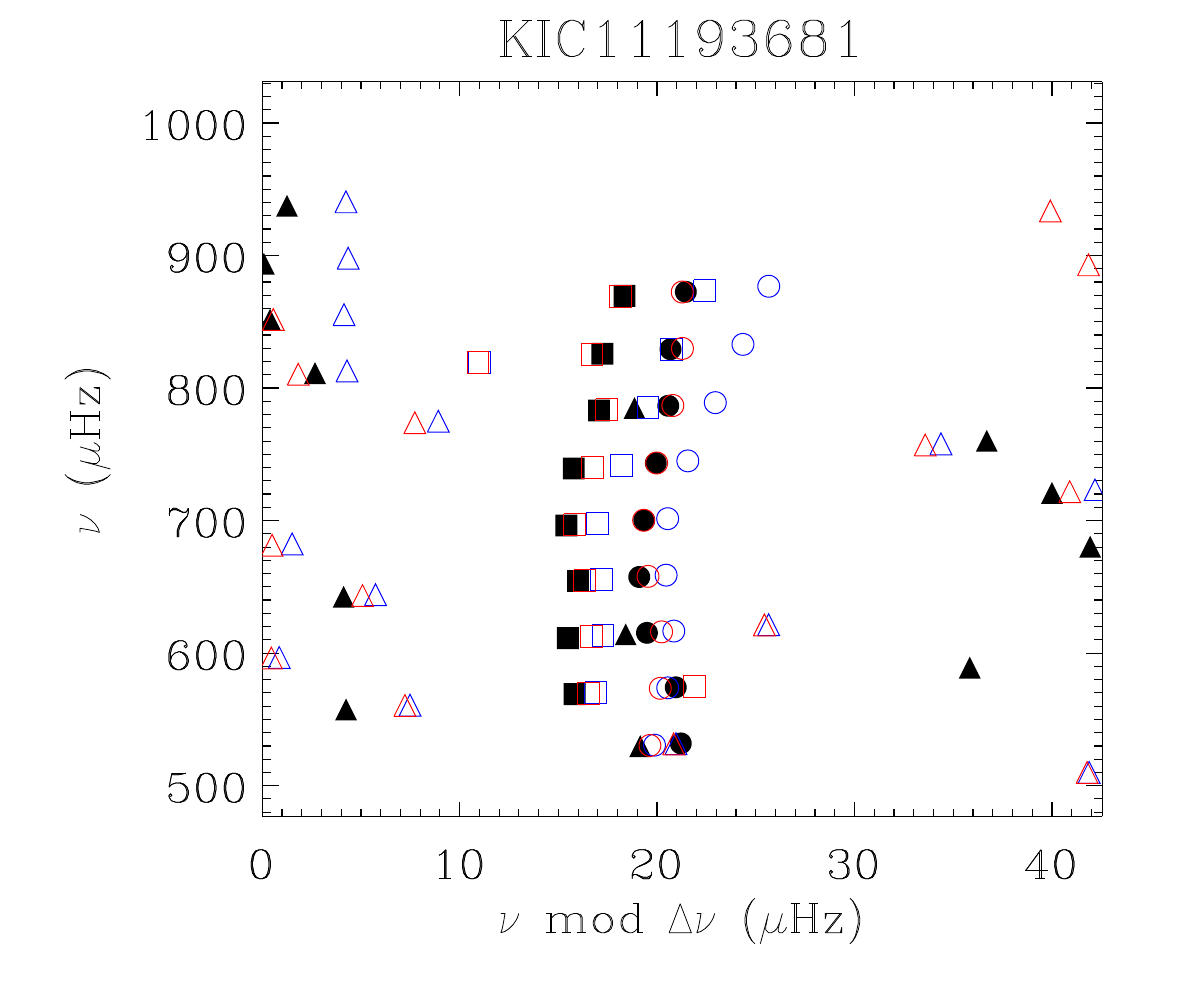} 
 \includegraphics[width=0.5\columnwidth]{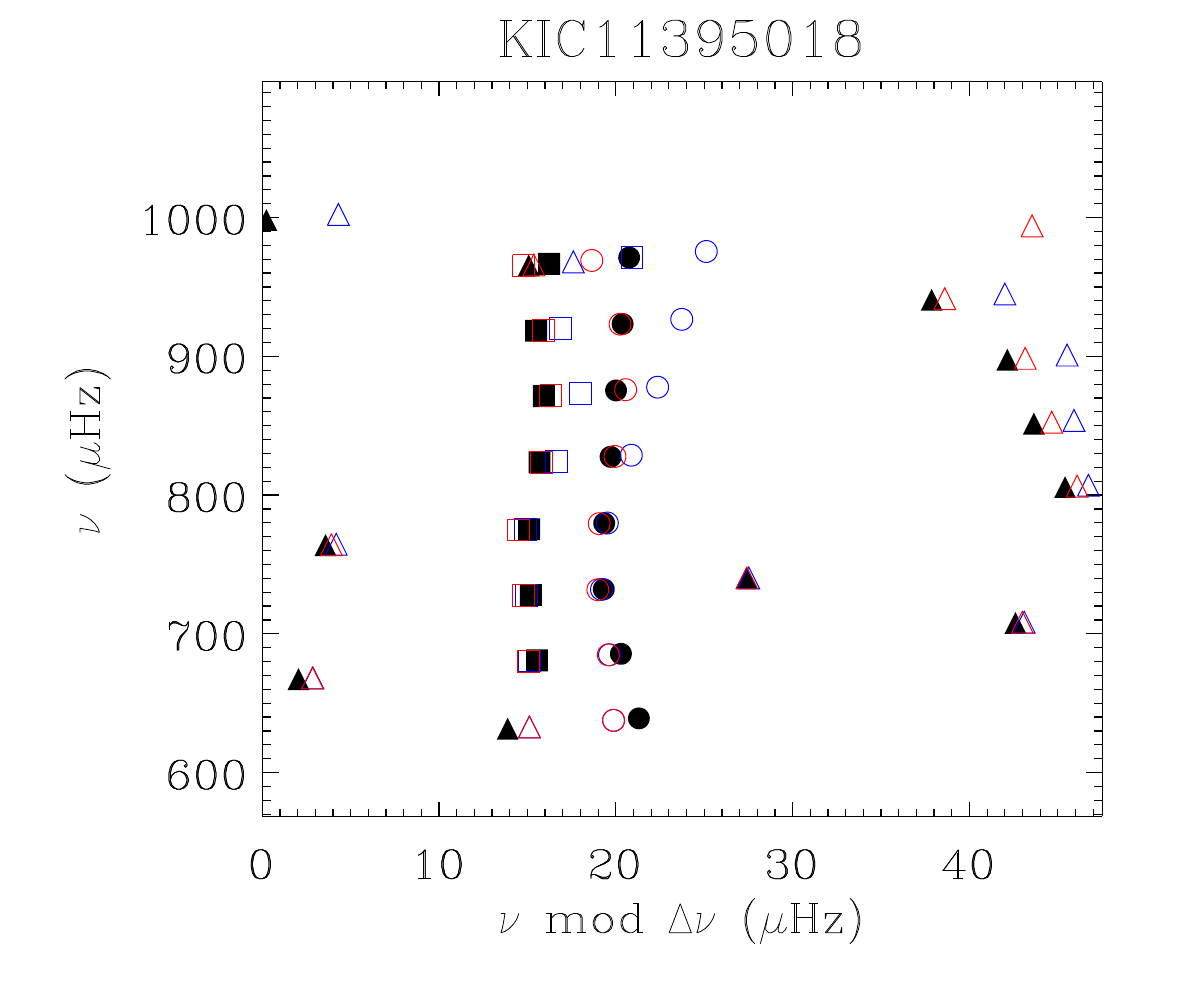} 
 \includegraphics[width=0.5\columnwidth]{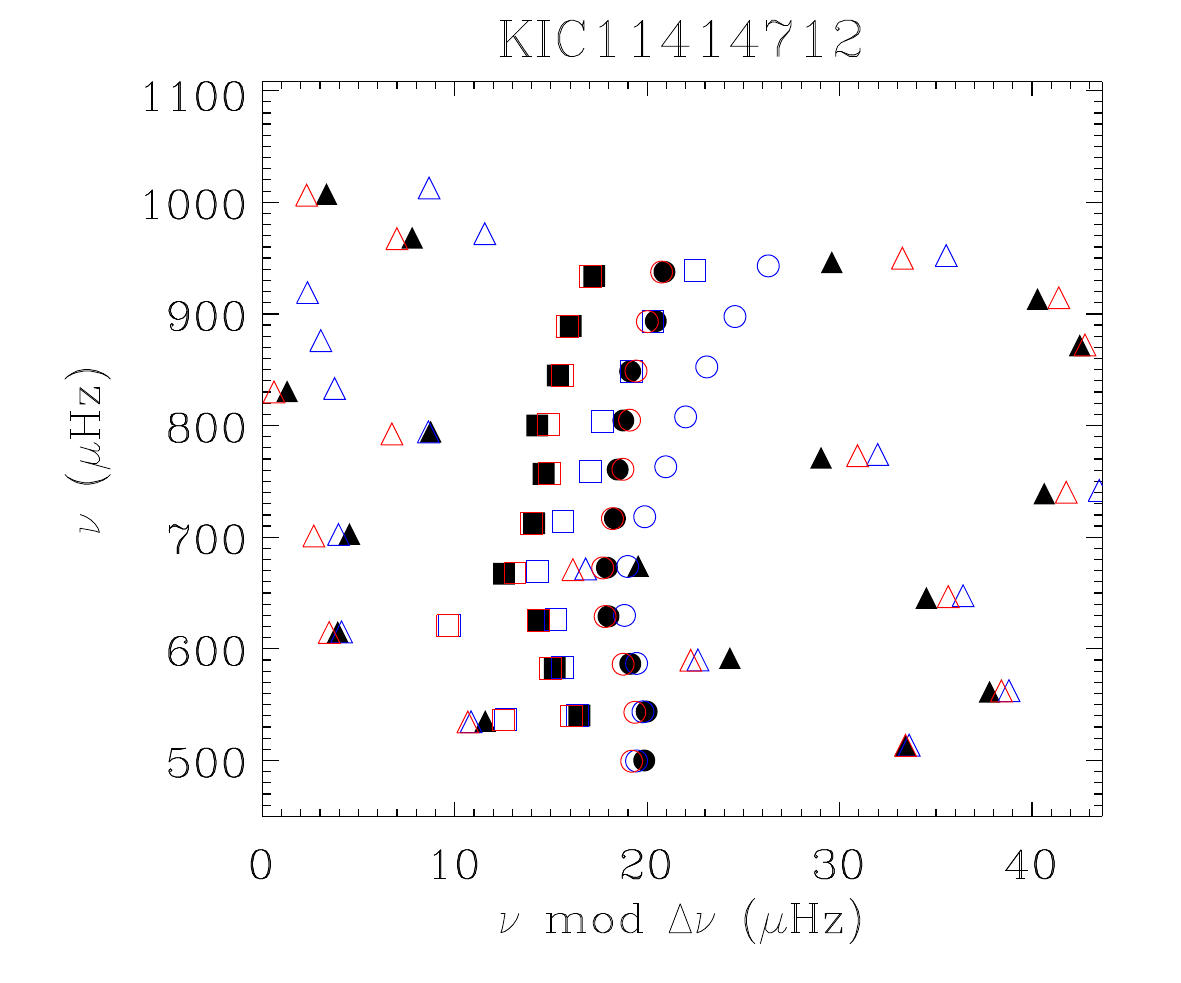} 
  \includegraphics[width=0.5\columnwidth]{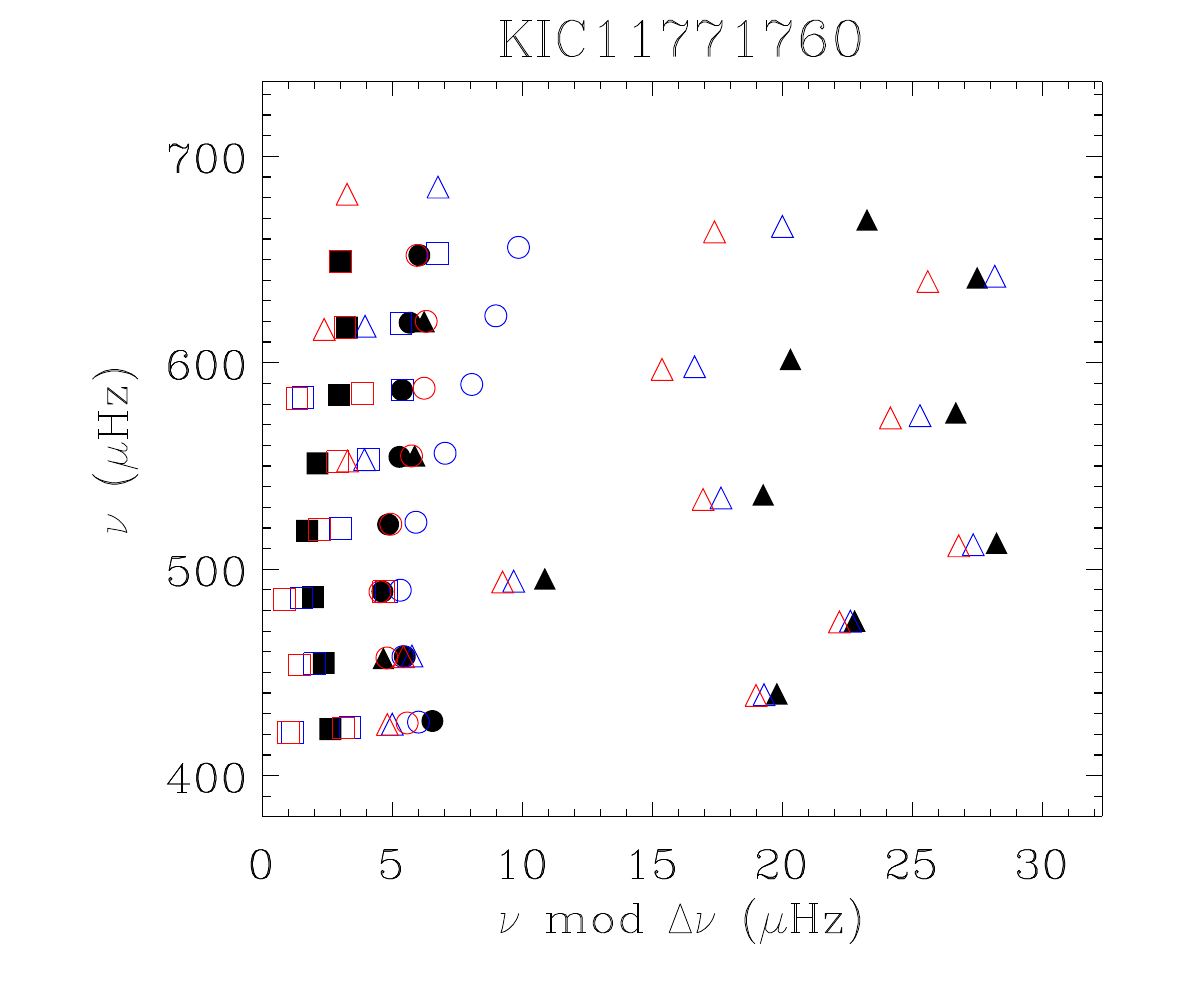} 
    \label{fig:echelle_5}
\end{figure}

 \begin{figure}
   \includegraphics[width=0.5\columnwidth]{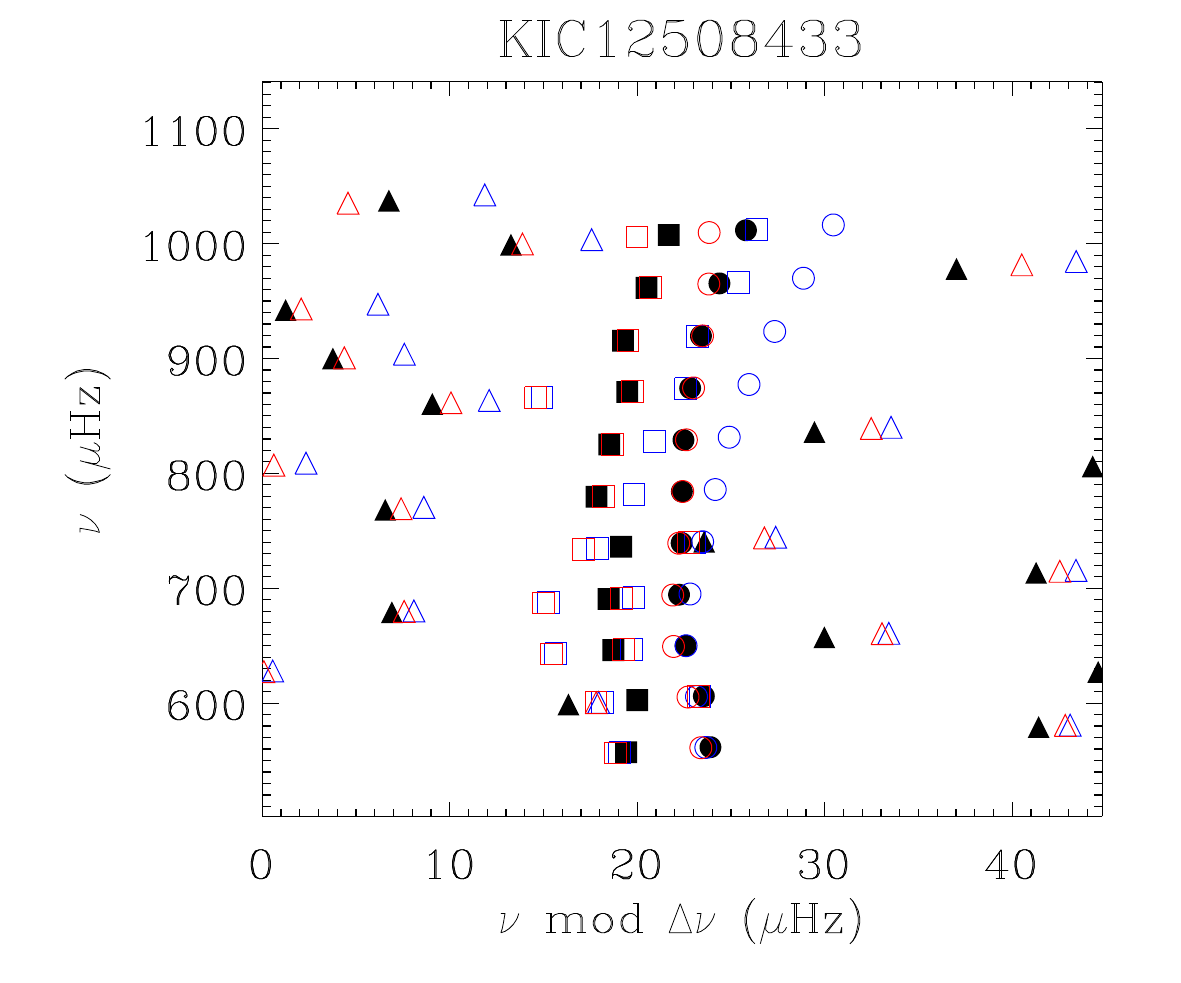} 
    \label{fig:echelle_6}
\end{figure}

\clearpage
\section{Probability Distributions of Masses and Ages of 31 Subgiants}\label{appB}
\clearpage

 \begin{figure}
 \centering
\includegraphics[width=0.75\textwidth]{mass_age_2d_KIC2991448-eps-converted-to.pdf}
\includegraphics[width=0.75\textwidth]{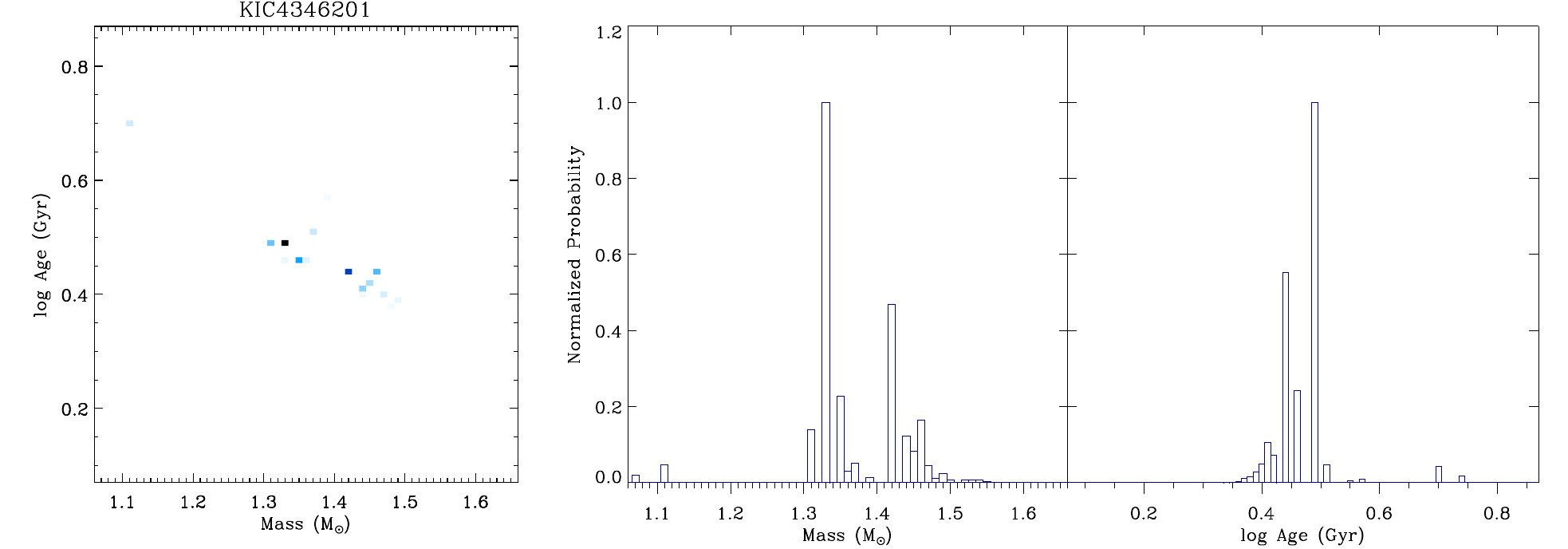}
\includegraphics[width=0.75\textwidth]{mass_age_2d_KIC5108214-eps-converted-to.pdf}
\includegraphics[width=0.75\textwidth]{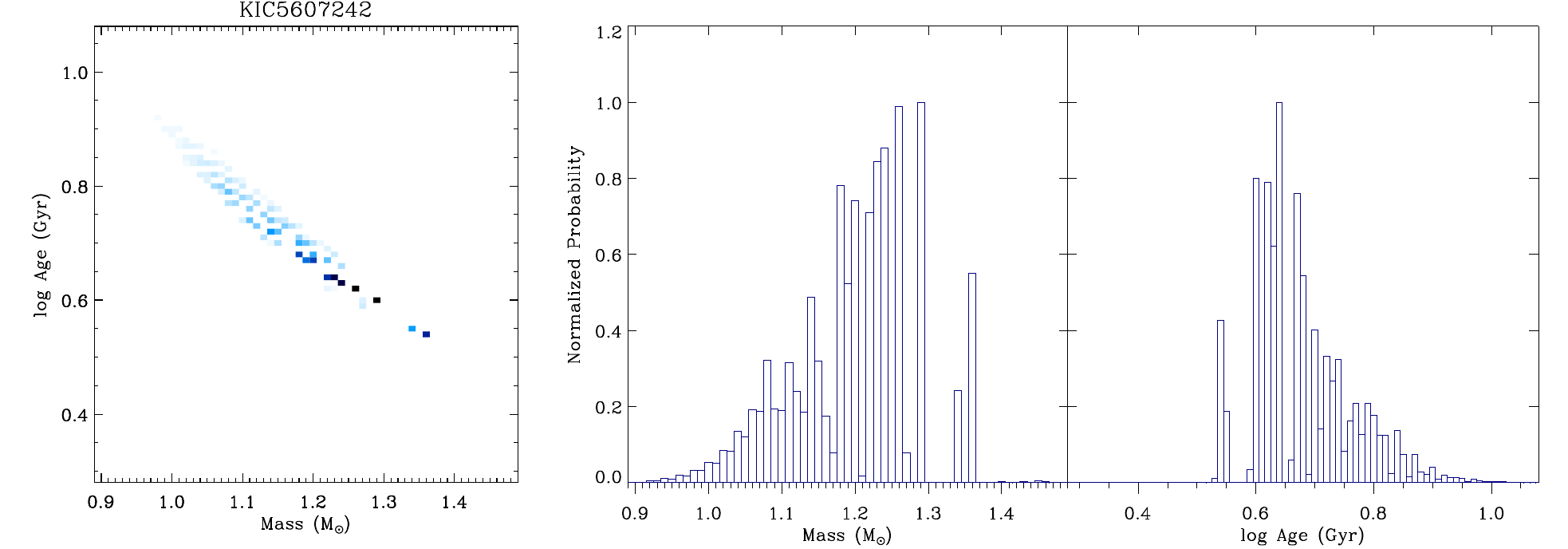}
\includegraphics[width=0.75\textwidth]{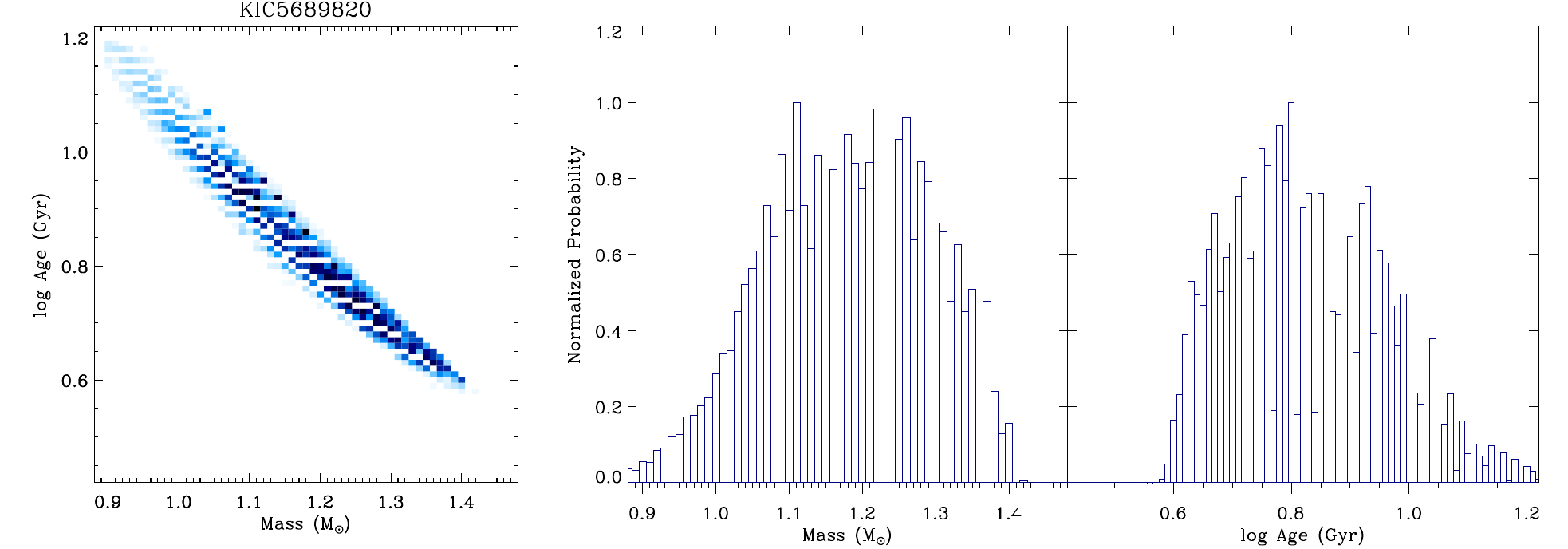}
  \label{fig:prob_1}
\end{figure}

 \begin{figure}
 \centering
\includegraphics[width=0.75\textwidth]{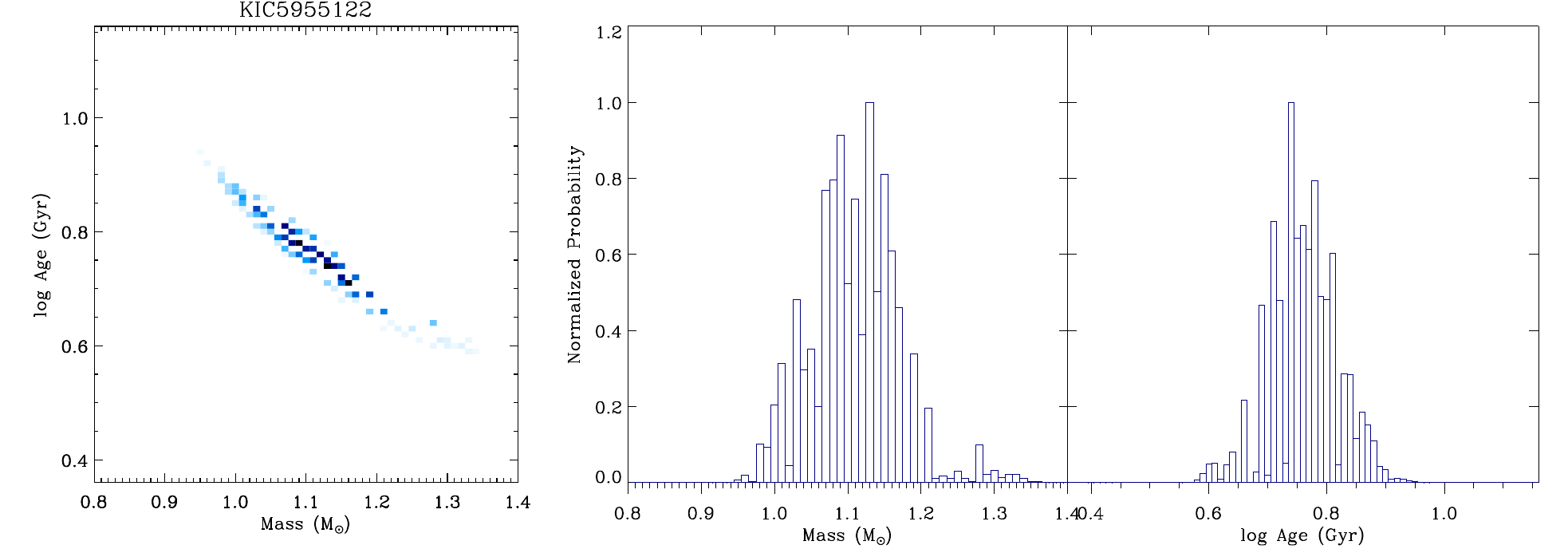}
\includegraphics[width=0.75\textwidth]{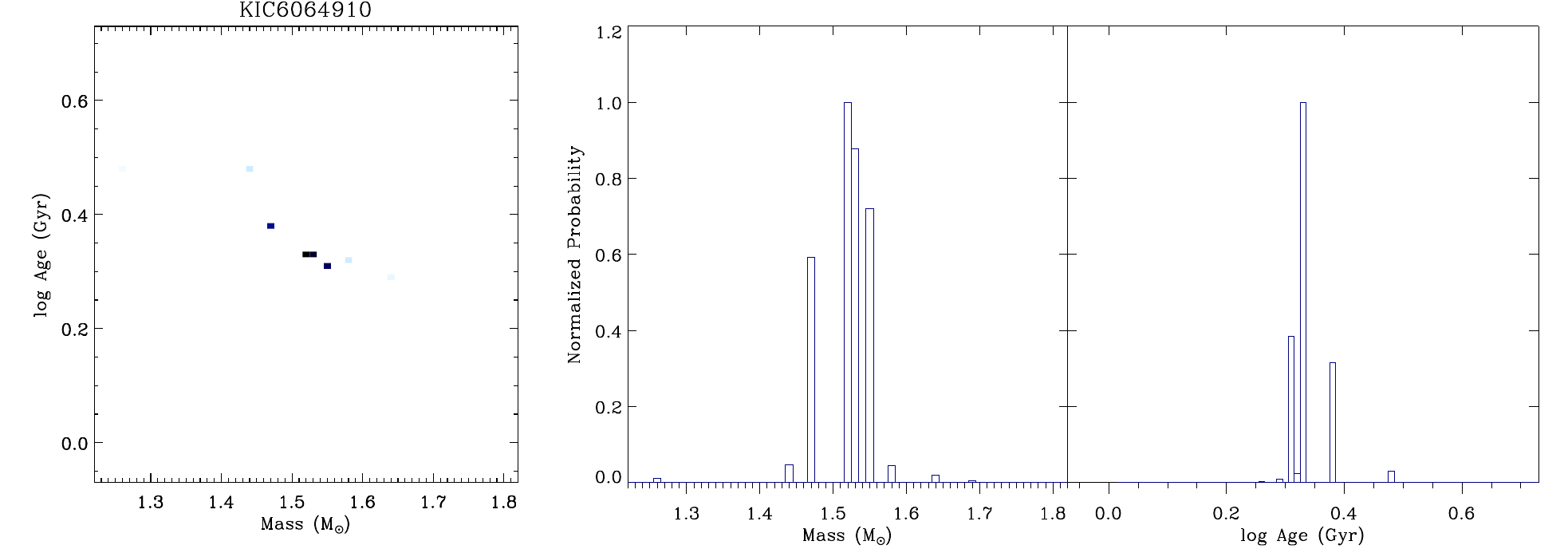}
\includegraphics[width=0.75\textwidth]{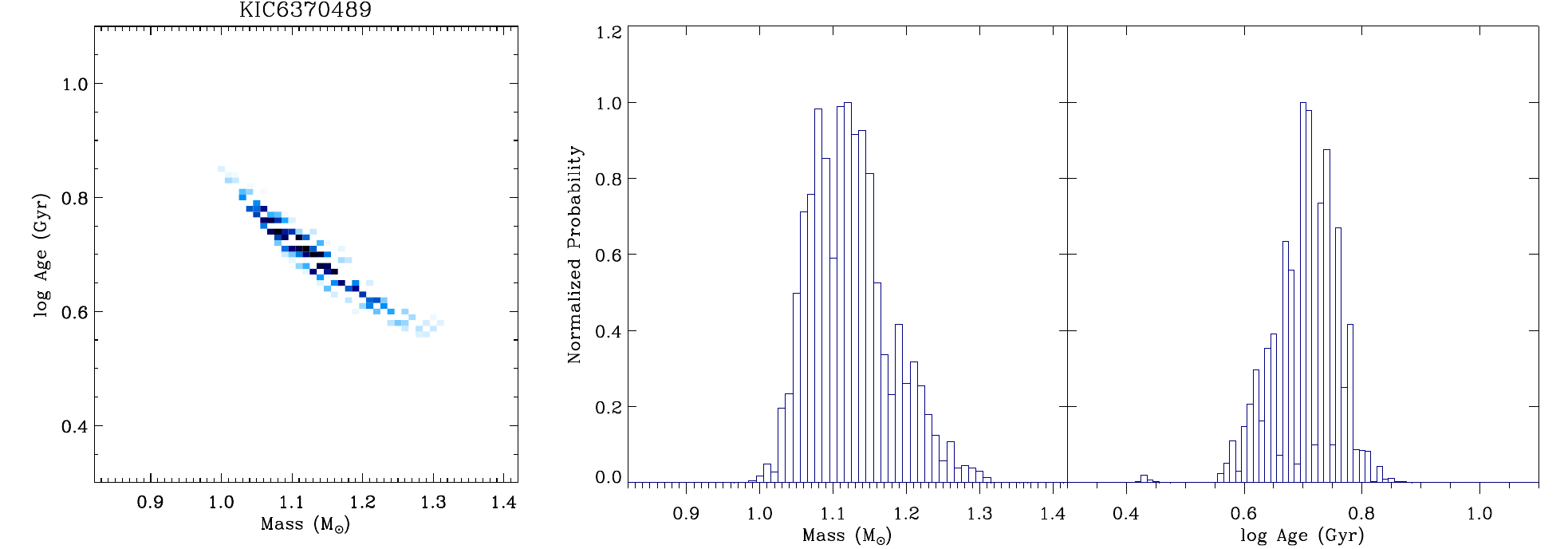}
\includegraphics[width=0.75\textwidth]{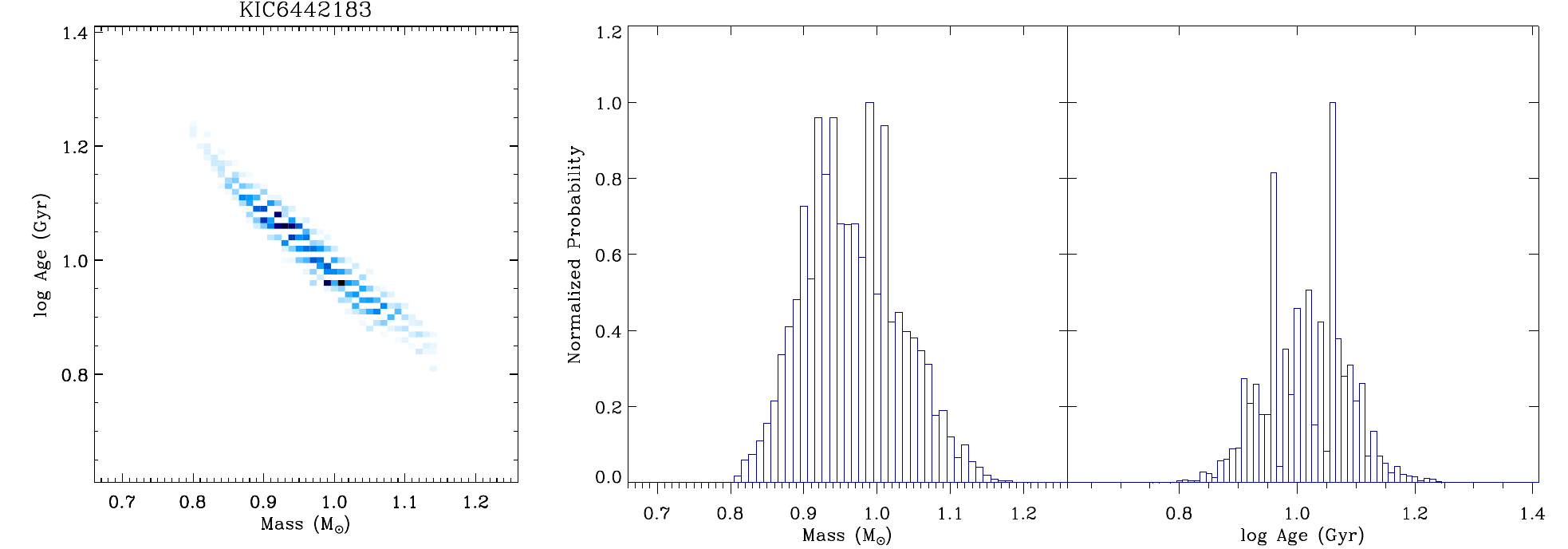}
\includegraphics[width=0.75\textwidth]{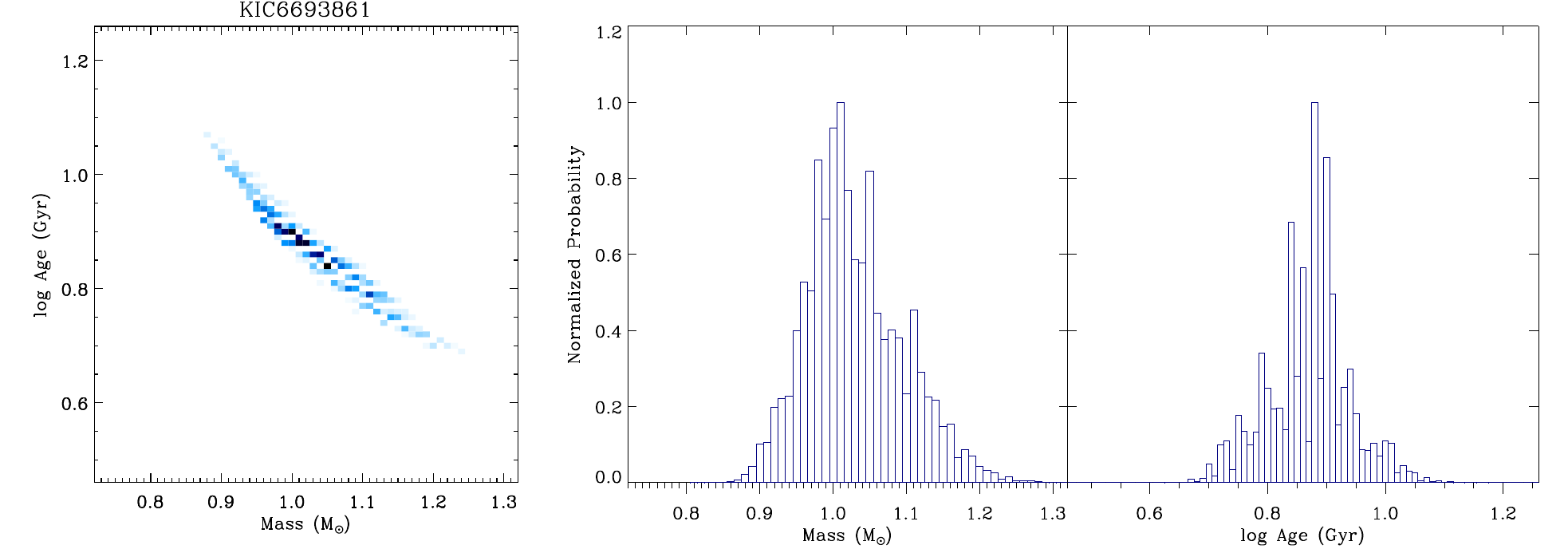}
  \label{fig:prob_2}
\end{figure}

 \begin{figure}
 \centering
\includegraphics[width=0.75\textwidth]{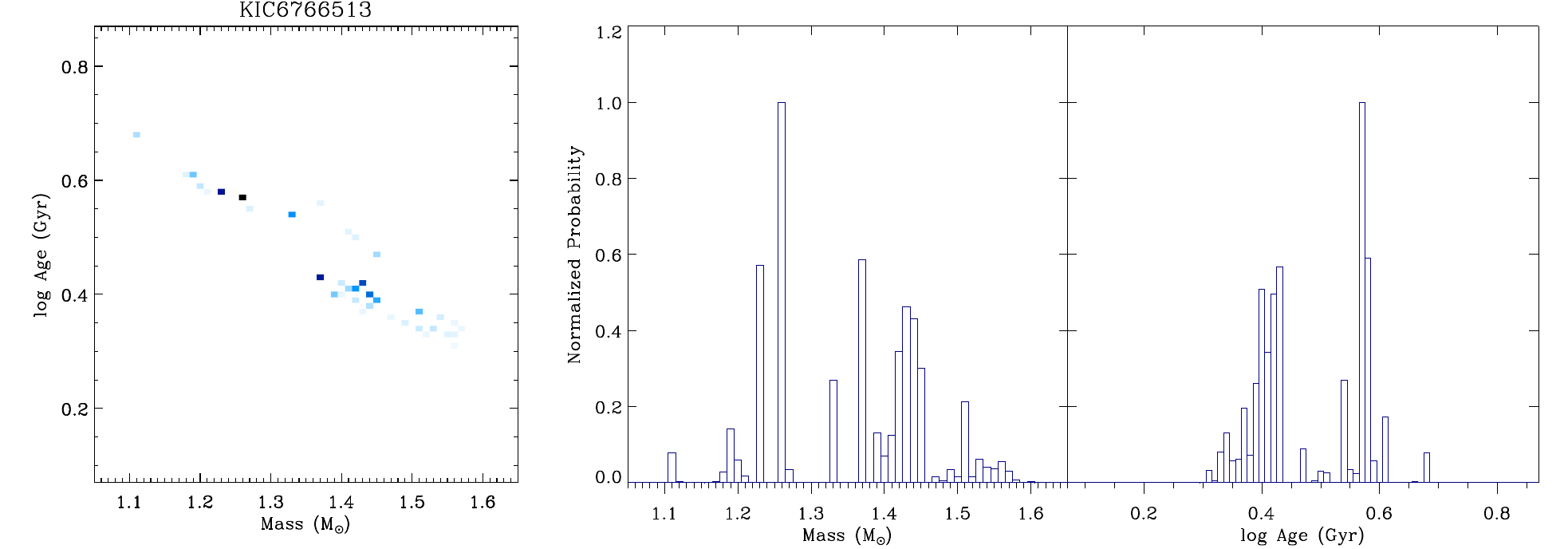}
\includegraphics[width=0.75\textwidth]{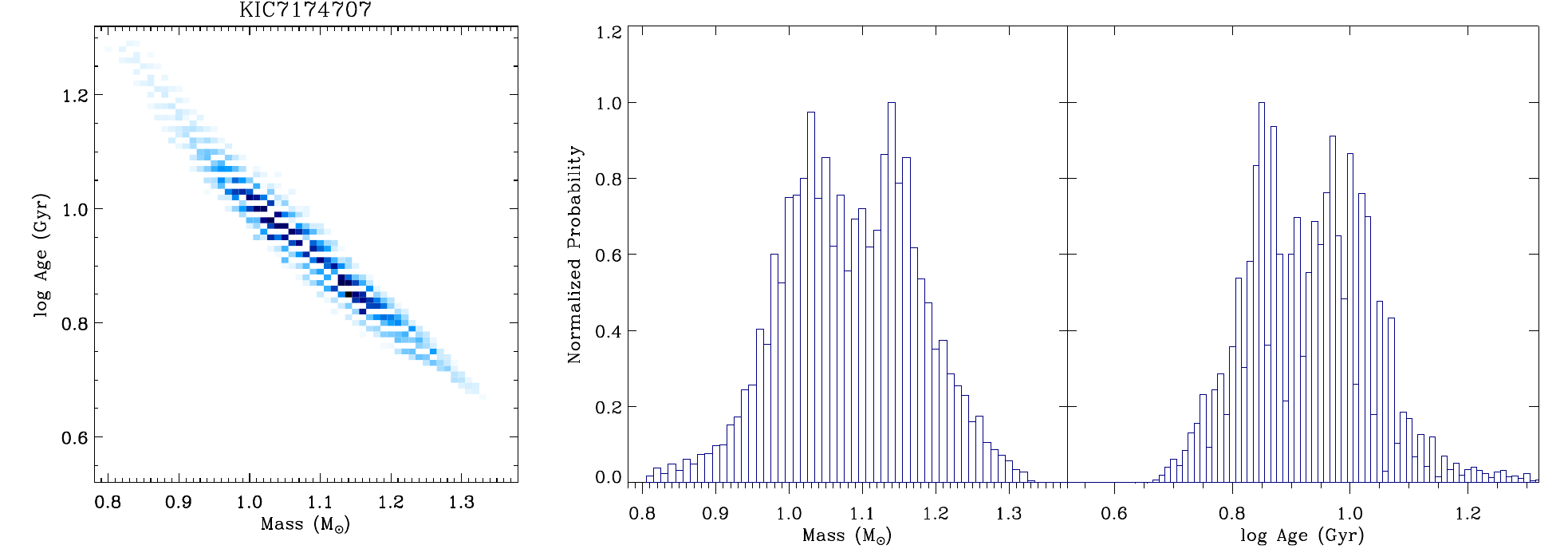}
\includegraphics[width=0.75\textwidth]{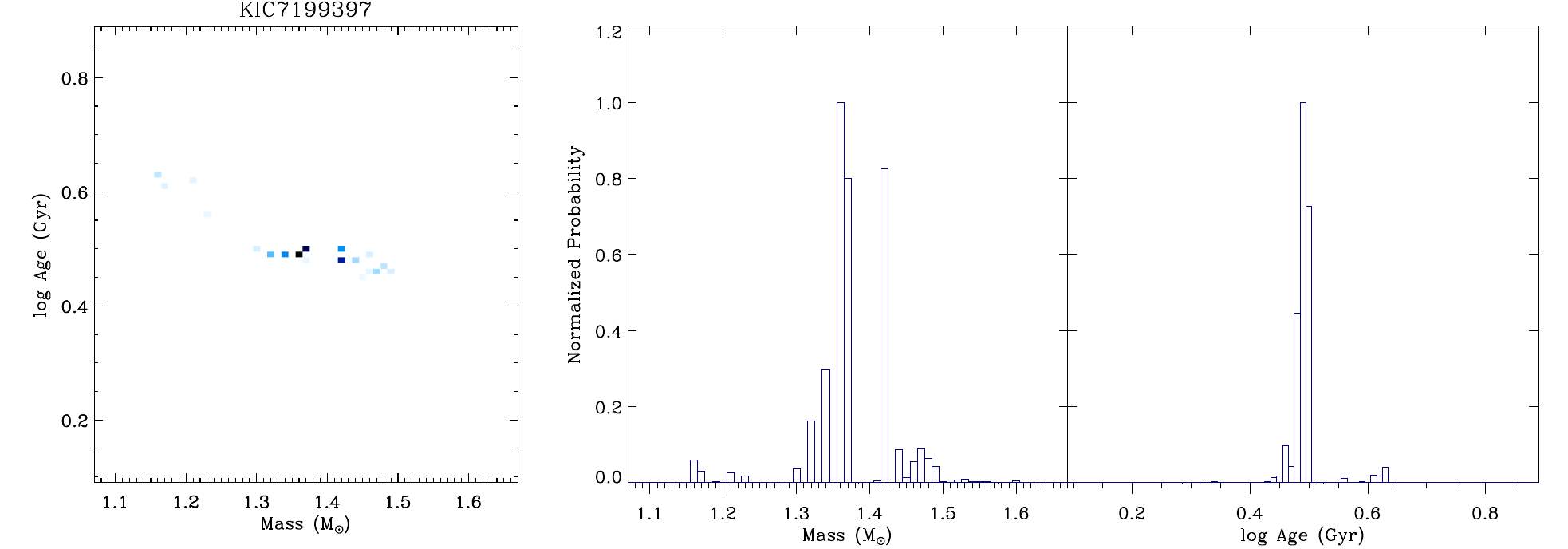}
\includegraphics[width=0.75\textwidth]{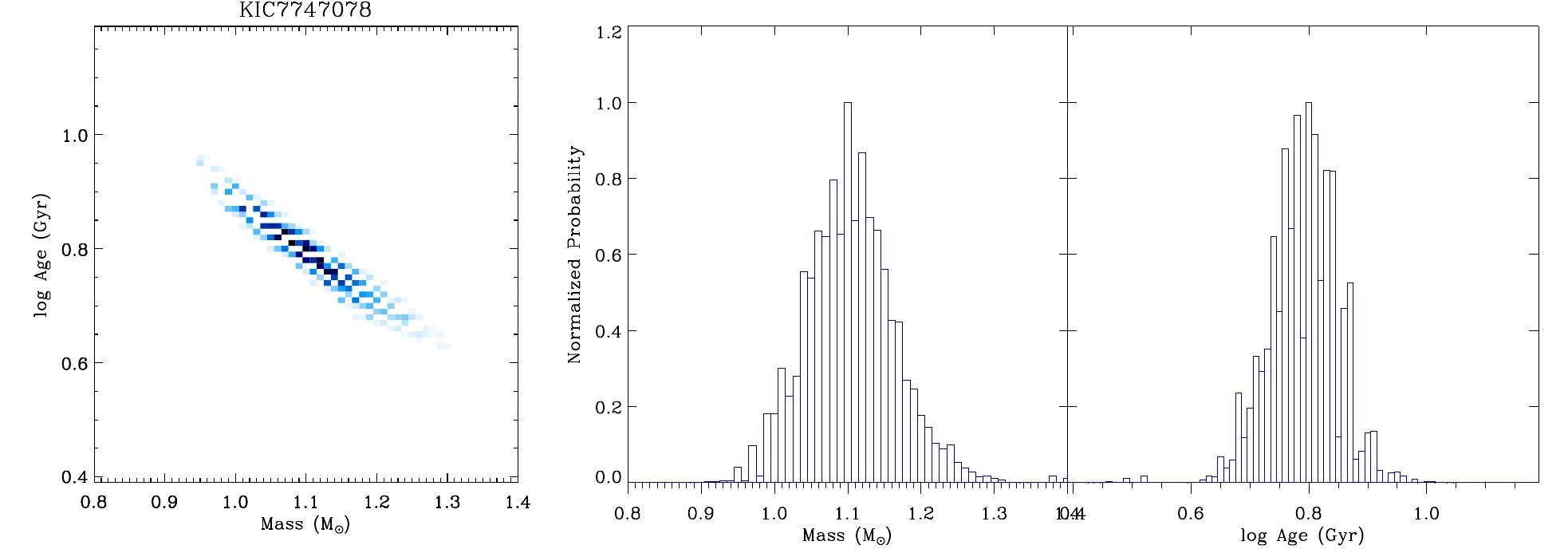}
\includegraphics[width=0.75\textwidth]{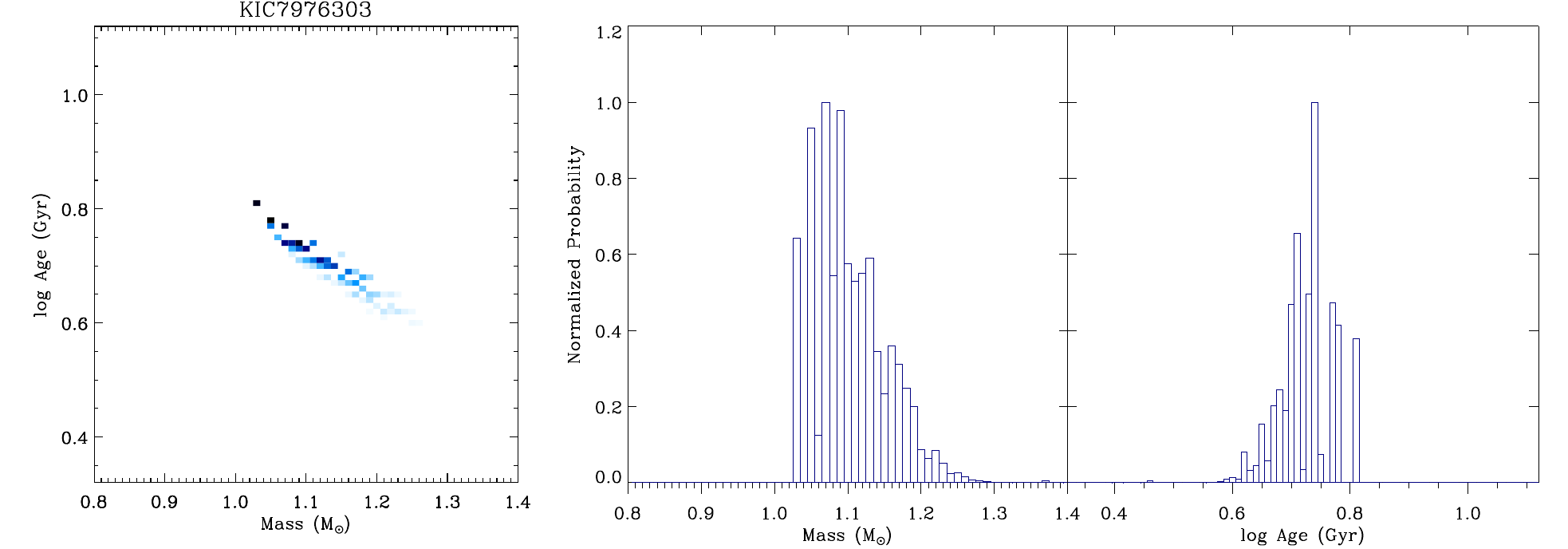}
  \label{fig:prob_3}
\end{figure}

 \begin{figure}
 \centering
\includegraphics[width=0.75\textwidth]{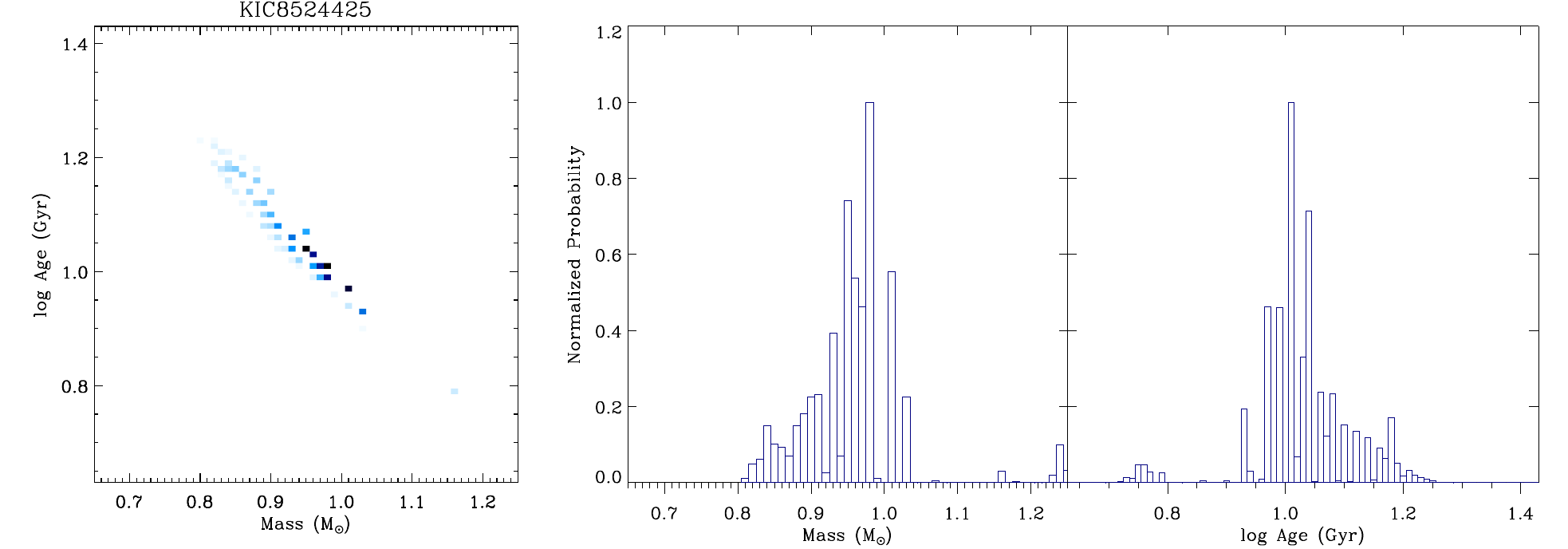}
\includegraphics[width=0.75\textwidth]{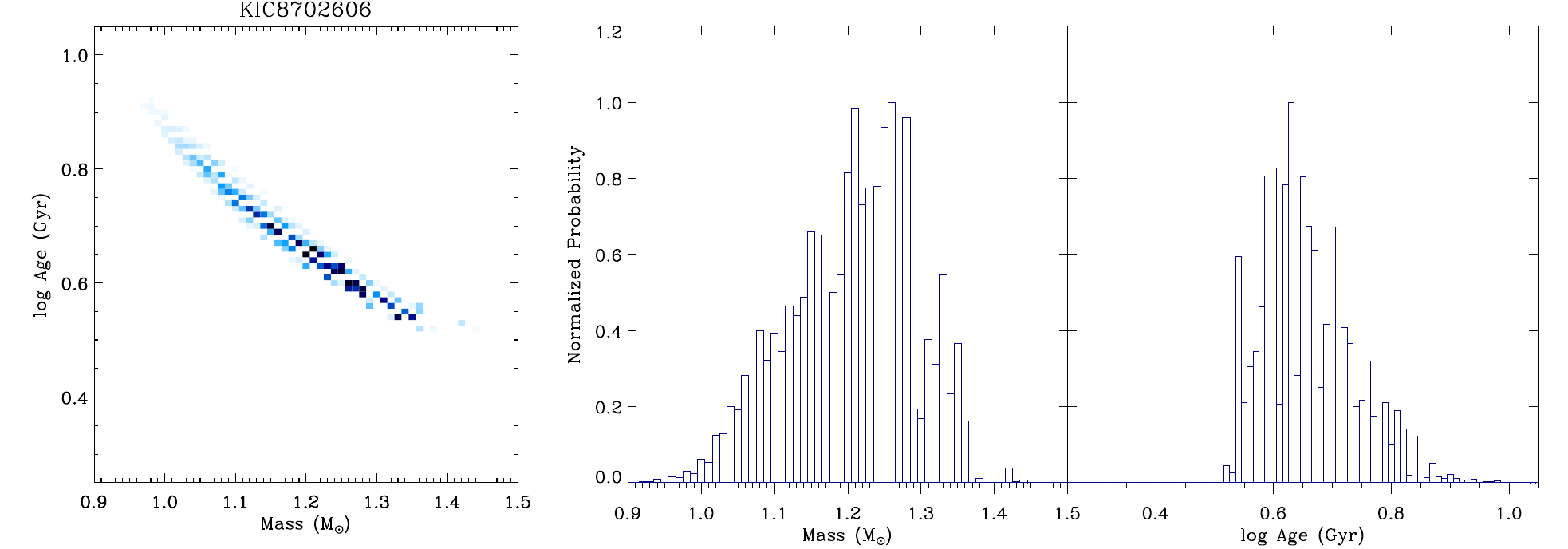}
\includegraphics[width=0.75\textwidth]{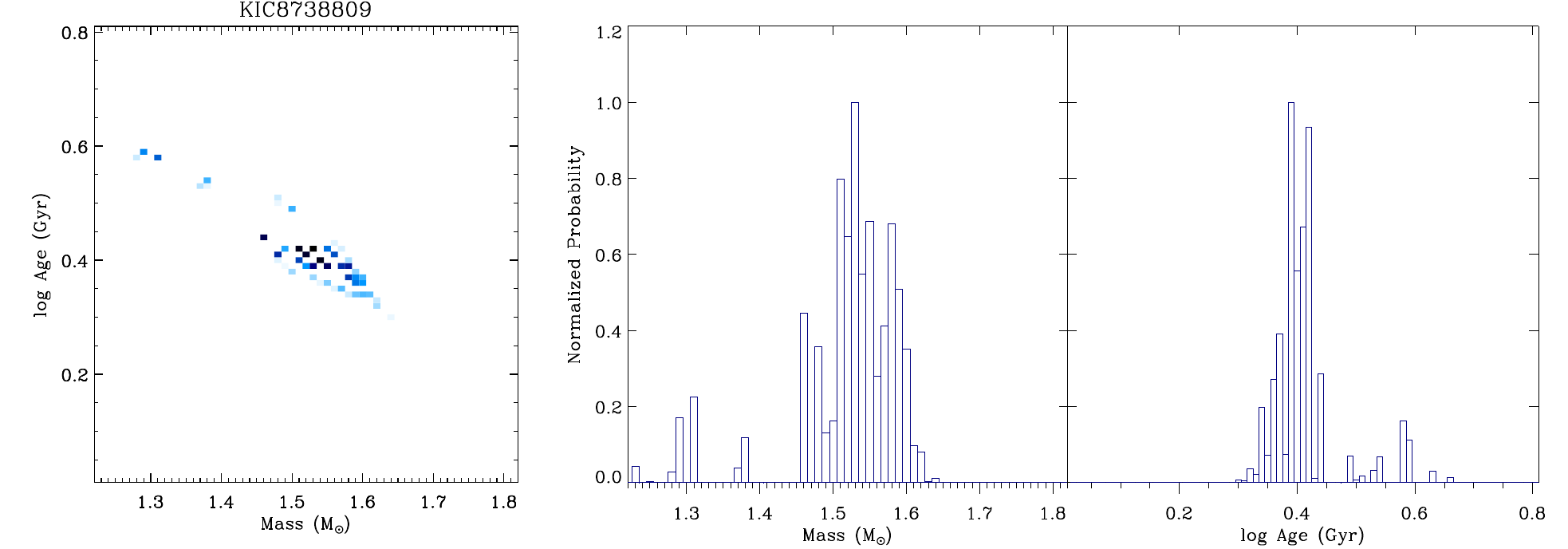}
\includegraphics[width=0.75\textwidth]{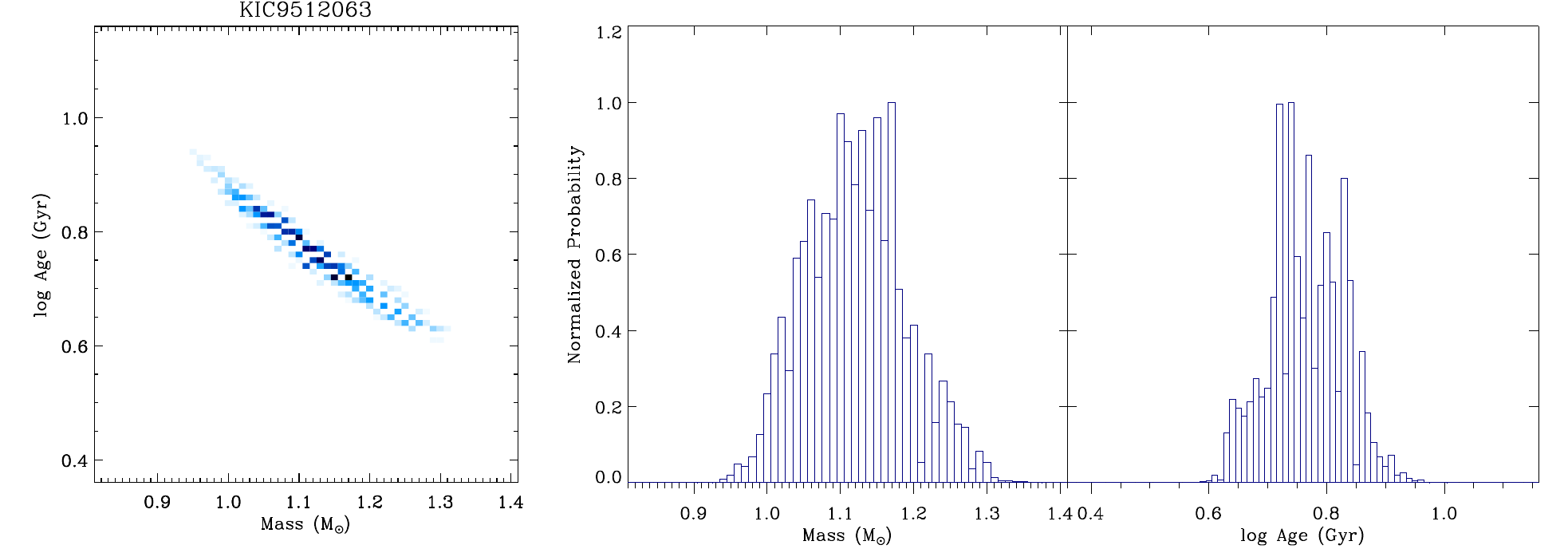}
\includegraphics[width=0.75\textwidth]{mass_age_2d_KIC10018963-eps-converted-to.pdf}
  \label{fig:prob_4}
\end{figure}

 \begin{figure}
 \centering
\includegraphics[width=0.75\textwidth]{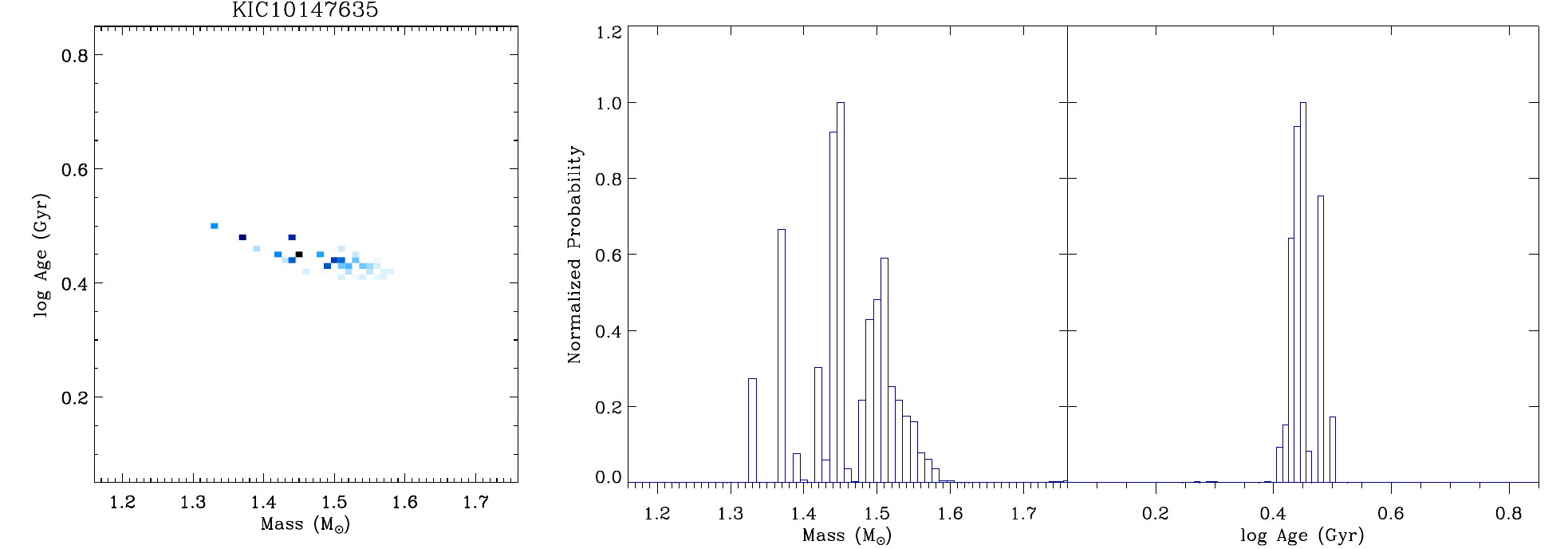}
\includegraphics[width=0.75\textwidth]{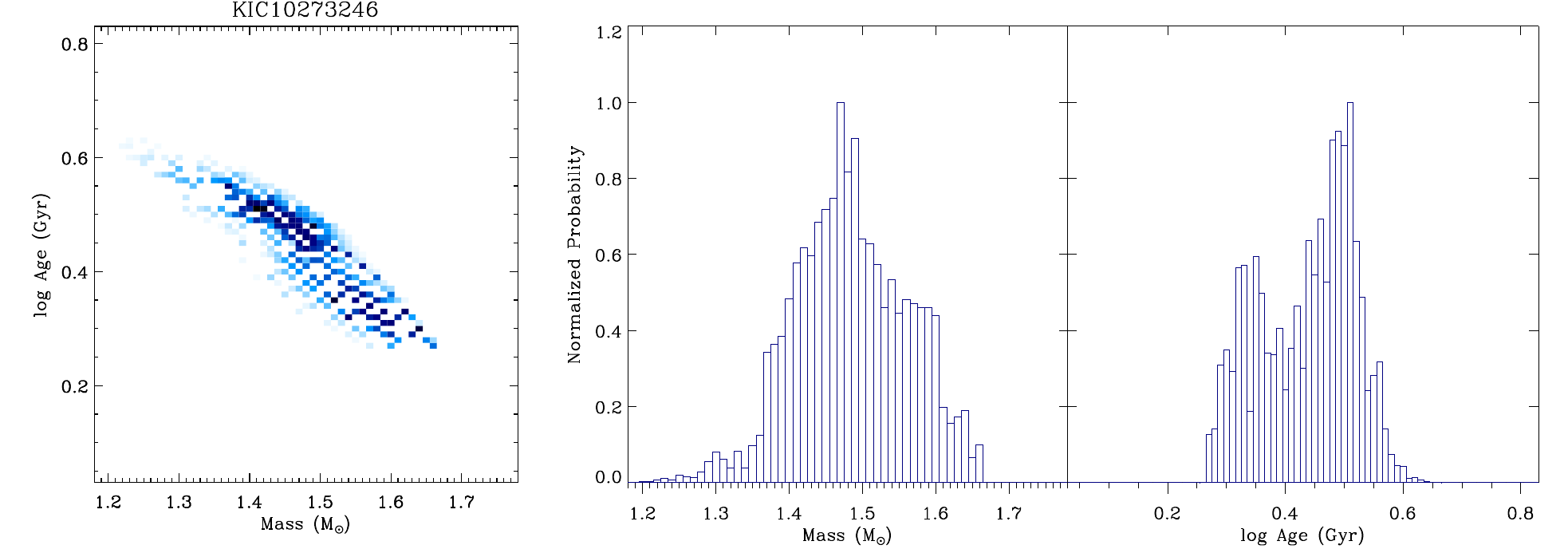}
\includegraphics[width=0.75\textwidth]{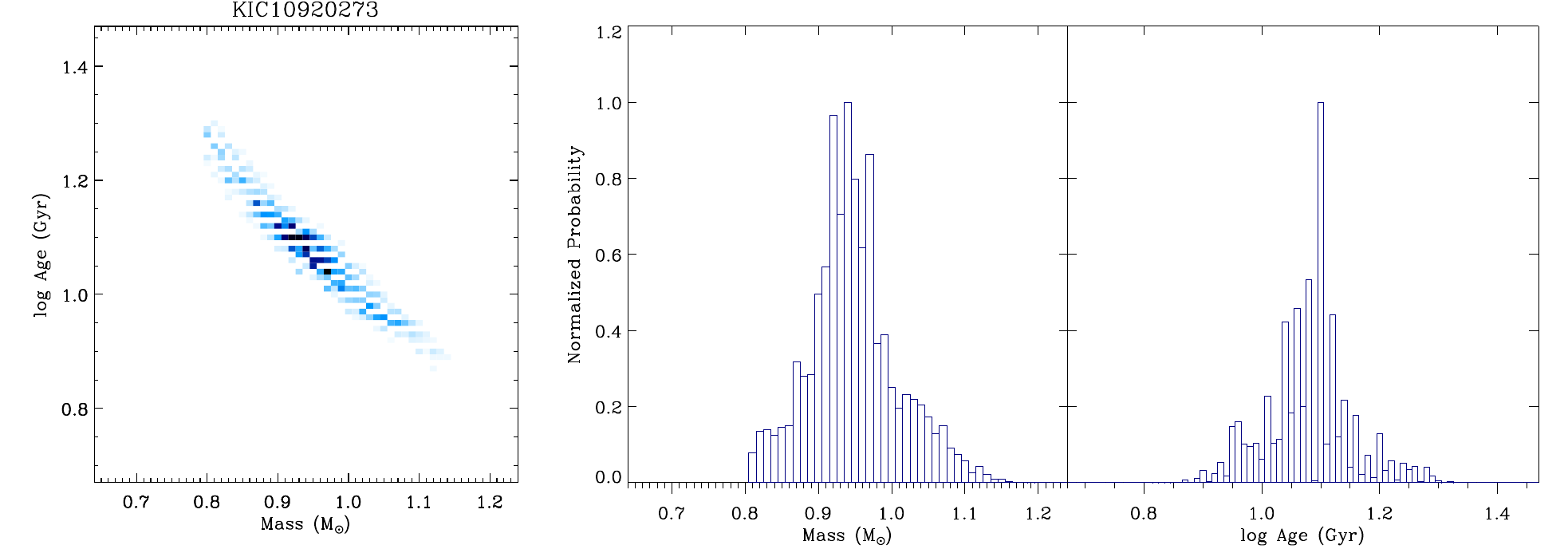}
\includegraphics[width=0.75\textwidth]{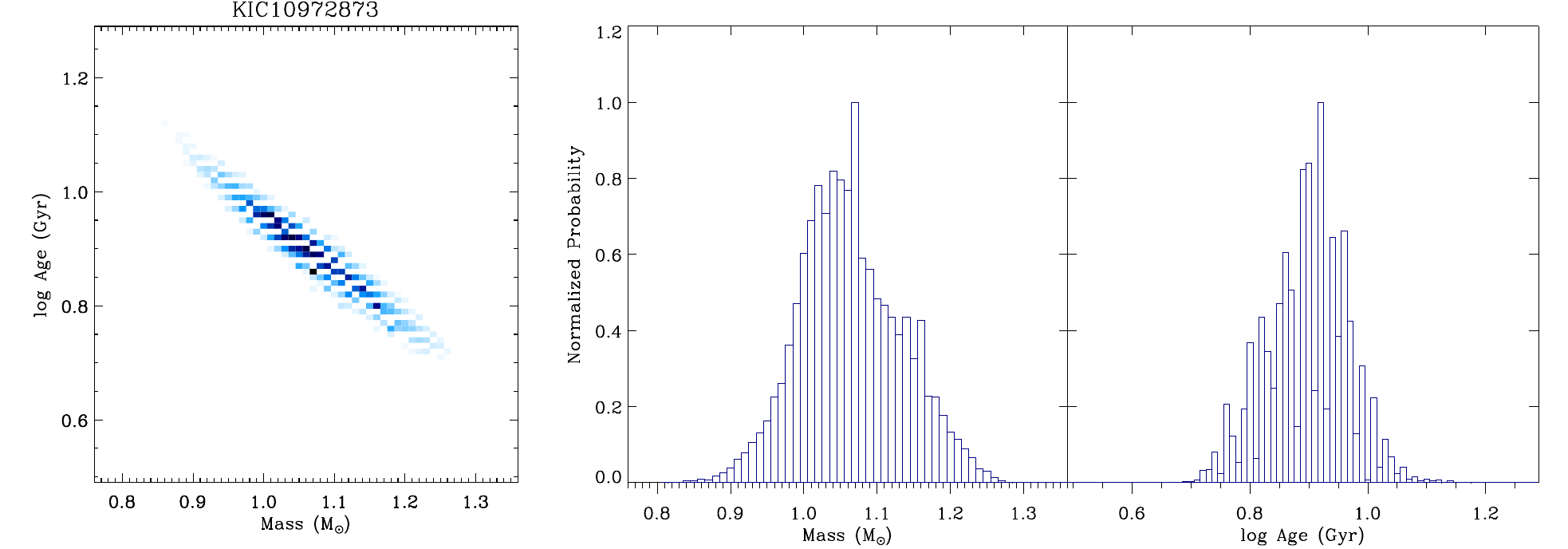}
 \includegraphics[width=0.75\textwidth]{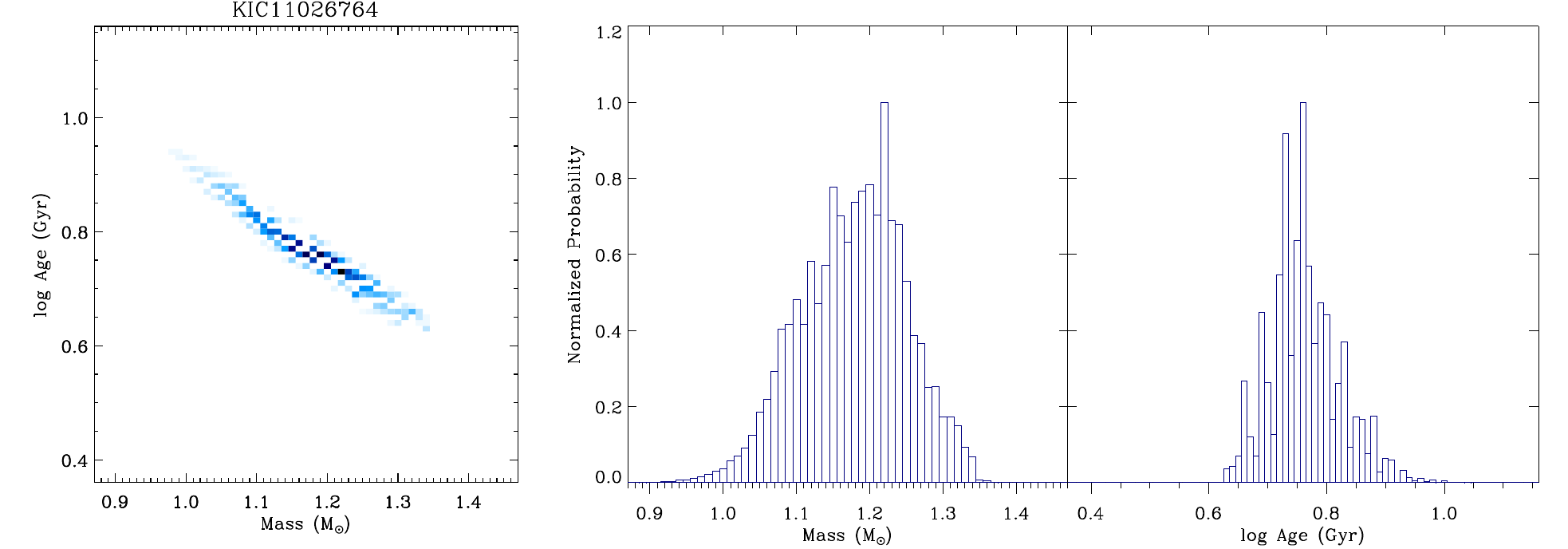}
  \label{fig:prob_5}
 \end{figure}

 \begin{figure}
 \centering
\includegraphics[width=0.75\textwidth]{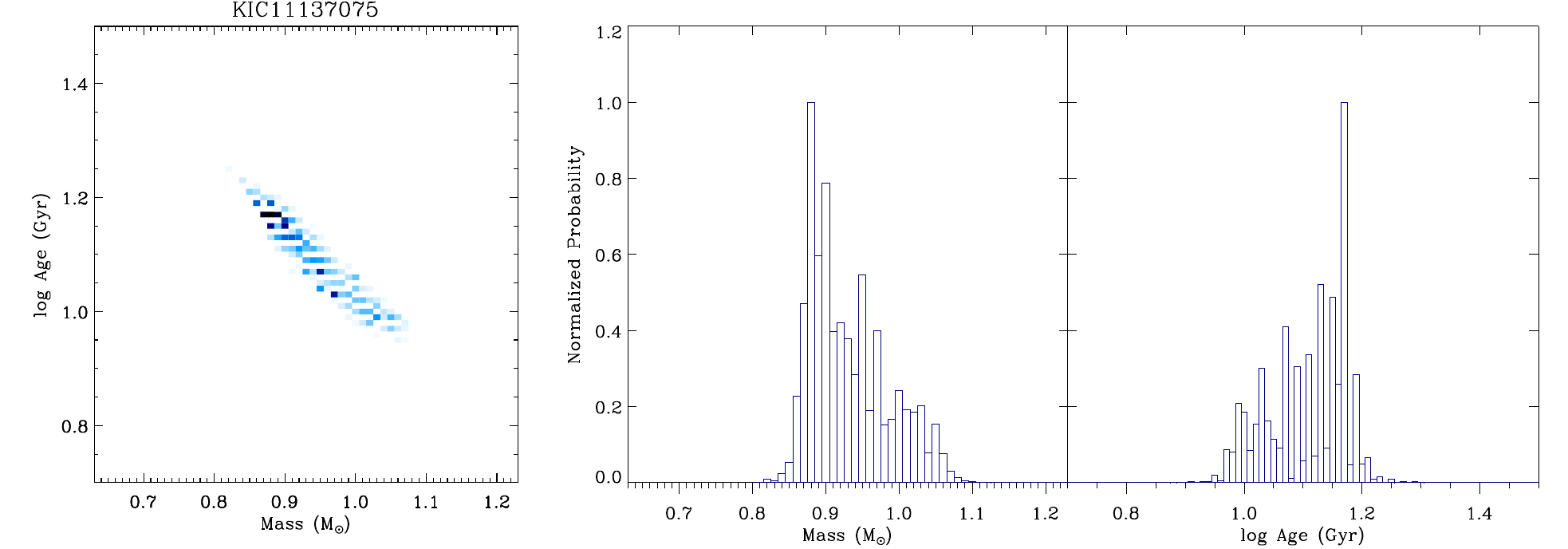}
\includegraphics[width=0.75\textwidth]{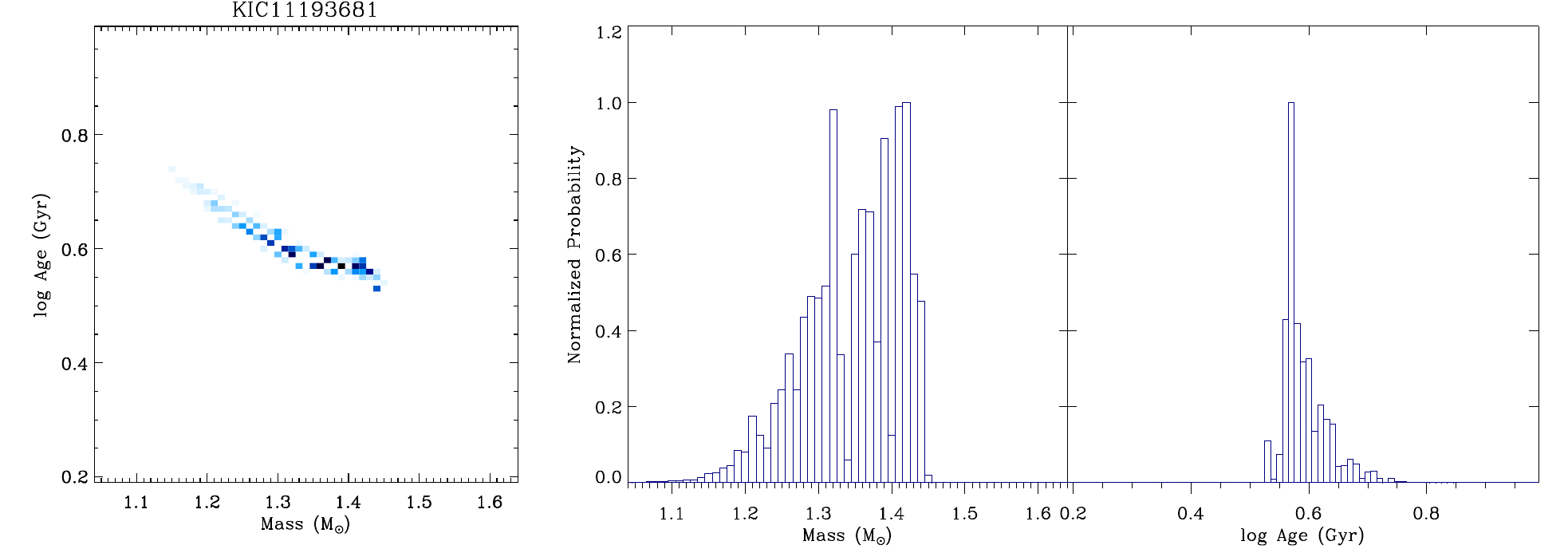}
\includegraphics[width=0.75\textwidth]{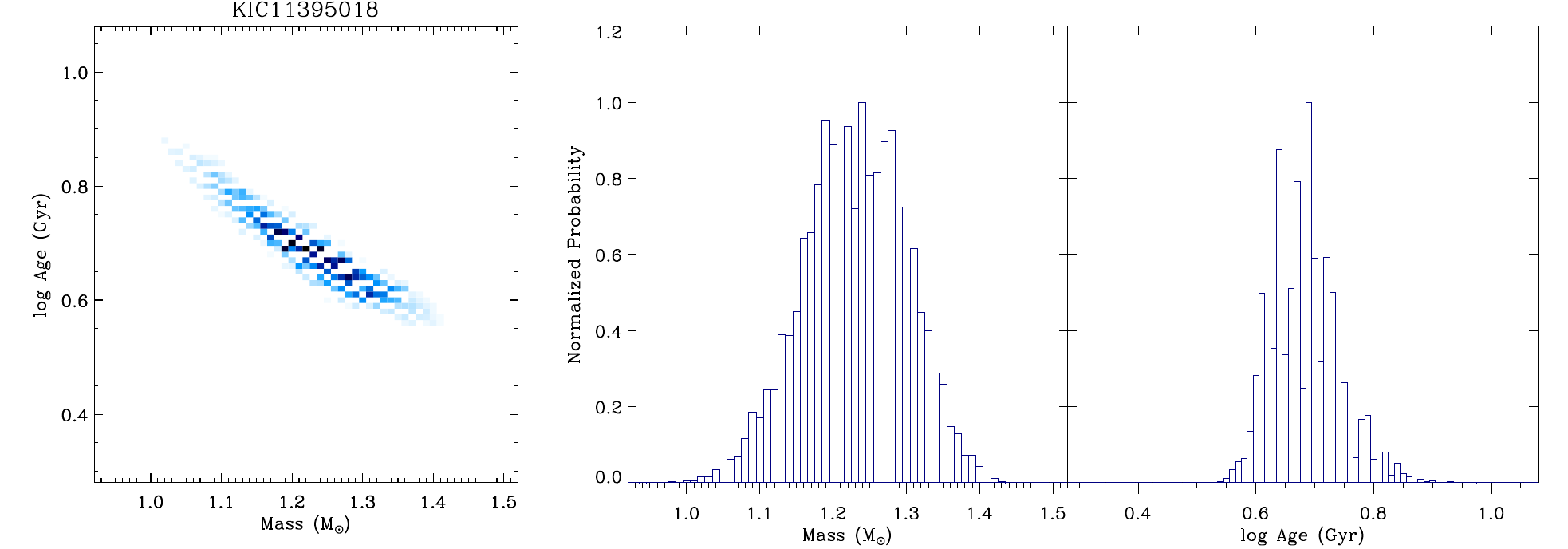}
\includegraphics[width=0.75\textwidth]{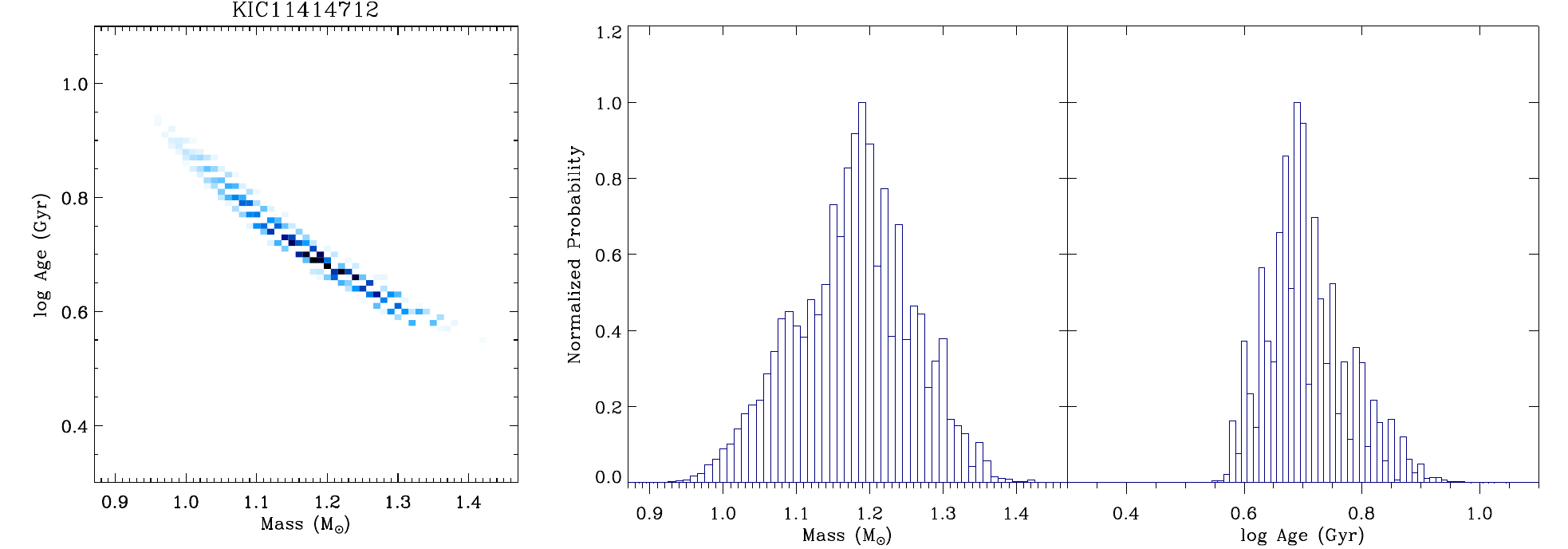}
\includegraphics[width=0.75\textwidth]{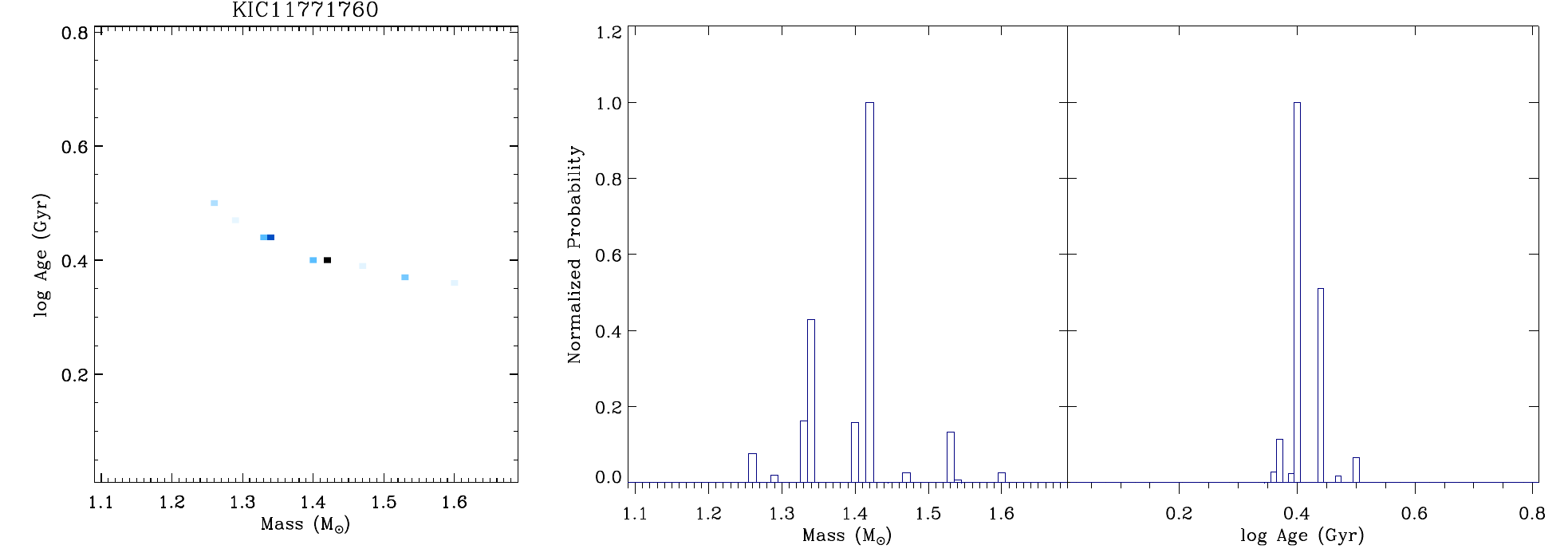}
  \label{fig:prob_6}
   \end{figure}
   
 \begin{figure}
 \centering
 \includegraphics[width=0.75\textwidth]{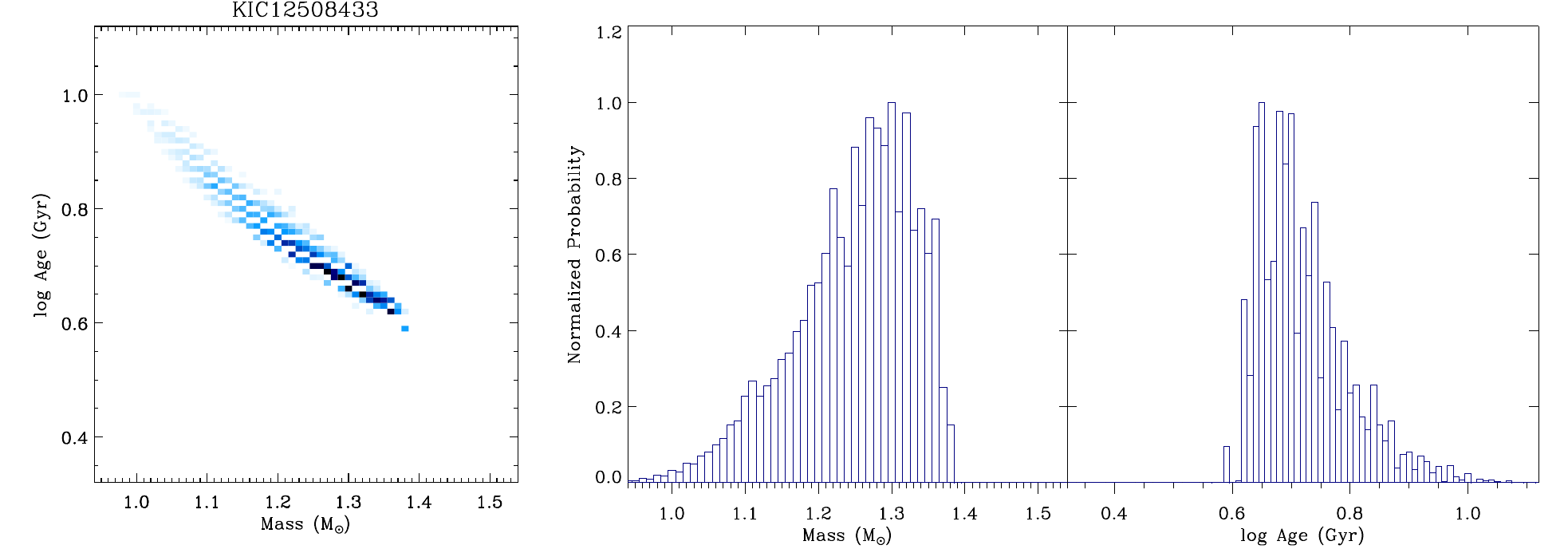}
  \label{fig:prob_7}
   \end{figure}

\clearpage


\bsp	
\label{lastpage}
\end{document}